\pdfoutput=1

\documentclass[11pt,openright,oneside,letterpaper,onecolumn]{report} 

\usepackage{graphicx}
\usepackage{verbatim}
\usepackage{amsfonts}
\usepackage{float}
\usepackage{subcaption}
\usepackage{rotating}
\usepackage[T1]{fontenc}
\usepackage[usenames,svgnames,table,dvipsnames]{xcolor}
\usepackage{amsmath,amssymb}
\usepackage{enumerate}

\newcommand{\thesistitle}{Population Genetics of Identity By Descent}
\newcommand{\thesisauthor}{Pier Francesco Palamara}
\newcommand{\thesismonth}{October}
\newcommand{\thesisyear}{2013}

\usepackage{amsmath}
\usepackage{named}
\usepackage{fancyhdr}
\usepackage{afterpage}

\usepackage[pdftitle={\thesistitle},pdfauthor={\thesisauthor},pdfpagemode={UseOutlines},bookmarks,bookmarksopen=true,pdfstartview={FitH},bookmarksnumbered=true,]{hyperref}

\newcommand{\singlespace}{\renewcommand{\baselinestretch}{1.15} \small \normalsize}

\newcommand{\doublespace}{\renewcommand{\baselinestretch}{1.5} \small \normalsize}
\newcommand{\normalspace}{\doublespace}
\newcommand{\thesistitlepage}{
    \normalspace
    \thispagestyle{empty}
    \begin{center}
        \textbf{\LARGE \thesistitle} \\[1cm]
        \textbf{\LARGE \thesisauthor} \\[8cm]
        Submitted in partial fulfillment of the \\
        requirements for the degree \\
        of Doctor of Philosophy \\
        in the Graduate School of Arts and Sciences \\[4cm]
        \textbf{\Large COLUMBIA UNIVERSITY} \\[5mm]
        \thesismonth~\thesisyear
    \end{center}
    \clearpage
}

\newcommand{\thesiscopyrightpage}{
    \thispagestyle{empty}
    \strut \vfill
    \begin{center}
      \copyright \thesisyear \\
      \thesisauthor \\
      All Rights Reserved
    \end{center}
    \cleardoublepage
}

\newcommand{\draft}{
    \renewcommand{\normalspace}{\singlespace}
    \normalspace
\clearpage }

\newcommand\independent{\protect\mathpalette{\protect\independenT}{\perp}}
\def\independenT#1#2{\mathrel{\rlap{$#1#2$}\mkern2mu{#1#2}}}

\begin{document}

\pagestyle{empty}

\thesistitlepage
\thesiscopyrightpage

    \thispagestyle{empty}
    \begin{center}
    \textbf{\LARGE ABSTRACT} \\[1cm]
     \textbf{\LARGE \thesistitle} \\[1cm]
     \textbf{\LARGE \thesisauthor} \\[1cm]
    \end{center}
    Recent improvements in high-throughput genotyping and sequencing technologies have afforded the collection of massive, genome-wide datasets of DNA information from hundreds of thousands of individuals. These datasets, in turn, provide unprecedented opportunities to reconstruct the history of human populations and detect genotype-phenotype association. Recently developed computational methods can identify long-range chromosomal segments that are identical across samples, and have been transmitted from common ancestors that lived tens to hundreds of generations in the past. These segments reveal genealogical relationships that are typically unknown to the carrying individuals. In this work, we demonstrate that such identical-by-descent (IBD) segments are informative about a number of relevant population genetics features: they enable the inference of details about past population size fluctuations, migration events, and they carry the genomic signature of natural selection.  We derive a mathematical model, based on coalescent theory, that allows for a quantitative description of IBD sharing across purportedly unrelated individuals, and develop inference procedures for the reconstruction of recent demographic events, where classical methodologies are statistically underpowered. We analyze IBD sharing in several contemporary human populations, including representative communities of the Jewish Diaspora, Kenyan Maasai samples, and individuals from several Dutch provinces, in all cases retrieving evidence of fine-scale demographic events from recent history. Finally, we expand the presented model to describe distributions for those sites in IBD shared segments that harbor mutation events, showing how these may be used for the inference of mutation rates in humans and other species.
    \cleardoublepage

\pagenumbering{roman}
\pagestyle{plain} 

\setlength{\footskip}{0.5in}

\setcounter{tocdepth}{2}
\renewcommand{\contentsname}{Table of Contents}
\tableofcontents
\cleardoublepage

\listoffigures
\cleardoublepage

\listoftables 
\cleardoublepage

~\\[1in]
\textbf{\Huge Acknowledgments}\\

\noindent 
I am deeply grateful to my advisor, Itsik Pe'er, for his endless guidance and support, for having spent five years teaching me so much, and for having given me the opportunity to work in such a stimulating environment. His enthusiasm has been a great source of inspiration. I certainly owe much to all members of the lab, past and present, and the students I worked with. I'm particularly grateful to Sasha, Snehit, Eimear, Vlada, Anat and Yufeng, for their friendship and for countless inspiring discussions. I am very grateful to my thesis committee members for their advice and the time dedicated to my work, and to the many collaborators of these years. I thank the great group of people I met back in my RoboCup days, with whom I made the first scientific steps that influenced me so much.

\vspace{4 mm}

\noindent 
To my family, Beatrice, Antonio, Gian Marco, and Susanna, for their constant support and their unlimited love.

\draft

\pagestyle{headings}
\pagenumbering{arabic}

\setlength{\textheight}{8.5in}
\setlength{\footskip}{0in}

\fancypagestyle{plain} {
\fancyhf{}
\fancyhead[C]{\thepage}
\fancyhead[C]{\itshape \leftmark}
\renewcommand{\headrulewidth}{0pt}
}
\pagestyle{plain}            

\chapter{Introduction}
\label{chap:intro}

In a famous paper published in 1965, Gordon Moore, currently co-founder and Chairman Emeritus of Intel Corporation, predicted that the number of transistors on integrated circuits would double approximately every two years, as a result of decreased production costs \cite{moore1965cramming}. During the past five decades of technological development, this prediction has been closely matched by empirical data, and Moore's law, as the conjecture is often referred to, is expected to last for a few more years. After the announced completion of the human genome project, in 2001 \cite{lander2001initial,venter2001sequence}, the development of DNA sequencing technologies has followed a similar trend, with the average cost for obtaining a full human genome DNA sequence dropping exponentially at a rate that closely matched Moore's law. With the transition from Sanger-based sequencing technologies to `next-generation' sequencing, in 2008, the cost of DNA sequencing had a further, dramatic drop, outpacing Moore's law and bringing the price of a single human genome from 2001's ${\sim}\$3$ billion to a few thousand dollars in little more than a decade \cite{NHGRI-seqCosts}.

While the speed at which large volumes of high-resolution DNA sequences are being produced exacerbates issues related to data handling (e.g. hardware storage, processing power), the availability of several fully sequenced individuals from multiple populations worldwide, together with phenotypic information, has enabled data-driven studies of the origins and diversification of human populations, including genomic signatures of evolutionary events \cite{pool2010population}, discovery of genetic markers responsible for the heritability of common traits \cite{hindorff2009potential}, and the development new tailor-made diagnostic and therapeutic tools based on an individual's genetic makeup \cite{hamburg2010path}. Achieving these tasks by analyzing such large volumes of data involves relying on statistical and computational methods to develop new specific tools that are simultaneously efficient, making minimal use of computational resources, and effective, successfully extracting and elaborating information for the question at hand. To this extent, a widespread analysis paradigm consists in working on specific ``features'', or summary statistics obtained from the DNA sequences of the analyzed cohort. These are chosen to succinctly capture the most relevant aspects of the data, while allowing efficient downstream analysis. Choosing the right genomic features is extremely important, as a particular summary statistic may not carry substantial information to address specific questions, while in other cases the relevant genomic features may be hard to access, or require intractable computational efforts.

In this thesis, we focus on developing new models and methodologies for genetic analysis that are based on a specific genomic feature that was recently made available due to technological and computational advances, namely the sharing of long-range haplotypes across purportedly unrelated individuals from a study cohort. These are chromosomal segments that are transmitted from the genome of common ancestors to sets of individuals. Such common ancestors may have lived a large number of generations in the past, so that the co-inheriting individuals may not be aware of their genetic relationship, being therefore purportedly unrelated. Since these segments are copied almost identical from the transmitting common ancestors, they are generally referred to as ``identical-by-descent'' (IBD) segments, although a small number of mutations and other rare genomic events may occur on the segments during the transmission process. The detection of IBD segments in large datasets of purportedly unrelated individuals (henceforth simply referred to as \emph{unrelateds}) was recently made possible due to (1) advances in the resolution and number of genomic sequences that modern technologies can produce (2) the development of computational methods that are able to phase (i.e. separate an individual's maternal and paternal copies of a diploid chromosome into two distinct sequences) and efficiently locate these IBD segments in a computationally tractable way. A more detailed introduction of the basic concepts underlying IBD segments is provided in Section \ref{intro:subsec:IBD_intro}, and recent review on the subject can be found in \cite{browning2012identity,thompson2013identity}.

The reminder of this chapter provides a brief overview of basic definitions and fundamental concepts of population genetics and identity-by-descent. Chapter \ref{chap:IBD_in_data} reports results of the analysis of several densely typed human datasets (HapMap 3, Jewish Hapmap), where descriptive statistics of IBD sharing across unrelateds were shown to capture relevant features of a population's recent evolutionary and demographic history. This preliminary analysis motivated investigating the formal link between IBD sharing and demographic history, which is introduced in Chapter \ref{chap:IBDmodel}, and used to infer population size fluctuations in several synthetic and real populations. In Chapter \ref{chap:migration}, the framework of Chapter \ref{chap:IBDmodel} is extended to allow for inferring recent demographic events in demographic models that include several demes, and migration across them. This extension is used to analyze recent demographic events using sequences of 250 families from several Dutch provinces (the Genome of the Netherlands Project). In Chapter \ref{chap:mutation}, the proposed model is further extended to include the occurrence of mutation events within IBD segments. These mutations are informative about the distance to transmitting common ancestors, and can be used in the study of mutation rates and several other applications. We finally provide a brief discussion of the presented work in Chapter \ref{chap:conclusions}.

\section{Population genetics}
\label{intro:sec:PopGenIntro}

Long before James Watson and Francis Crick presented the double helical structure of DNA \cite{watson1953structure}, statisticians of the past century had laid the theoretical foundations of population genetics, which is aimed at providing mathematical support to describing the dynamics of key genetic quantities resulting from the interbreeding of organisms in a sexual population. The pioneering work of Sewall Wright, John B. S. Haldane and Ronald A. Fisher, generally considered the fathers of population genetics, has now been further developed for more than a century, and theoretical predictions of these models have recently been extensively validated by empirical evidence in thousands of genome sequences from diverse populations in different species. In this section, we provide a brief introduction of the basic concepts of population genetics that will be used in the remainder of this thesis, namely the coalescent process and identity-by-descent. Comprehensive introductions to the concepts here briefly illustrated can be found in textbooks such as \cite{hartl1988primer,hartl1997principles,hein2004gene,wakeley2009coalescent}. The presented overview is in some cases a summary of the material that can be found in these books.

\subsection{Basic definitions}
\label{intro:subsec:BasicDef}

Deoxyribonucleic acid, or DNA, is hereditary material coded using an alphabet of four chemical bases: adenine (A), cytosine (C), guanine (G), and thymine (T). Each base couples with its complement (adenine with thymine and cytosine with guanine), forming \emph{base pairs} which are attached to a sugar and a phosphate molecules to form \emph{nucleotides}. A sequence of nucleotides is arranged in a double helix structure which coils around proteins called \emph{histones} to form \emph{chromosomes}, basic physical units found in the nucleus of cells. Humans have 23 such chromosomes, of which 22 are of the same kind in males and females (\emph{autosomes}), while one, the sex chromosome, may differ. Two copies of each chromosome are stored, one inherited from each parent, making humans a \emph{diploid} organism (as opposed to \emph{haploid}, where one copy of each chromosome is stored, or \emph{polyploid}, which may have multiple copies). Diploid individuals produce \emph{gametes} (egg and sperm cells) for sexual reproduction. These contain a single copy of each chromosome formed by mixing the two existing copies during the process of \emph{meiosis}. For the purpose of this thesis, two main events occurring during meiotic division will be discussed: \emph{mutation} and \emph{recombination}.

Mutation occurs when errors are randomly made during the copying of genetic material when the haploid gametes are formed. Mutations involving the change of a single base pair are called \emph{point mutations}. While mutations can occur at other stages of the cell life cycle, those occurred during the production of germ cells, which will be passed down to offsprings, are called \emph{germline} mutations. As these mutations are not present in the parents, they are often referred to as \emph{de novo} mutations. Point mutations are extremely rare, with an estimated genome wide rate of ${\sim1.1\times 10^{-8}}$ \cite{roach2010analysis} per nucleotide, per generation (with variation that may depend, among other things, on the father's age at conception \cite{kong2012rate,sun2012direct}). Since a haploid copy of the genome is composed of ${\sim}3,200,000,000$ bases (or $3.2$ giga base pairs), however, the average diploid genome is expected to harbor around $70$ de novo mutations. We note that several other types of rare alterations may occur during meiosis (e.g. insertion, deletion, inversion of genetic material), however these are not relevant for this thesis work, and will not be discussed.

Abstracting from biological mechanisms, a germ cell is created during meiosis by copying consecutive base pairs of a randomly chosen copy of each chromosome (maternal or paternal), until the chromosome end is met or a recombination event occurs. The occurrence of a recombination event between two adjacent nucleotides interrupts the copying process of the currently chosen chromosome (maternal/paternal), and starts the copying of DNA material from the other chromosome of the diploid individual (paternal/maternal) to the haploid gamete, thus potentially creating a patchwork of the original two chromosomes. In a population, recombination results in   the shuffling of genetic variation which is created by mutation events. Similarly to mutations, meiotic recombination events are rare, occurring at an average rate\footnote{average rate computed from autosomal genetic map of the $1,000$ genomes project available at \url{http://mathgen.stats.ox.ac.uk/impute/data_download_1000G_phase1_integrated_SHAPEIT2.html}} of ${\sim}1.3\times 10^{-8}$ between pairs of neighboring nucleotides. The probability of a recombination event occurring is far from uniform across the genome, as specific genomic regions may harbor increased recombination rates (hotspots), while others may have little or no recombination occurring (coldspots). The reconstruction of a mapping between physical genomic location and recombination probability (genetic map) has been extensively studied in both families and using population-level datasets of unrelated individuals \cite{hudson2001two,kong2002high}. The length of genetic maps is measured in Morgans (M), or centimorgans (cM). A centimorgan is defined as $1\%$ chance of observing a recombination event during a meiosis (one generation).

In the remainder of this work, a specific genomic location may be referred to as a \emph{site} or, equivalently, a \emph{locus} (plural: \emph{loci}), or a \emph{gene} (the latter typically indicating a region whose DNA content encodes a protein). Due to the occurrence of mutations, different versions of a locus or of a gene may exist in a population. These are referred to as \emph{alleles}. When several loci are simultaneously considered and they all belong to a single chromosome (e.g. maternal/paternal), these constitute a \emph{haplotype}. Haplotypes need not be adjacent sites, and may consist of a sparse subset of loci from a genomic region. Datasets will be distinguished in \emph{SNP array} data and \emph{whole-genome sequencing} data. SNP array data results from genotyping technologies that do not read the entire genome of the analyzed individuals, but rather only focus on subsets of genomic sites that are known to harbor single-point mutations that reached high frequency in certain human populations, and are  informative for medical genetics purposes or to discriminate genomic variation across individuals. These mutations are called \emph{SNPs}, short for \emph{single nucleotide polymorphisms}. In this work, we will ignore those rare polymorphisms for which more than two alleles are present in the population. We will only deal with polymorphisms where two alleles are present: the \emph{wild type}, or the \emph{reference allele}, and the mutated allele. Whole-genome sequencing data results from the more recent high-throughput sequencing technologies, and typically results in the complete reading of a human genome. It is to be noted that both genotyping and sequencing technologies typically do not provide information on the maternal/paternal haplotypes of the analyzed individuals. Rather, for a biallelic locus they provide a \emph{genotype}, i.e. the count of nucleotide copies that differ from the human genome reference sequence at a specific location. For a diploid individual, these counts take values $0, 1$ or $2$. The process of reconstructing haplotypes from genotype information is called \emph{phasing}, or \emph{haplotyping}. While phasing approaches are not directly discussed in this work, the ability to correctly phase genotypes into haplotypes is fairly important for the material presented in this thesis, and a review of methods for computational phasing can be found in \cite{browning2011haplotype}.

\subsection{Population models}
\label{intro:subsec:PopModels}

The distribution of genetic variability found in modern day populations is strongly influenced by demographic history. Events such as migrations and population size fluctuations determine the rate at which new mutations spread, and the frequencies of these mutations may differ substantially across different cohorts. Several idealized populations models have been developed in order to study quantities such as the frequency and distribution of genetic variation. In all cases, the goal is to simplify the relevant biological processes to achieve mathematical tractability while maintaining the highest level of realism.

The Wright-Fisher model \cite{fisher1930genetical,wright1931evolution} is arguably the most important and widely used population model. A number of assumptions are made in a Wright-Fisher population:

\begin{enumerate}
\item Generations do not overlap. All individuals in the population die at the same time, and a new generation is created.
\item The population size remains constant in time. At each generation the number of individuals is the same as in the previous generation.
\item All individuals in a population have a single chromosome, and do not need another individual to reproduce to the next generation (asexual, haploid).
\item There is no recombination (or only one site is considered). When reproduction occurs, the entire genetic material is copied to an individual of the next generation.
\item Equality of fitness and lack of population structure. At each generation an individual may reproduce to the next generation with the same probability of all other individuals.
\end{enumerate}

\begin{figure*}
    \centering
    \begin{subfigure}[b]{0.32\textwidth}
            \centering
            \includegraphics[width=\textwidth]{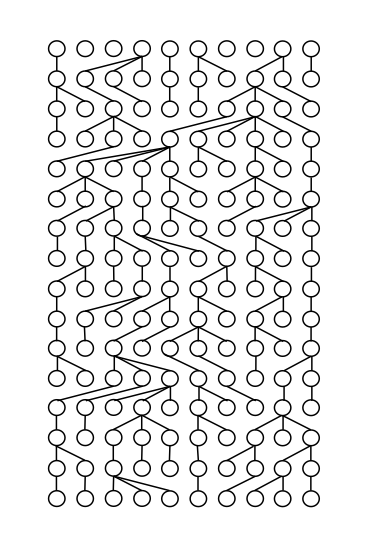}
            \caption{A genealogy in the Wright-Fisher model.}
            \label{fig:intro:WF2}
    \end{subfigure}
    ~
    ~
    ~
    ~
    ~
    \begin{subfigure}[b]{0.32\textwidth}
            \centering
            \includegraphics[width=\textwidth]{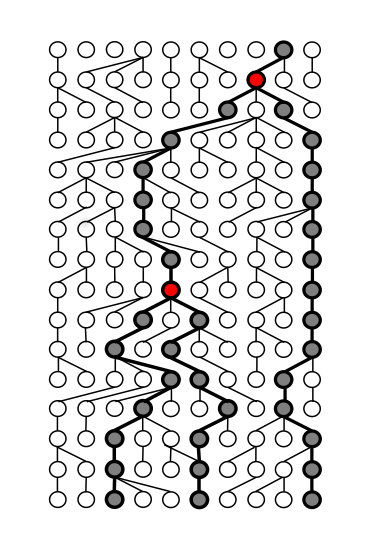}
            \caption{Highlighting the ancestry of three samples.}
            \label{fig:intro:WF3}
    \end{subfigure}
    ~
	\caption{Details from a sample genealogy in a Wright-Fisher model (figures adapted from \protect\cite{hein2004gene}).}
	\label{fig:intro:WF}
\end{figure*}

To create a new generation for a population of size $N$, an individual is sampled from the previous generation, with replacement. The sampling is repeated $N$ times, until the new generation is fully defined, and the previous generation dies. An example of this process is shown in Figure \ref{fig:intro:WF}. This model allows calculating several quantities of interest. First, it is possible to compute a distribution for the number of offspring that an individual has in the following generation. Since all individuals have the same chance $1/N$ of being chosen at each draw of a new individual for the next generation, the distribution for the number $n$ of offspring at the next generation will be binomial, with mean $1/N$:

\begin{equation}
\begin{split}
P(n = k|N) = {N \choose k} \left(\frac{1}{N}\right)^k \left(1-\frac{1}{N}\right)^{N-k}
\end{split}
\label{eq:intro:binomialchildren}
\end{equation}

It follows that the expectation and variance for the number of offspring of an individual are

\begin{equation}
\begin{split}
\operatorname{E}[n|N]=N\left(\frac{1}{N}\right)=1
\end{split}
\label{eq:intro:expChildren}
\end{equation}
\begin{equation}
\begin{split}
\operatorname{Var}[n|N]=N\left(\frac{1}{N}\right)\left(1-\frac{1}{N}\right)=1-\frac{1}{N}
\end{split}
\label{eq:intro:varChildren}
\end{equation}

A multinomial distribution can be used to model the joint distribution for the number of children of two individuals from a population, and, using standard properties of multinomials, we can obtain their 
covariance as

\begin{equation}
\begin{split}
\operatorname{Cov}[n_1n_2|N]=-N\left(\frac{1}{N}\right)^2=-\frac{1}{N}
\end{split}
\label{eq:intro:covChildren}
\end{equation}

As expected the covariance decreases as $N$ is increased, since an individual that has a large number of children does not strongly affect the number of children another individual may have if the population is not constrained to be small. If an allele is carried by $i$ individuals in the population, the chance of finding $k$ copies at the next generation can be computed using analogous reasoning, but the probability of a single draw in the binomial distribution will now be $p=\frac{i}{N}$. The expectation and variance will therefore be 

\begin{equation}
\begin{split}
\operatorname{E}[n|N,i]=N\left(\frac{i}{N}\right)=i
\end{split}
\label{eq:intro:expChildrenFreq}
\end{equation}
\begin{equation}
\begin{split}
\operatorname{Var}[n|N,i]=N\left(\frac{i}{N}\right)\left(1-\frac{i}{N}\right)=i\left(1-\frac{i}{N}\right)
\end{split}
\label{eq:intro:varChildrenFreq}
\end{equation}

One important quantity that may be calculated in this model is the probability that out of two sampled individuals one carries an allele and the other does not, given that the population is of size $N$ and that the allele has frequency $i$. Under the assumption that the two chromosome copies of a diploid individual are randomly sampled from a population of haploid individuals, this is the probability of finding a heterozygous site along the genome of an individual (\emph{heterozygosity}). Assuming an allele can be of the kinds \emph{A} or \emph{a}, and that there are $i$ copies of the allele \emph{A} at generation $0$, the initial frequency of \emph{A} is $p_0 = i/N$, and the chance of sampling (with replacement) two different copies out of $N$ individuals is

\begin{equation}
\begin{split}
H_0 = P(A,a|N,i)+P(a,A|N,i) = 2p_0(1-p_0)
\end{split}
\label{eq:intro:heterozygosity0}
\end{equation}

Using the random variable $P_1$ to represent the frequency of the allele at generation $1$, the expected heterozygosity at the next generation can be computed using equations \ref{eq:intro:expChildrenFreq} and \ref{eq:intro:varChildrenFreq}

\begin{equation}
\begin{split}
\operatorname{E}[H_{1}|N,i] &= \operatorname{E}[2P_{1}(1-P_{1})] \\
&=2(\operatorname{E}[P_{1}]-\operatorname{E}[P_{1}^2]) \\
&=2(\operatorname{E}[P_{1}]-\operatorname{E}[P_{1}]^2-\operatorname{Var}[P_{1}]) \\
&=2p_{0}(1-p_{0})\left(1-\frac{1}{N}\right) \\
&=H_{0}\left(1-\frac{1}{N}\right)
\end{split}
\label{eq:intro:heterozygosity1}
\end{equation}

Indicating that heterozygosity is expected to decrease, and it is expected to do so faster in small populations. Note that, applying the result of equation \ref{eq:intro:heterozygosity1} recursively for $g$ generations, heterozygosity is expected to have an exponential decay

\begin{equation}
\begin{split}
\operatorname{E}[H_{g}|N,i] &=H_{0}\left(1-\frac{1}{N}\right)^g \approx H_0 \ e^{-g/N}
\end{split}
\label{eq:intro:heterozygosityG}
\end{equation}

We note that while the Wright-Fisher model is the most widely adopted, other models have been proposed and are in some cases more convenient in terms of realism or mathematical tractability. One notable example is the Moran model \cite{moran1958random,moran1962statistical}, which will be however omitted as not relevant for this work.

\subsection{The coalescent}
\label{intro:subsec:Coalescent}

In a series of papers published in $1982$, Kingman has shown that a stochastic process named \emph{the coalescent} is able to describe the genealogical dynamics emerging from several idealized population models, including the Wright-Fisher model \cite{kingman1982coalescent,kingman1982genealogy,kingman1982exchangeability}. In the coalescent, the ancestral lineages of a set of considered individuals from a population are traced backwards in time, allowing for a quantitative description of key genealogical events that only requires keeping track of such subset of lineages.

\subsubsection{The basic coalescent}

If we trace the ancestral lineages of two individuals from a Wright-Fisher population back in time, repeatedly sampling a random ancestor from the previous generation, a common ancestor will be found when both individuals happen to sample the same parent (i.e. these lineages \emph{coalesce}, as in the example of Figure \ref{fig:intro:WF3}). The chance a parent is chosen by one of the individuals is $N^{-1}$, and since both individuals choose independently, the chance both individuals choose the same parent is $N^{-2}$. Since there are $N$ parents to choose from, the chance a common ancestor will be found at a given generation is $N \times N^{-2} = N^{-1}$. The waiting time (in generations) to the most recent common ancestor (\emph{TMRCA}) can therefore be expressed using a geometric distribution with parameter $N^{-1}$

\begin{equation}
\begin{split}
P(g=k|N) = \left(1-\frac{1}{N}\right)^{k-1}\frac{1}{N}
\end{split}
\label{eq:intro:coalGeom}
\end{equation}

If we are tracing $n$ individuals from the current generation, a total of $n \choose 2$ pairs of ancestral lineages are followed, and we are interested in the time to the first coalescence of such lineages. The chance that no coalescence occurs during one generation is now

\begin{equation}
\begin{split}
\frac{N-1}{N}\frac{N-2}{N}\dots\frac{N-n+1}{N} = \prod_{i=1}^{n-1} \left(1-\frac{i}{N}\right)=1-\sum_{j=1}^{n-1}\frac{j}{N}+\mathcal{O}\left({\frac{1}{N^2}}\right)
\end{split}
\end{equation}

If we ignore the term in $\mathcal{O}\left({\frac{1}{N^2}}\right)$, which is negligible for large population sizes, the probability of a coalescent event in the previous generation is then ${n \choose 2} \frac{1}{N}$ and again, using a geometric distribution

\begin{equation}
\begin{split}
P(g=k|N) \approx \left[1-{n \choose 2} \frac{1}{N}\right]^{k-1} {n \choose 2} \frac{1}{N}
\end{split}
\end{equation}

Note that we can switch to a continuous time approximation, using the exponential distribution in lieu of the geometric distribution.

\begin{equation}
\begin{split}
P(T=t|N) \approx {n \choose 2} \frac{1}{N} e^{-{n \choose 2} \frac{1}{N}}
\end{split}
\end{equation}

Simulating the genealogy for a sample of $n$ individuals in a population of size $N$ is easy using this formulation, and it involves repeatedly sampling coalescent times from exponential distributions with parameters ${n_t \choose 2} \frac{1}{N}$, for $n_t=n,n-1,\dots,2$, reflecting the decreasing number of ancestral lineages as pairs of individuals find common ancestors.

Again, a number of relevant genealogical quantities can be expressed in this model. Since we are using an exponential distribution, the expected time to the first coalescence event for these samples is $\operatorname{E}[T_n]=\left[{n \choose 2}\frac{1}{N}\right]^{-1}=\frac{2N}{n(n-1)}$, and the variance is $\operatorname{Var}[T_n]=\left[{n \choose 2 }\frac{1}{N}\right]^{-2}=\frac{4N^2}{n^2(n-1)^2}$. We can now compute the expected TMRCA for all these samples by summing the expected times for the occurrence of $n-1$ coalescence events, which occur with linearly decreasing rate as pairs of lineages coalesce

\begin{equation}
\begin{split}
\operatorname{E}[T] = \sum_{i=2}^{n}\operatorname{E}[T_i]=2N\sum_{i=2}^{n} \frac{1}{i(i-1)}=2N\left(1-\frac{1}{n}\right)
\end{split}
\end{equation}

And the variance can be similarly computed by summing the independent variances of each coalescence event. Using similar principles, we may also compute the expected total branch length for a tree representing the genealogy of these samples as

\begin{equation}
\begin{split}
\operatorname{E}[L] = \sum_{i=2}^{n}i\operatorname{E}[T_i]=2N \sum_{i=1}^{n-1} \frac{1}{i} \approx 2N\log{n}
\end{split}
\label{eq:intro:expLenTree}
\end{equation}

\subsubsection{The coalescent with mutation}
\label{subsubsec:intro:coalMut}

The coalescent process is suitable to include mutation events, and therefore study the distribution of genetic variation in idealized populations. The \emph{infinite sites assumption}, due to Motoo Kimura \cite{kimura1969number}, allows to simplify calculations in this context. Under the infinite sites assumption, whenever a mutation occurs, it always results in a new mutated site (i.e. it is impossible that a site that is already mutated in the population mutates again). Since the chance of two mutations affecting the same site is inversely proportional to the number of available sites, assuming an extremely large genome results in an infinitesimal probability for this event. This assumption is justified by the observation that the number of mutated sites in human populations is relatively small compared to the number of available sites in the genome (i.e. human DNA sequences are largely identical).

Consider $n$ individuals who have a genome of $s$ sites, and a genealogical tree of total length $L$ representing the coalescent history of these individuals along their entire sequence (as later described, the occurrence of recombination events may result in different trees for different genomic regions, but no recombination is assumed for now). A mutation may occur independently, with a small probability $\mu$ at any meiotic copy of each nucleotide. In this scenario, the total number of mutation events can be modeled as a Poisson distributed random variable, with mean $\mu sL$. Furthermore, due to the infinite sites assumption, each mutation event occurring along the genealogy is harbored by a distinct site. The number of total mutation events will therefore be equivalent to the number of mutated sites. Using Equation \ref{eq:intro:expLenTree} to express the expected volume of the genealogical tree and defining $\theta = 2N\mu$, the following estimator can be obtained:

\begin{equation}
\begin{split}
\Hat{\theta}_s = \frac{m}{\sum_{i=1}^{n-1} \frac{1}{i}}
\end{split}
\label{eq:intro:watterson}
\end{equation}

Where $m$ is the observed number of mutated sites in the analyzed sample. The parameter $\theta$ is referred to as the scaled mutation rate, as it includes the value of the population size $N$. Such an estimator, often referred to as Watterson's estimator, allows inferring the size of the population based on the observed number of mutated sites in a group of sequences, assuming a Wright-Fisher population model, and for a given value of $\mu$. Because a Wright-Fisher model is only approximating the real genealogical process that results in the observed distribution of mutation events, the recovered population size has to be viewed as a projection of the true genealogical process onto the idealized Wright-Fisher population. A population size inferred using similar estimators is generally referred to as \emph{effective population size} (\cite{wright1931evolution}). Using classical estimators such as Watterson's, the effective population size of all humans has been inferred to be $N_e\approx 20,000$ haploid individuals \cite{takahata1993allelic}. However note that several possible definitions of effective population size exist \cite{ewens2004mathematical}, depending, among other things, on which summary statistics are used to match real and idealized populations (e.g. the number of segregating sites in the case of Watterson's estimator). Inferring the effective population size will be a central task in the remainder of this work, and new estimators of $N_e$ will be derived in Chapter \ref{chap:IBDmodel}. The assumption of constant population size used in the Wright-Fisher population model will often be relaxed (thereby describing effective population sizes as a function of the considered genealogical time), and new summary statistics obtained from the genetic data will be employed to achieve higher resolution into the recent past of a studied cohort.

\subsubsection{The coalescent with mutation and recombination}

To conclude this overview of the coalescent process, we include the modeling of recombination events along the sequences during the genealogical process, introduced in \cite{hudson1983properties}. As in the case of mutations, recombination events may occur between any pair of sites at any transmission of the genetic material (provided the recombination rate between these sites is positive). While mutation does not affect the tree structure of the genealogy, however, recombination does.

\begin{figure*}
    \centering
	\includegraphics[scale=0.5]{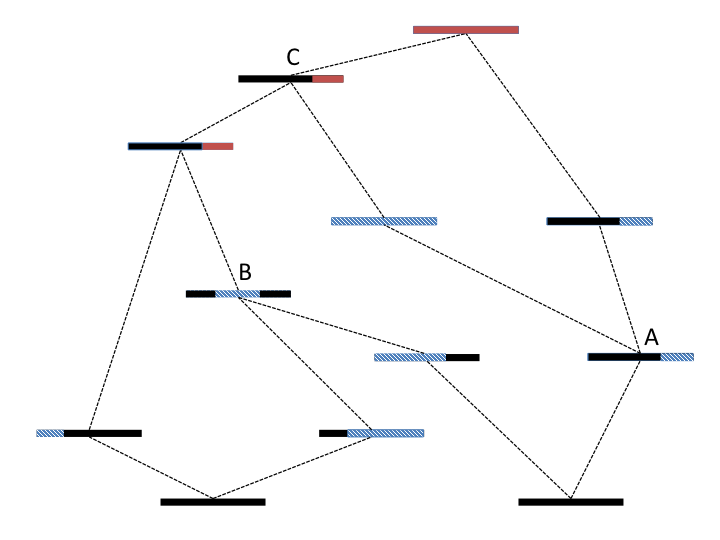}
	\caption{A sample genealogical history including recombination events (figure adapted from \protect\cite{hein2004gene}).}
    \label{fig:intro:coalRec}
\end{figure*}

Again, consider $n$ sequences of length $s$ sites. Assume recombination occurs at the same rate for all pairs of sites, and that a sequence has a total chance of recombining of $\rho$ per generation. Under these conditions, if a recombination event occurs, the exact location can be randomly sampled along the sequence. It is possible that, for long chromosomal regions that have high recombination rates, more than one recombination occurs during one generation. Again, however, we measure time in the continuous space, so that only one recombination event is allowed to occur at a time, but the number of recombination events occurring during a unit interval of time may be greater than one. The effect of a recombination event occurring between the sites $s_i$ and $s_{i+1}$ is to break one of the ancestral lineages that we are tracing backwards in time. This creates two lineages, one harboring the ancestral material in the range $[1,s_i]$, and the other carrying the ancestral material in  $[s_{i+1},s]$. After a recombination event occurs, the number of ancestral lineages being traced increases by one. This turns the genealogical structure representing the cohort's genetic history from a tree into a graph, as shown in Figure \ref{fig:intro:coalRec}. This graph structure, which may assume very complex forms for large sample sizes and long genomic regions, is called the \emph{ancestral recombination graph} (ARG), introduced in \cite{griffiths1997ancestral}.

While some quantities may still be derived analytically, the ARG is a fairly complex mathematical object, and it often requires the use of numerical sampling for its use in quantitative analyses. A possible sampling algorithm for the ancestral recombination graph operates as follows:

\begin{enumerate}
\item Initialize the number of ancestral sequences to $k = n$, the samples from the current generation.
\item Recombination occurs with rate $k\rho$, while coalescence occurs with rate $\frac{{k \choose 2}}{N}$, and the time distribution to the first event is exponential in both cases. To sample the time to the first occurrence of an event (either recombination or coalescence), sample from an exponential distribution with rate $k\rho + \frac{{k \choose 2}}{N}$.
\item The event is a recombination with probability $Nk\rho/{k \choose 2}$, a coalescent otherwise. Draw a uniform value between $0$ and $1$ to select the type of event.
\item Handle the sampled event: if it is a coalescent, randomly choose two lineages and merge them; update $k=k-1$. If it is a recombination, sample a random lineage and break it at a uniformly chosen point along the genome; update $k=k+1$.
\item If $k>1$, go to step 2.
\end{enumerate}

Figure \ref{fig:intro:coalRec} shows an example of running such an algorithm. Note that although the number of traced lineages may grow through recombination events, the algorithm is expected to converge, since individuals are eliminated through coalescent events at a rate that is quadratic in $k$, and created through recombination at a rate that is linear in $k$. As in the case of no recombination, sequences can be generated after having sampled an ancestral recombination graph, by introducing mutations over the graph edges using the same procedure that was discussed in Section \ref{subsubsec:intro:coalMut}. 

\subsubsection{Approximations of the coalescent}
\label{subsec:intro:approx}

The algorithm shown in the previous section for sampling from the coalescent with recombination process may be improved in several ways. A first possible improvement follows from the observation that some lineages that are created and traced during the sampling process are not affecting the final sequences. Consider for example the ARG of Figure \ref{fig:intro:coalRec}. The event marked with the letter ``A'' is a recombination that creates a lineage that does not contain any genetic material inherited by present-day individuals. This lineage will increase the coalescent rate, and will eventually be absorbed during the coalescent event marked with the letter ``C''. The creation and the absorption of this lineage have no effect on the genealogy, and may be omitted. Since the exponential distributions that are used to model the timing of these events are memoryless, it turns out that omitting these events does not affect the distribution of the sampled ARG structures. To avoid tracing these lineages, therefore, it is sufficient to modify the algorithm so that if a recombination would produce a lineage that carries no ancestral material, no action is taken.

A number of additional improvements can be developed for this basic algorithm, an extensive discussion is beyond the scope of this work. It is however worth mentioning an approximation of the ARG generation algorithm that resulted in substantial further development. The algorithm described in the previous section operates backwards in time (``vertical algorithm''), starting from a set of individuals in the present generation and sampling ancestors or splitting recombinant lineages until a single common ancestor is found. Alternatively, it is possible to sample from the same space of ancestral recombination graphs by moving along the chromosome (``horizontal algorithm''), rather than backwards in time. Such horizontal algorithm, which was developed in \cite{wiuf1999recombination} and is here omitted for brevity, has a computational complexity that is comparable to that of the horizontal version (depending on which improvements to the basic version are considered). It is however appealing because several methods in computational genetics analyze DNA sequences moving from left to right (or right to left), assuming an underlying Markovian process and relying on computational machinery such as Hidden Markov Models to perform inference of relevant features. The version introduced in \cite{wiuf1999recombination}, however, violates Markovian properties, as ARGs are intrinsically not Markovian when analyzed horizontally. This is due to the presence of nodes such as the one marked with letter ``B'' in the example of Figure \ref{fig:intro:coalRec}, where a lineage with a ``gap'' is created from the coalescence of two lineages whose ancestral material does not overlap. The existence of this kind of coalescent events requires keeping track of the entire history of genealogical events in an algorithm that moves horizontally across the genome, therefore violating a key Markovian property that requires the distribution of future states to be only dependent on recent states. In a seminal paper by Gil McVean \cite{mcvean2005approximating}, it was noted that the effects caused on commonly used summary statistics by the coalescence of lineages that with ``gaps'' in their ancestral material are negligible. The sequentially Markovian coalescent (SMC), introduced in \cite{mcvean2005approximating}, provides an approximation of Wiuf and Hein's horizontal algorithm that substantially simplifies the computation of ARGs. This approach has been recently used in a variety of genomic applications, some of which found application in the reconstruction of demographic events, and will be briefly discussed in Chapter \ref{chap:conclusions}. Many of the methods described in this thesis are related to the SMC model, depending on the definition of IBD (see Section \ref{intro:subsec:IBD_def}).

We conclude by noting that approximations of the vertical algorithm have also been developed. In \cite{parida2011minimal}, for instance, a similar approximation is made to limit coalescent events to those lineages that have an overlapping region of ancestral material, preventing the formation of gaps as the one seen in the example of Figure \ref{fig:intro:coalRec}.

\subsection{Identity by descent}
\label{intro:subsec:IBD_intro}

In this section we will introduce the basic concepts related to the co-inheritance of identical-by-descent (IBD) haplotypes that are relevant to the development of this work.

\begin{figure*}
    \centering
	\includegraphics[scale=0.6]{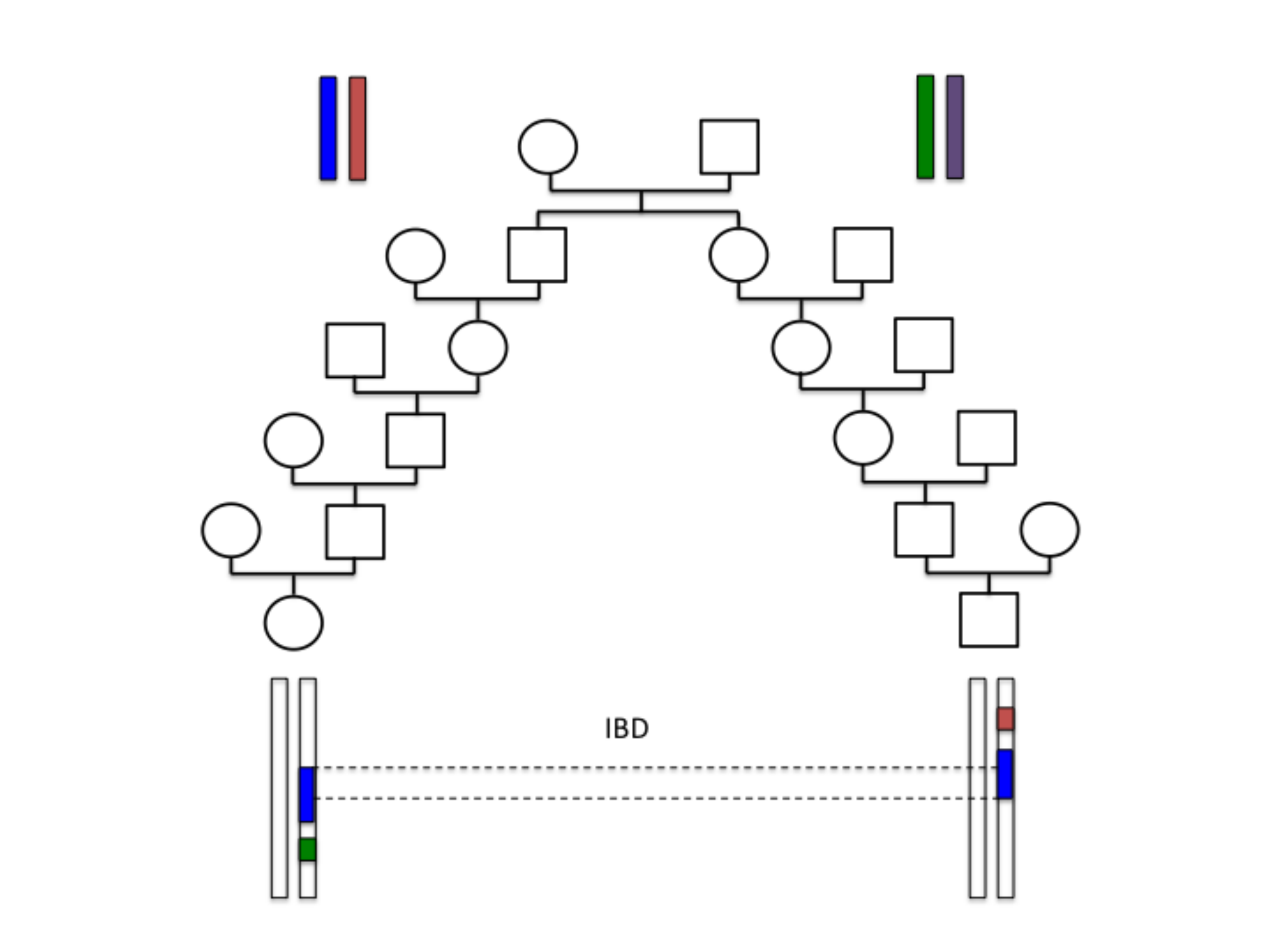}
	\caption{A pedigree structure where two fourth degree cousins co-inherit an IBD segment from ancestors that lived five generations in the past (figure adapted from \protect\cite{browning2012identity}).}
    \label{fig:intro:pedigree}
\end{figure*}

Consider the structure represented in Figure \ref{fig:intro:pedigree}. In this sample pedigree a pair of fourth degree cousins share two common ancestors that lived five generations in the past. These diploid ancestors each have two copies of their autosomal chromosomes, represented using colored bars. At each generation, the offspring inherit a chromosome copy from each of their two parents. Such inherited copies result from the meiotic events that generate germ cells, during which recombination may break down and mix the original chromosome copies present in the diploid parents. In the depicted pedigree, individuals from the population mate with individuals that are direct descendants of the pair of common ancestors living five generations in the past. It is assumed that the genetic material of these external individuals (founders) is unrelated to that of the pair of ancestors. After five generations, the pair of extant fourth degree cousins happen to both inherit stretches of the colored chromosomes from their common ancestors. The blue stretch of chromosome, in particular, overlaps in a region, which constitutes an identical-by-descent segment, or haplotype.

Identical-by-descent haplotypes have been extensively studied in the context of pedigree structures, particularly in early genotype-phenotype association studies, which generally involved information about the family structure of the analyzed samples (\cite{spielman1993transmission}), therefore several quantities regarding IBD haplotypes can be derived from pedigrees. Some basic quantities can be easily derived as follows. Consider a pair of siblings sharing two common ancestors (their parents) one generation in the past, and a single nucleotide on a haplotype along their genome. Such nucleotide may have been co-inherited by both individuals from the same copy of their parental genome, with probability $1/2$ (if the copies of the father are, for instance, $A$ and $a$, the two offspring will co-inherit the same copy if both choose $A$, or both choose $a$, and the same reasoning holds for the copy they inherit from the maternal side). Now consider a pair of first degree cousins descending from these siblings. One chromosomal copy for these first degree cousins will be inherited from a parent chosen from the general population. As previously assumed, these are completely unrelated individuals, and such chromosome will not harbor an IBD locus. Focusing on the chromosome that is inherited through the lineage leading the their shared common ancestors, the probability of being IBD is $1/4$. This is due to the fact that each cousin will inherit one of the four possible copies present in their grand parents, and will choose the same with probability $1/4^2\times4=1/4$. Recursively computing this probability for the following generations, we obtain that the chance that two $(k-1)$-th degree cousins that share two diploid common ancestors $k$ generations in the past are IBD at a chosen genomic location is $(1/4)^{k-1}$. Due to the linearity of the expectation operator, this also corresponds to the expected fraction that a pair of $(k-1)$-th degree cousins will share IBD. Note that this quantity decreases exponentially in $k$, and indeed after a relatively small number of generations it is very common that no IBD sharing exists at all. If IBD sharing exists, however, this typically occurs through the sharing of relatively long IBD haplotypes. If a genomic locus is shared IBD by a pair of individuals, the flanking positions along the genome are in fact typically also shared IBD, because the haplotypes that are transmitted from common ancestors are delimited by recombination events. As shown in the previous section, a recombination event may occur during meiosis between any two consecutive nucleotides. These recombination events are rare and independent, and their occurrence can therefore be modeled through a Poisson process with exponentially distributed waiting times between arrivals. The length of an IBD haplotype that has been transmitted from common ancestors that lived $k$ generations in the past is therefore exponentially distributed, averaging $100/(2k)$ centimorgans. The number of IBD segments that are expected to be found for $(k-1)$-th degree cousins can be similarly computed. After $2k$ generations that separate the two cousins in the pedigree, a chromosome of genetic length $l$ Morgans is expected to be broken into $2lk$ distinct haplotypes, each representing a potential IBD segment. The probability that one such segment is co-inherited is $(1/4)^{k-1}$, resulting in an average of $2lk(1/4)^{k-1}$ IBD segments. Again, these are rare and independent, and their number can be modeled as a Poisson distributed random variable.

\subsubsection{Definition of IBD}
\label{intro:subsec:IBD_def}

Despite the name, IBD segments need not be identical. Mutations in IBD segments may in fact arise during transmission from a common ancestor to her descendants, as detailed in Chapter \ref{chap:mutation}. Because the number of mutations per base pair is proportional to the distance, in generations, to the common ancestor, the genomic segments transmitted to a set of individuals from very recent common ancestors will be almost identical, while regions that are co-inherited from very remote ancestors will tend to have a larger number of differences per base pair. Analyses of IBD sharing in pedigrees are usually concerned with the transmission of long IBD segments through common ancestors that span a small number of generations. These segments are therefore typically long and almost identical, and short IBD segments transmitted from very remote ancestors from the general population, which are not reported in the pedigree and are not considered members of the family, are neglected. However, when IBD sharing is detected in unrelated individuals from a population, as we do in this work, haplotypes may be co-inherited from common ancestors that lived several generations in the past, and harbor a relatively higher number of mutations. Based on these considerations, we may consider several definitions of an IBD segment:

\begin{enumerate}[(a)]
\item A chromosomal region transmitted from a common ancestor that lived at most $t_0$ generations in the past (e.g. see \cite{chapman2003model}).
\item A chromosomal region of length at least $u$ cM that is transmitted from a common ancestor that lived at any time in the past.
\item A chromosomal region of length at least $u$ cM that is transmitted \emph{without recombination} from a common ancestor that lived at any time in the past.
\end{enumerate}

While (a) is suitable in cases where $t_0$ is known (e.g. pedigrees) or where the focus is on modeling the descent of a known set of individuals founding a population $t_0$ generations in the past, this definition becomes impractical in the general case of IBD segments detected in a set of unrelateds. IBD detection in unrelated individuals usually results in a list of segments that have been discovered with a relatively high level of confidence. Often times these segments will be detected on the basis of being more similar (e.g. identical by state, IBS) compared to surrounding genomic regions. These segments will typically be transmitted from one common ancestor, generally delimited by recombination events, but their length alone is insufficient to determine the age of these segments, which has large variance for all but the very long shared haplotypes. Definitions (b) and (c) are therefore more suitable for the analysis of these segments, as no value of $t_0$ is assumed. In practice, current IBD detection algorithms are typically only able to reliably detect segments that are longer than a certain centimorgan length threshold, which can be accommodated in definitions (b) and (c).

In the remainder of this thesis, we use definition (c), i.e. we require that an IBD segment is transmitted from a common ancestor and is delimited by any recombination events along the lineages connecting modern day individuals to the common ancestor. Note, however, that several neighboring chromosomal regions may be merged together while still being transmitted from the same common ancestor, in which case definition (b) and (c) may not entirely overlap, depending on several factors such as population size and distance to the shared ancestor. When computing distributions of IBD sharing in chapters \ref{chap:IBDmodel}, \ref{chap:migration} and \ref{chap:mutation}, we will rely on definition (c) to derive analytical results. When using coalescent simulations to create synthetic datasets used to compare predicted and observed IBD values, however, we will compute IBD segments using definition (b), i.e. we will only require that a chromosomal region is co-inherited from the same common ancestor, without restrictions on the occurrence of recombination along these lineages, unless otherwise specified. It is evident that when very short IBD segments are considered as defined in (b) or (c), these may have a fairly large number of differences due to mutations arising along the lineages leading to extant individuals. We will still refer to these segments as IBD, although the ``I'' of identical may be inappropriate in this case. As we consider shared segments that are transmitted from ancestors that lived a large number of generations ago, it may be more appropriate to refer to these regions as \emph{non-recombinant}, when definition (c) is adopted.

\chapter{IBD sharing in contemporary human populations}
\label{chap:IBD_in_data}
As introduced in the previous chapter, the co-inheritance of long IBD haplotypes is usually a sign of recent genetic relatedness across individuals. If the most recent common ancestor of a pair of individuals is relatively remote, the chance of finding IBD segments is very small. A pair of seventh degree cousins, for instance, will typically share no IBD segments at all. If such sharing occurs, however, the IBD haplotypes tend to be relatively long (for seventh degree cousins, for instance, IBD segments are expected to be $6.25$cM long, or ${\sim}4.8 \times 10^6$ base pairs, assuming a recombination rate of ${\sim}1.3$cM/Mb). Furthermore, if a large number of individuals is analyzed, the chance of finding IBD segments may become significant. When $n$ individuals are analyzed, there are in fact $n \choose 2$ possible pairs of IBD sharing individuals. This motivated the development of several algorithms that allow detecting IBD haplotypes in large cohorts of unrelated individuals \cite{purcell2007plink,gusev2009whole,browning2010high,browning2011fast,browning2013improving}. At the time the work presented in this chapter was developed, a number of large SNP array datasets comprising individuals from several human populations became available. The goal this work was to mine the presence of IBD segments in such cohorts, aiming to answer questions such as

\begin{itemize}
\item Are IBD haplotypes commonly found in purportedly unrelated individuals?
\item Does IBD sharing reflect modern day geographic origins, and does it provide more information than other available summary statistics of genetic similarity?
\item Can haplotype sharing be used to investigate a population's demographic history?
\item Is the signature of natural selection visible in the distribution of IBD segments?
\end{itemize}

As discussed in the remainder of this chapter, IBD sharing was found to be pervasive in large cohorts of unrelated individuals, and was shown to be informative about both demographic and evolutionary events in human populations.

\section{World-wide sharing of IBD segments}

This section reports the results of IBD analysis performed on several large SNP array datasets, namely the HapMap 3 dataset \cite{frazer2007second}, the Hebrew University Genetic Resource \cite{HUGR}, and the InTraGen Population Genetics Database (Idb, \cite{mitchell2004new,duerr2006genome}). Abbreviations for the distinct populations contained in these datasets can be found in Table \ref{tbl:IBDdata:sharingStat}. The results reported in this chapter, together with additional details on other analyses and the utilized datasets can be found in \cite{gusev2012architecture}. The work reported in this section was performed in close collaboration with Alexander Gusev.

\subsection{IBD detection}

IBD sharing was detected in the analyzed datasets using the GERMLINE software package \cite{gusev2012architecture}. Before analyzing the available real datasets, we assessed the accuracy of GERMLINE's IBD detection using synthetic datasets obtained using the GENOME rapid coalescent-based whole-genome simulator \cite{liang2007genome}. We measured the accuracy of GERMLINE's IBD discovery using standard measures of precision (fraction of discovered segments that correspond to real IBD segments) and recall (fraction of real IBD segments retrieved). A ground-truth set for IBD segments is obtained considering all identical segments in the set of simulated haplotypes. Haplotypes were merged to form synthetic genotypes, discarding phase information. GERMLINE's \emph{haplotype} and \emph{genotype} extension modes were tested on both perfectly phased and computationally phased data. Discovered segments of $3$ cM or longer were reported. To compute recall, GERMLINE's, IBD discovery was compared with true segments longer than 3 cM. A measure of false-positive segments was computed comparing the obtained IBD matches with segments $\geq1$ cM long in the ground-truth set.

Comparing the accuracy of both \emph{haplotype} and \emph{genotype} extensions on simulated data, the haplotype extension mode was found to have extremely good performance on perfectly phased data, while its recall deteriorated when computational phasing was used, as a result of unreliably reconstructed haplotypes. The genotype extension mode, on the other hand, showed a high rate of false positive IBD segment (${\sim}30\%$ of the total) and an almost perfect recall rate. The genotype extension mode was also found to be robust to variation in the simulated demographic parameters, which, as further analyzed in Chapter \ref{chap:IBDmodel}, have an impact on phasing accuracy and therefore on the performance of the haplotype extension mode. Based on these results, and because the datasets analyzed in this work included individuals from heterogeneous populations, often with small sample sizes resulting in phasing uncertainty, GERMLINE's genotype extension mode was used for IBD detection in all reported results.

\subsection{IBD-based graph clustering recapitulates populations structure}

\begin{table*}	
	\begin{subtable}{.45\linewidth}
		\centering
	    \begin{tabular}{|c|c|c|c|c|c|}
	    Population & Samples & \begin{tabular}[x]{@{}c@{}}Average shared \\ genome (\%) \end{tabular} & \begin{tabular}[x]{@{}c@{}}Average segment \\ length (cM) \end{tabular}& \begin{tabular}[x]{@{}c@{}} \% of pairs \\ sharing IBD \end{tabular} & \begin{tabular}[x]{@{}c@{}}Cryptic \\ relatives \end{tabular} \\ \hline
	    Ashkenazi Jews (AJ) & 397 & 1.73 & 5.51 & 96.9 & 3 \\
	    \end{tabular}
		\caption{Samples in the HUGR dataset.}
	\end{subtable}
	
	\begin{subtable}{.45\linewidth}
		\centering
	    \begin{tabular}{|c|c|c|c|c|c|}
	    Population & Samples & \begin{tabular}[x]{@{}c@{}}Average shared \\ genome (\%) \end{tabular} & \begin{tabular}[x]{@{}c@{}}Average segment \\ length (cM) \end{tabular}& \begin{tabular}[x]{@{}c@{}} \% of pairs \\ sharing IBD \end{tabular} & \begin{tabular}[x]{@{}c@{}}Cryptic \\ relatives \end{tabular} \\ \hline
	    Ashkenazi Jews (AJ) & 389 & 1.43 & 5.52 & 99.3 & 2 \\
	    Europeans (EU) & 514 & 0.05 & 4.11 & 36.6 & 3 \\
	    \end{tabular}
		\caption{Samples in the Idb dataset.}
	\end{subtable}

	\begin{subtable}{.45\linewidth}
		\centering
		\small
	    \begin{tabular}{|c|c|c|c|c|c|}
	    Population & Samples & \begin{tabular}[x]{@{}c@{}}Average shared \\ genome (\%) \end{tabular} & \begin{tabular}[x]{@{}c@{}}Average segment \\ length (cM) \end{tabular}& \begin{tabular}[x]{@{}c@{}} \% of pairs \\ sharing IBD \end{tabular} & \begin{tabular}[x]{@{}c@{}}Cryptic \\ relatives \end{tabular} \\ \hline
		\centering
		African Americans (ASW) & 42 & 0.14 & 7.08 & 0.3078 & 4 \\
		Europeans (CEU) & 109 & 0.48 & 3.77 & 0.9886 & 1 \\
		Han Chinese (CHB) & 82 & 0.46 & 3.66 & 0.9913 & 0 \\
		Metropolitan Chinese (CHD) & 70 & 0.46 & 3.66 & 0.9896 & 2 \\
		Gujarati Indians (GIH) & 83 & 0.78 & 4.26 & 0.9245 & 5 \\
		Japanese (JPT) & 82 & 0.77 & 3.71 & 0.9997 & 0 \\
		Luhya in Kenya (LWK) & 83 & 0.80 & 4.98 & 0.9924 & 11 \\
		Mexicans (MEX) & 45 & 0.96 & 3.87 & 0.9939 & 4 \\
		Maasai in Kenya (MKK) & 143 & 1.06 & 8.58 & 0.9379 & 94 \\
		Tuscans in Italy (TSI) & 77 & 0.4 & 4.23 & 0.9679 & 0 \\
		Yoruba in Ibadan (YRI) & 108 & 0.11 & 4.19 & 0.6333 & 2 \\
	    \end{tabular}
		\caption{Samples in the HMP3 dataset.}
	\end{subtable}
\caption{Description of the samples contained in the analyzed datasets and summary of IBD sharing.}
\label{tbl:IBDdata:sharingStat}
\end{table*}

Although the analyzed datasets were composed entirely of purportedly unrelated individuals, IBD segments were found to be ubiquitous between and across populations, as shown in Table \ref{tbl:IBDdata:sharingStat}. To allow for population-wide analysis of IBD sharing, we built a graph model where each individual is represented as a vertex, and the amount of IBD sharing between two individuals corresponds to a single weighted edge. Building such graph for the Idb dataset results in the formation of a large connected component of individuals. The occurrence of such large connected component is extremely unlikely to occur by chance, and it indicates the presence of underlying structure in the graph (p value $<10^{-100}$ under a hypergeometric distribution). The cohort is indeed structured, and the node membership in the connected component is highly correlated with self identification as Ashkenazi Jews (99.7\% of Ashkenazi individuals are spanned by the connected component, constituting 91.5\% of the component's nodes). Overall, the total genome-wide sharing for an average pair of AJ samples ($54.25$ cM) is considerably higher than that of EU samples ($1.81$ cM).

\begin{figure*}
    \centering
	\includegraphics[width=0.8\textwidth]{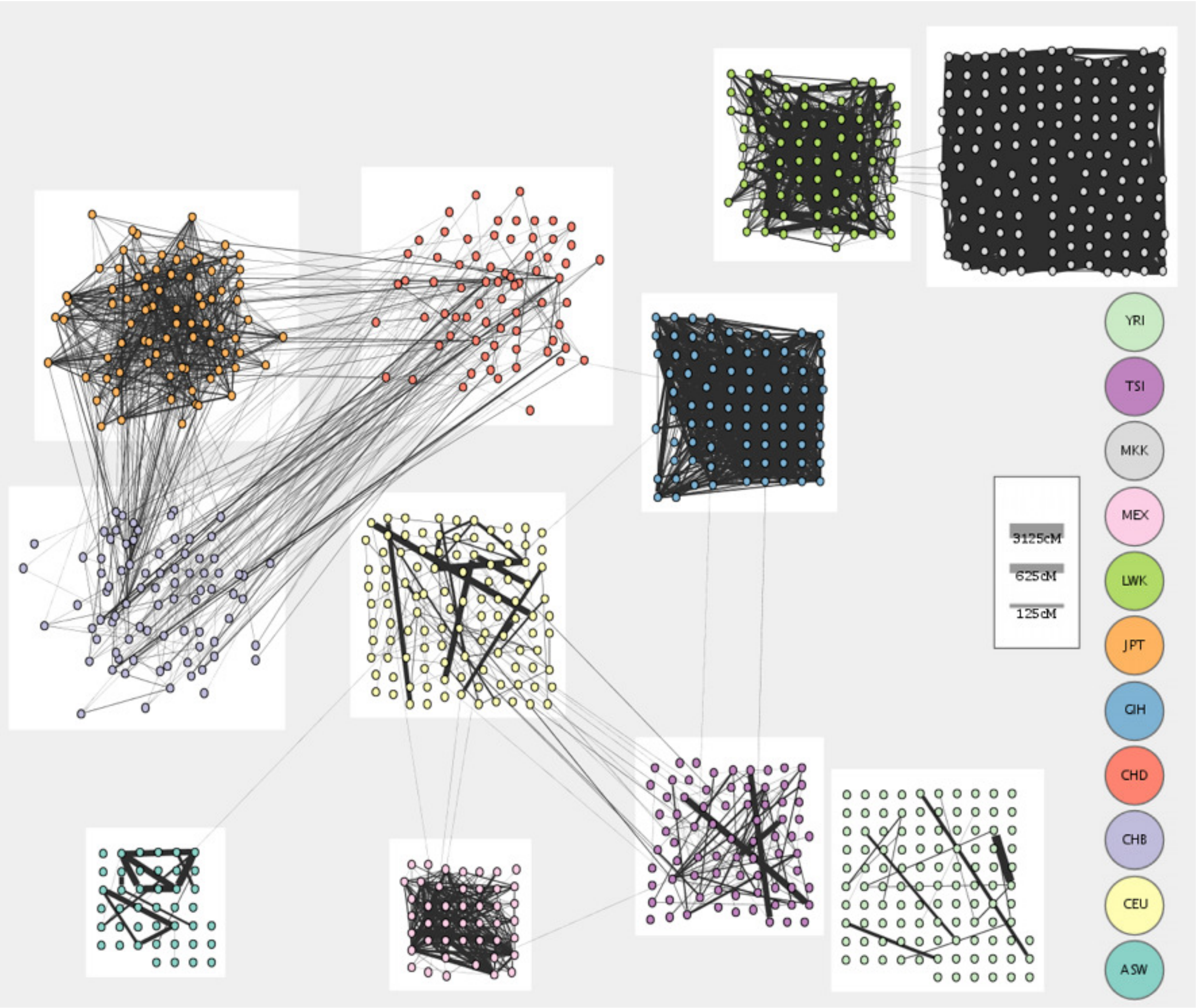}
	\caption{The IBD sharing graph for HMP3 samples.}
    \label{fig:IBDdata:HMP3}
\end{figure*}

We set out to verify the presence of similar structure in IBD sharing graphs for the HMP3 dataset. The network of shared segments in HM3 (Figure \ref{fig:IBDdata:HMP3}) is dense within populations and geographic regions and sparse between them. We can immediately observe an abundance of recent sharing within the cohorts, particularly in the MKK and LWK Africans; the GIH Indians. Moreover, this high level of sharing is homogeneous across most of the population and not suggestive of individual cryptic relatives. Several pairs of close relatives (defined as pairs of individuals sharing at least $1,700$cM of their genome) are found within the Maasai sample. This unexpected finding will be further discussed in Chapter \ref{chap:IBDmodel}. Looking across populations, only the JPT, CHD, and CHB East Asian groups exhibit a large number of shared segments, particularly between the two Chinese populations. The few remaining segments are also overwhelmingly within continental groups, particularly between CEU and TSI.

To investigate the ability to recapitulate population structure using the observed IBD sharing, we refined the construction of the IBD graph to allow downstream clustering analysis. In the constructed IBD graph, the weight of an edge between a pair of individuals is proportional to the sum of the length (in centiMorgans) of the IBD segments shared between the individuals. To account for the higher informativeness of rarely shared regions, the sum is normalized by the region-specific frequency of sharing in the entire population. More formally, given a set of $n$ ordered SNPs $s \in \{1\dots n\}$, we define a function to represent the normalized length of an interval between two SNPs as follows:

\begin{equation}
\begin{split}
	F(s) =
	\begin{cases}
	\frac{l(s,s+1)}{\pi(s,s+1)} & \mbox{if } \pi(s,s+1) \neq 0, \\
	0 & otherwise. \\
	\end{cases}
\end{split}
\end{equation}

where $l(s,s+1)$ is the length of the segment $[s,s+1]$, and $\pi(s,s+1)$ is the number of individuals sharing the segment $[s,s+1]$. The maximum normalized length (all SNPs being shared by a pair of individuals) is then:

\begin{equation}
\begin{split}
	W_{tot} = \sum_{s=1}^{n} F(s)
\end{split}
\end{equation}

For each pair of individuals $i$ and $j$ sharing a set of segments $K$, we compute a raw edge weight normalizing the total shared length by the maximum normalized segmental length:

\begin{equation}
\begin{split}
	W_{ij} = \frac{1}{W_{tot}} \sum_{r \in K} \ \sum_{t=k_{i,r}}^{k_{e,r}} F(t)
\end{split}
\end{equation}

Where $k_{i,r}$ and $k_{e,r}$ are the first and the last SNPs in the segment $r$.

The obtained value is representative of the total sharing between the two individuals and ranges between $0$ (i.e., no sharing) and $1$ (i.e., sharing of the whole genome). To account for the exponential decrease in the segmental length that occurs with the number of meioses, we use the weight $w_{ij}=\log(W_{ij})$ on the edges in our clustering calculations.

After constructing such graph, we performed graph clustering using the Markov Cluster Algorithm (MCL), detailed in \cite{van2000graph}. MCL detects clusters based on the recurrence of a random walk across a weighted graph. We run MCL with default parameters as well as the \emph{force-connected} flag which adjusts the output clusters to ensure that they are connected components. We performed the clustering in an iterative procedure that seeks to find the underlying population structure as well as identify genetic regions that are shared between clusters. The procedure starts considering all shared segments longer than $3$ cM and performs the following analysis in each iteration:

\begin{enumerate}
\item Compute the sharing graph from the current set of shared segments. This weighted graph is then provided as input for MCL, which identifies clusters of increased relatedness.
\item Calculate the probability that a genomic locus is shared across the identified clusters, and identify any region enriched for cross-cluster sharing (1 standard deviation above the genome-wide mean).
\item Excise all enriched cross-cluster regions as well as any affected matches that overlapped these regions and were shortened below 3 cM. The un-excised data are used as input for the next iteration.
\end{enumerate}

\begin{figure*}
    \centering
	\includegraphics[width=1\textwidth]{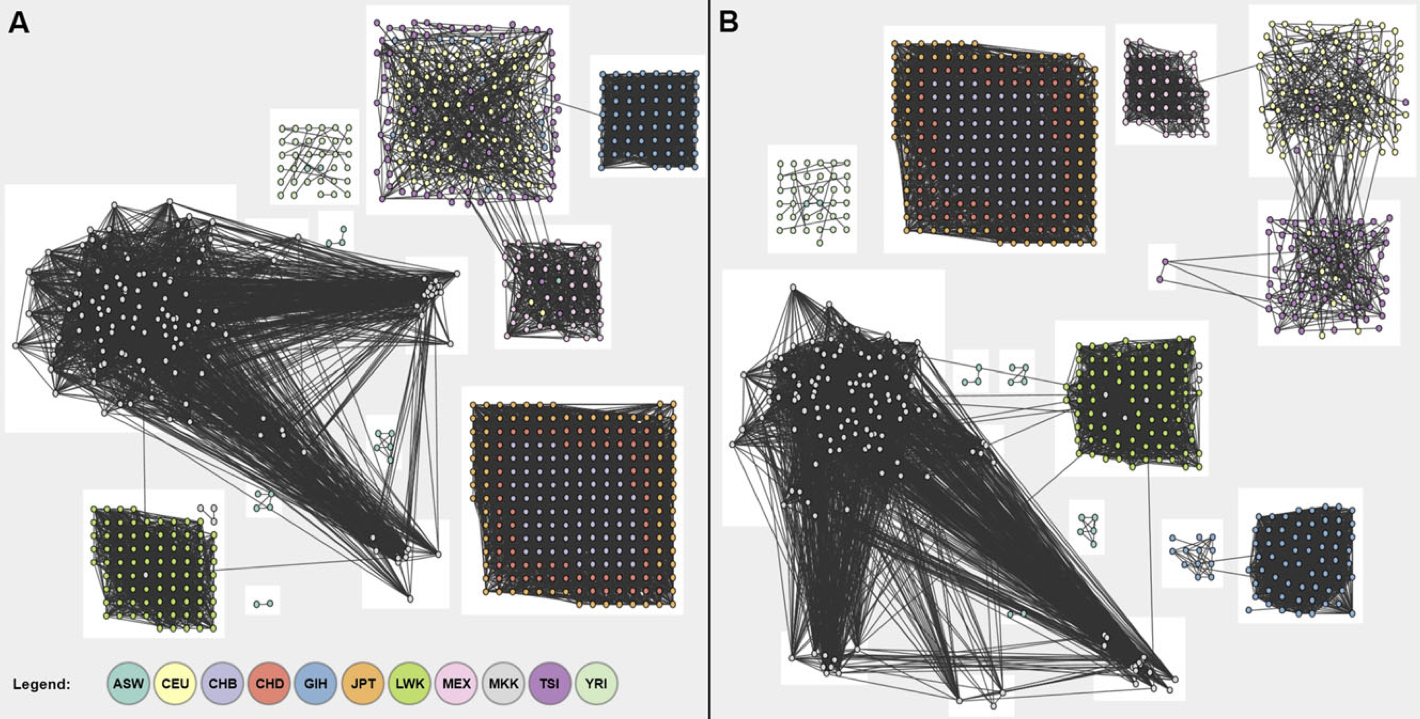}
	\caption{Clusters emerging from the IBD sharing graph in the HMP3 dataset reflect population structure. (A) Initial clusters from unfiltered sharing, where \{GIH\},\{LWK\},\{JPT,CHD,CHB\},\{CEU,TSI\} segregate. (B) Final clusters after cross-cluster edges have been iteratively removed, where \{TSI\},\{CEU\} newly segregated.}
    \label{fig:IBDdata:HMP3_clust}
\end{figure*}

This iterative process eventually converges when no further excision is made. Applying this procedure to the IBD sharing graph of the HMP3 dataset, we indeed recover underlying population structure. The final clusters demonstrate improved resolution between populations, with six cross-cluster regions remaining, as shown in Figure \ref{fig:IBDdata:HMP3_clust}.

\subsection{IBD sharing provides insight into recent demographic history}

\begin{figure*}
    \centering
	\includegraphics[width=1\textwidth]{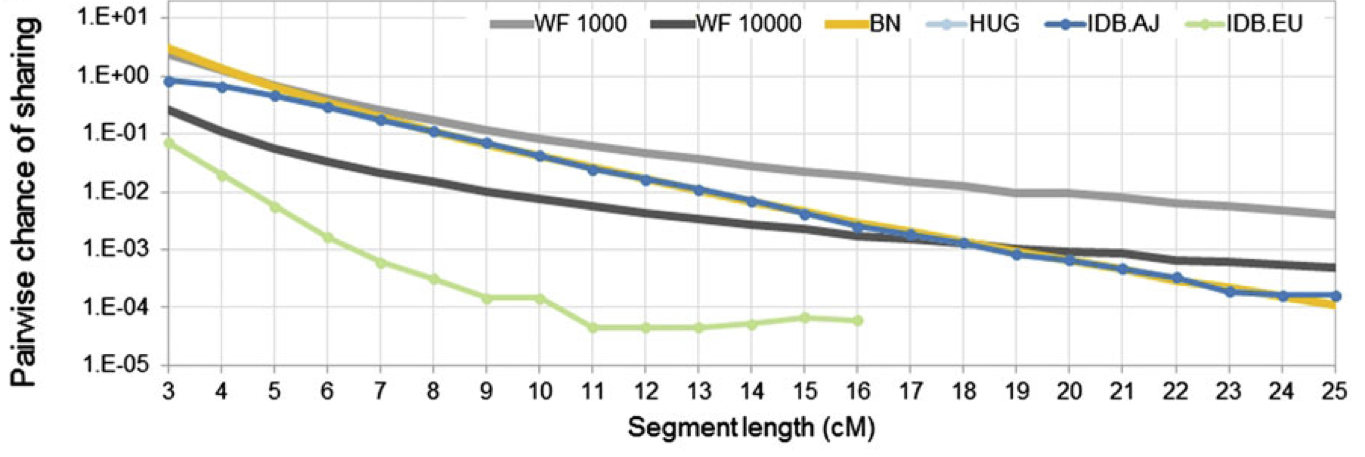}
	\caption{Frequency of IBD sharing as a function of genetic length in the AJ and CEU cohorts, and comparison to synthetic datasets.}
    \label{fig:IBDdata:AJ_decay}
\end{figure*}

Further investigating the substantial IBD sharing in the Ashkenazi Jewish cohort, we examined the frequency distribution of shared IBD segments as a function of their genetic length (Figure \ref{fig:IBDdata:AJ_decay}). Based on simulations, we noticed that the slope of such distribution is not compatible with the slope obtained in populations of constant size (Wright-Fisher populations). A population expansion, however, results in a steep exponential decrease compatible with what is observed in the AJ cohorts.

To obtain an initial rough estimate of an expansion rate that is compatible with the one observed in the AJ data, we considered an idealized extreme bottleneck-expansion scenario where a population is formed by one individual $G$ generations before present, and infinite individuals from generation $G$ to present. In such a scenario, all coalescent events happen at generation $G$. For a population that underwent an extreme bottleneck-expansion at generation $G$, two contemporary individuals are expected to share a number of segments of length $l$ proportional to $p(1-p)^{2Gl}$, where the length is expressed in centiMorgans, and $p=0.01$ represents the chance of a recombination event along one unit of length for a shared segment at each generation. $G$ can be computed from $N_l$ and $N_{l+1}$ as:

\begin{equation}
\begin{split}
	\frac{N_{l+1}}{N_l} = 0.99^{2G}
\end{split}
\end{equation}

therefore

\begin{equation}
\begin{split}
	G = \frac{\log(\frac{N_{l+1}}{N_l})}{2\log(0.99)}
\end{split}
\end{equation}

The observed exponential decay of $0.671$ per cM (std $0.055$) is consistent in this model with a bottleneck-expansion event occurred around $20$ generations before present. We refined this estimate using extensive simulations, performing grid search in a richer parameter space (timing of the bottleneck, ancestral population size, and current population size) using a demographic model of exponential expansion (for details on these simulations, see \cite{gusev2012architecture}). We observe the effect of the ancestral population size are mostly noticeable on the frequency of short IBD segments, whereas the current population size mainly affects the longer segments. The timing of the bottleneck affects the entire distribution, with stronger effects on midrange segments. Our grid search suggests a rapid expansion of about $950$ diploid individuals $23$ generations before present to current hundreds of thousands. More complex models than those tested in this analysis may be required to explain the deviation observed for segments shorter than 5 cM (see Chapter \ref{chap:IBDmodel}). The estimated timing is compatible with a model of AJ population structure inferred from historical data in \cite{slatkin2004population} and can be reconciled with previous analysis of rare mutations \cite{risch2003geographic} and mithocondrial data \cite{behar2006matrilineal}. Although significant admixture can be shown to influence the sharing distributions, our use of a single-population model seems reasonable due to the limited amount of recent sharing observed between European and Ashkenazi samples and by the strong similarity of the length distributions for AJ individuals sampled in Israel and USA (Idb.AJ and HUGR, see Materials and Methods). In other populations, the number of shared-segment pairs is smaller (Table \ref{tbl:IBDdata:sharingStat}) and does not yet allow for robust inference of demography.

\begin{figure*}
    \centering
    \begin{subfigure}[b]{0.7\textwidth}
		\vspace{-2mm}
            \centering
            \includegraphics[width=\textwidth]{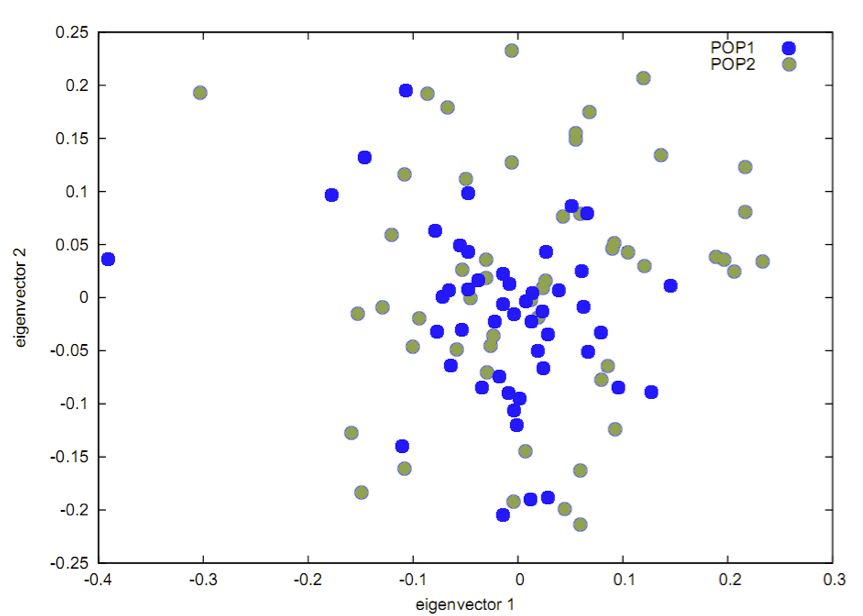}
            \caption{Principal component analysis.}
    \end{subfigure}

    \begin{subfigure}[b]{0.48\textwidth}
            \centering
            \includegraphics[width=\textwidth]{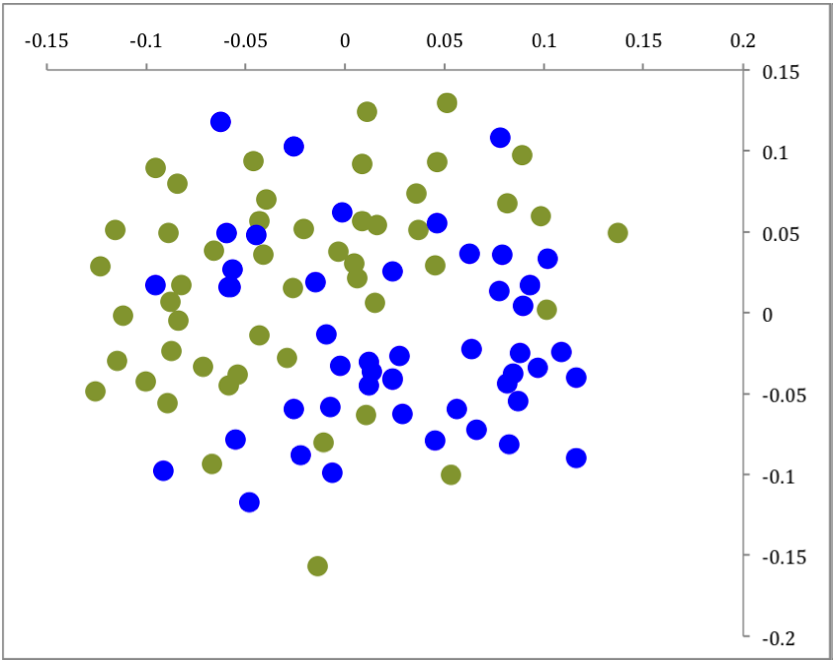}
            \caption{Multidimensional scaling using IBS kernel.}
    \end{subfigure}
    ~
    \begin{subfigure}[b]{0.465\textwidth}
            \centering
            \includegraphics[width=\textwidth]{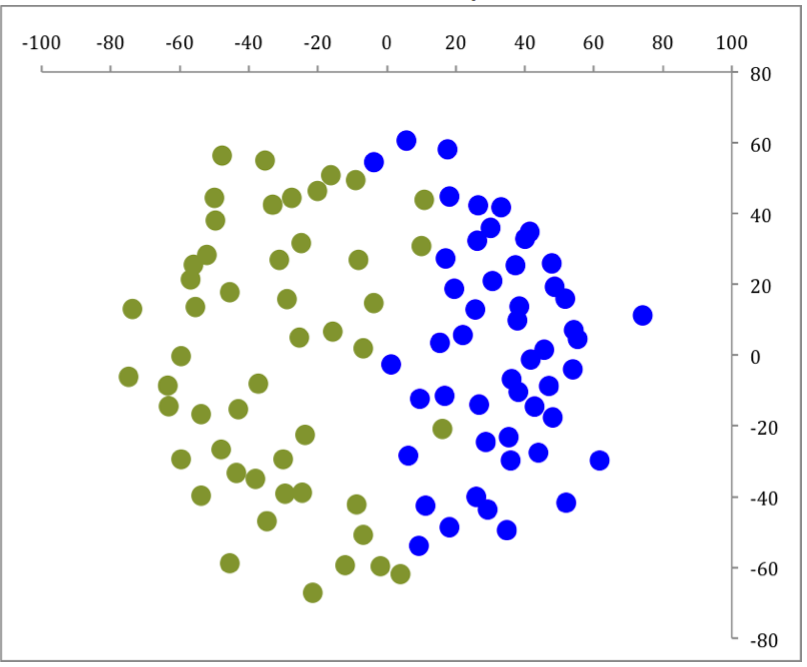}
            \caption{Multidimensional scaling using IBD kernel.}
    \end{subfigure}
    ~
	\caption{Comparison of principal component analysis and multidimensional scaling using IBS and IBD kernels for a recent split of two populations (represented by blue and green colors).}
	\label{fig:IBDdata:PCA-MDS}
\end{figure*}

The analysis of demographic events that occurred in the very recent history of the AJ population suggested that summary statistics of IBD sharing are informative about extremely recent demographic events. To test whether these insights may also be obtained using other methods available at the time this study was performed, we simulated a population split occurring $50$ generations before present. A population of $50,000$ individuals splits into two groups of $49,000$ and $1,000$ individuals. The smaller group then exponentially expands to reach size $5,000$ individuals. We sampled 50 diploid individuals from each of these two modern groups, and analyzed realistic genotype data using several methods to investigate population structure (Figure \ref{fig:IBDdata:PCA-MDS}). When principal component analysis was used to obtain a lower dimensionality projection of the data (\cite{price2006principal}), little or no population structure became evident. We subsequently built a matrix representing the relatedness of individuals based on their identity-by-state (IBS), and performed multidimensional scaling using such matrix. While the subdivision of the two groups starts being visible in this case, a clear distinction is only obtained when the similarity matrix is built using IBD sharing, indicating that methods relying on summary statistics of haplotype sharing may in some cases outperform methods based on other classical genomic features.

\subsection{Regions of increased IBD sharing are enriched for structural variation and loci implicated in natural selection}

\begin{figure*}
	\vspace{-2mm}
    \centering
    \begin{subfigure}[b]{0.9\textwidth}
		\vspace{-2mm}
            \centering
            \includegraphics[width=\textwidth]{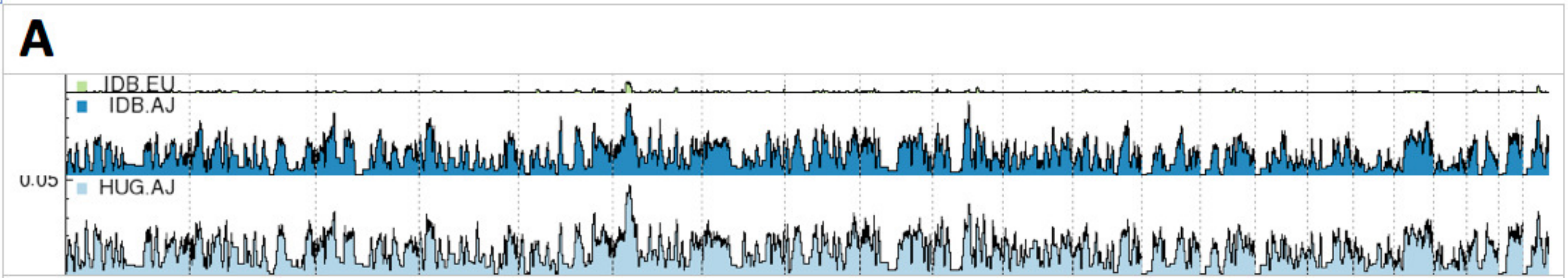}
    \end{subfigure}

    \begin{subfigure}[b]{0.9\textwidth}
		\vspace{-2mm}
            \centering
            \includegraphics[width=\textwidth]{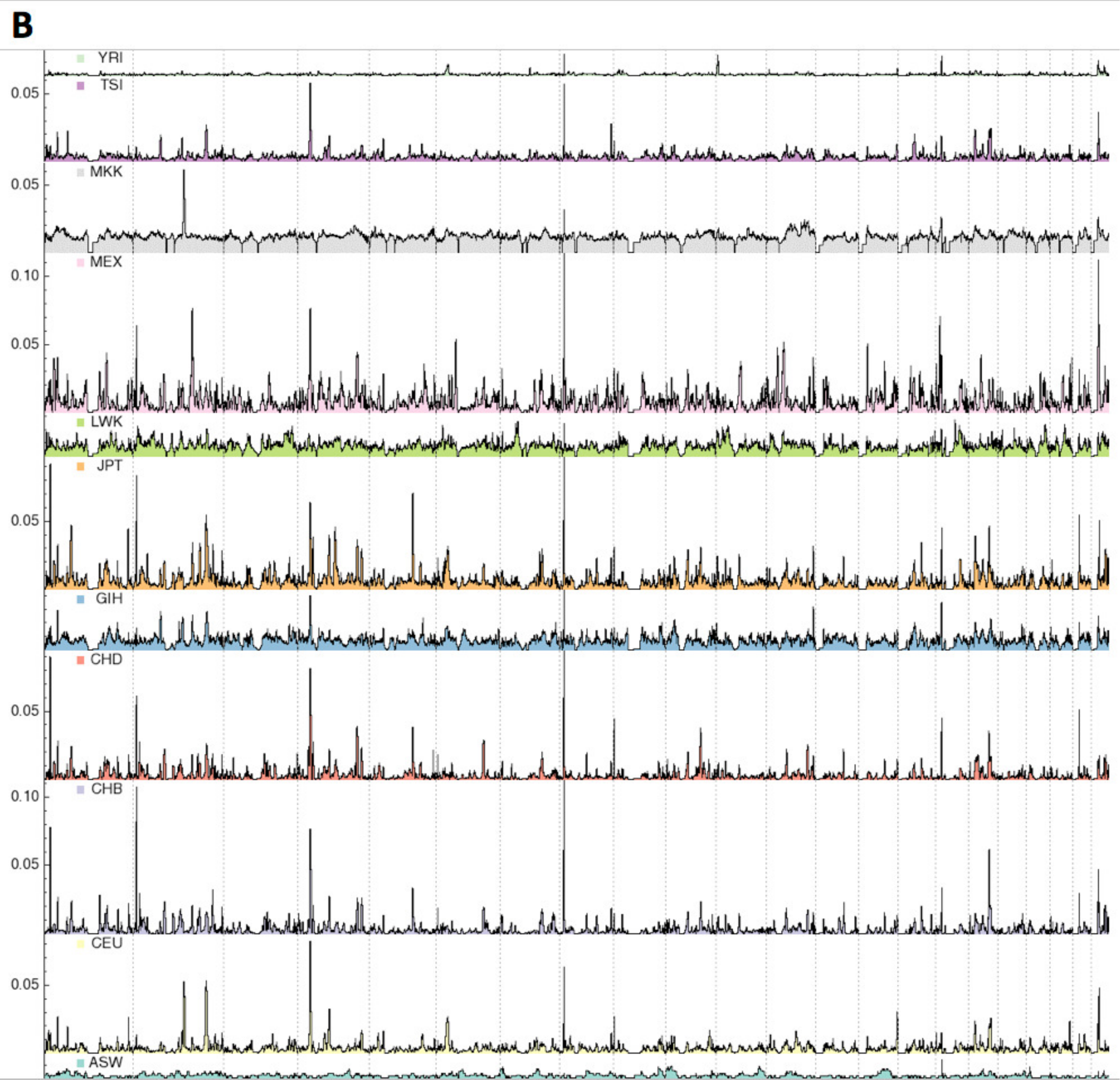}
    \end{subfigure}

    \begin{subfigure}[b]{0.9\textwidth}
		\vspace{-2mm}
    		\centering
    		\includegraphics[width=\textwidth]{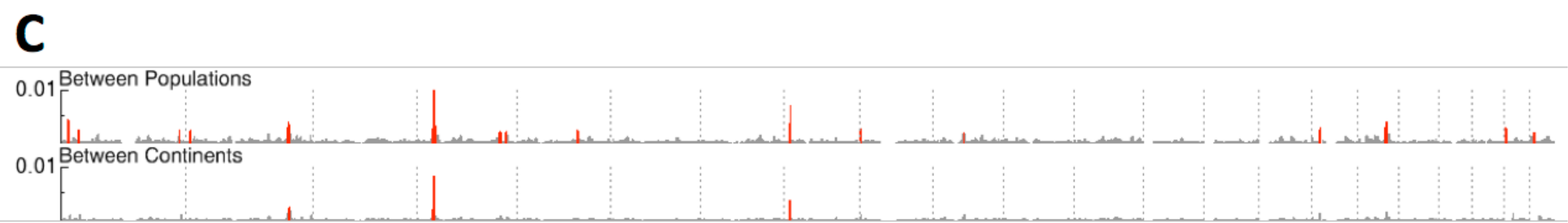}
    \end{subfigure}

	\caption{The physical distribution of IBD sharing along the genome within populations of the Idb dataset (A), the HMP3 dataset (B) and across continents/populations of the HMP3 dataset (using a different scale).}
	\label{fig:IBDdata:sel}
\end{figure*}

In order to examine locus-specific phenomena, we focus our analysis on local segment sharing due to intermediate and remote relatedness rather than genome-wide sharing between close relatives. IBD sharing is detected everywhere along the genome, averaging population-specific background levels (Figure \ref{fig:IBDdata:sel}). We analyzed the physical distribution of IBD sharing within and across populations, observing regions with a much higher amount of sharing than expected. Analyzing AJ samples, the most prominent such region is the human leukocyte antigen (HLA) locus. The entire segment of chromosome $6$, between $25$ and $35$ Mb, is shared among individuals unrecombined at least $4$-fold more than any other region in the genome ($4.2$-fold in Idb, $5.1$-fold in HUGR). This is in accordance with previous observations of complex haplotype structure along the HLA locus \cite{de2006high}.

Examining the regions of intense sharing within HM3 populations, HLA still exhibits a very high sharing density for some of the populations: Western Europeans (CEU), Gujarati Indians (GIH), Luhya Kenyans (LWK), and Yoruba Nigerians (YRI). Additional regions along the genome exhibit notably high sharing densities within populations. Interestingly, many of these tend to also recur across unrelated individuals of different geographical origin. Segments at the recurrently shared regions in chromosomes $2$, $4$, and $8$ are shared even across different continents of origin. Of particular interest may be the most commonly shared region, on chromosome $8p23.1$, overlapping $5$ Mb of a common inversion polymorphism, the third longest reported structural variant in the entire genome \cite{iafrate2004detection}.

In total, the 16 cross-population commonly shared regions span only $<35$ Mb ($<0.92$\%) of the genome but account for $9.6\%$, $16.1\%$, and $18.1\%$ of sharing within populations, between populations, and between continents, respectively. We note that these regions are not correlated to SNP density and would be unaffected by slight changes in the information content filtering. Although sharing of a region may indicate recent common ancestry, the agglomeration of shared segments at 16 loci is highly nonrandom. Biological factors or recent positive selection are possible causes of the observed reduction in haplotype diversity. Some of the identified loci correspond to previously reported regions of recent positive selection. In particular, 8 of the 16 regions were reported: $1p34.3$, $2q32.3$ \cite{voight2006map}; $4p15$ \cite{voight2006map,sabeti2002detecting,pickrell2009signals}; $4q32.1$, $17q22$ \cite{sabeti2002detecting,pickrell2009signals}; $10q21.1$, $21q21.1$, $22q11.22$ \cite{pickrell2009signals}; an overlap not expected by chance ($p<0.0017$ based on permutations). Further evidence for biological retention of unrecombined ancient haplotypes, rather than random retention of new ones, comes from examining annotation for these 16 commonly shared segments. Seeking commonalities, we observe 12 of these segments to overlap structural variants that are common and long enough to have been detected in the HapMap by CGH (\cite{iafrate2004detection,perry2008fine}). Such overlap is not expected by chance ($p<0.00052$ in 100 longest based on permutations).

\section{Reconstructing demographic events of the Jewish diasporas}

The descriptive statistic of IBD sharing and the methods to analyze them that were developed in the previous section outline the potential of relying on shared haplotypes to gain insight into recent demographic events. In a series of three papers \cite{atzmon2010abraham,campbell2012north,velez2012impact}, we used these and other methods to study the signature of recent demographic variation in SNP array datasets comprising individuals from the Jewish Diaspora. The demographic events that shaped relatedness in these groups are expected to have occurred during recent millennia, and individuals from Jewish cohorts are expected to share increased IBD sharing as a result of cultural isolation following the diaspora events, motivating this analysis. In this section, we report main results and methodological development of these works, limiting the discussion to analyses of IBD sharing in these datasets. Additional analyses may be found in \cite{atzmon2010abraham,campbell2012north,velez2012impact}.

\subsection{Jewish communities of the Mediterranean}

Participants for this study were recruited from the Iranian (IRN, $28$ samples), Iraqi (IRQ, $37$ samples), Syrian (SYR, $25$ samples), Ashkenazi (ASH, $34$ samples), Greek Sephardic (GRK, $42$ samples), Turkish Sephardic (TUR, $34$ samples) and Italian (ITJ, $37$ samples) Jewish communities, and included only if all four grandparents came from the same Jewish community. Subjects were excluded if they were known first- or second-degree relatives of other participants or were found to have  $\hat{\pi}\geq.30$ by analysis of microarray data using the PLINK software \cite{purcell2007plink}. Genotyping was performed with the Affymetrix Genome-Wide Human SNP Array 6.0 (Affy v 6). In addition to these groups, we sometimes included in the analysis a subset of populations extracted from the Human Genome Diversity Panel (HGDP), and the PopRes datasets.

IBD segments were detected with the GERMLINE algorithm in Genotype Extension \cite{gusev2012architecture}. The output of GERMLINE was used to detect unreported close relatives, who were omitted from the analysis. Two individuals were considered cryptic relatives if their total sharing was observed larger than $1,500$ cM and if the average segment length was more than $25$ cM, suggesting an avuncular or closer relationship. The output was also used to produce sharing densities, sharing graphs, and sharing statistics.

GERMLINE output was filtered to ensure consistency across genotyping platforms and to remove noise by filtering out regions of low information content. SNP density in sliding, non-overlapping blocks across the genome was used to filter shared segments that spanned SNP-sparse regions, particularly the edges of the centromere and telomere. Specifically, regions that presented less than $100$ SNPs per megabase or $100$ SNPs per centimorgan were identified and excised and, subsequently, shared segments that were shorter than $3$ cM were removed.

The amount of sharing for the analyzed data set was visualized with the ShareViz software, developed in \cite{gusev2012architecture}. As described in the previous section, individuals were represented as nodes, grouped into populations of origin. The thickness of the edges between nodes represent the total amount of sharing (in centimorgans) between each pair of individuals. For presenting populations geographically, planar quasi-isometric embedding (ISOMAP \cite{tenenbaum2000global}) was used, where distances between populations were defined as inverse of the populations' pairwise average.

To compute the average total sharing between populations I and J, the following expression was used:
\begin{equation}
\begin{split}
	W_{IJ} = \frac{\sum_{i\in I}\sum_{j \in J} W_{ij}}{nm}
\end{split}
\end{equation}
where $W_{ij}$ is the total sharing between individuals $i$ and $j$ from populations $I$ and $J$, respectively, and $n$ and $m$ are the number of individuals in populations $I$ and $J$. The average lengths of the shared segments across populations were computed through the arithmetic mean of the shared segments for each pair of populations.

IBD between Jewish individuals exhibited high frequencies of shared segments (Table \ref{tbl:IBDdata:IBDJ}). The median pair of individuals within a community shared a total of $50$ cM IBD (quartiles: $23.0$ cM and $92.6$ cM). Such levels are expected to be shared by $4$th or $5$th cousins in a completely outbred population. However, the typical shared segments in these communities were shorter than expected between $5$th cousins ($8.33$ cM length), suggesting multiple lineages of more remote relatedness between most pairs of Jewish individuals.

\begin{sidewaystable*}
\centering
\tiny
\begin{tabular}{|c|c|c|c|c|c|c|c|c|c|c|c|c|c|c|c|c|c|}
	 \hline
	 & N\footnote{number of samples} & IRN & IRQ & SYR & ASH & ITJ & GRK & TUR & N\_Italian & Sardinian & French & Basque & Adygei & Russian & Palestinian & Druze & Bedouin \\ \hline
	IRN & 29 & 41.95 & 4.91 & 1.00 & 0.75 & 0.61 & 0.56 & 0.75 & 0.68 & 0.67 & 0.50 & 0.58 & 0.60 & 0.47 & 0.53 & 0.66 & 0.57 \\ \hline
	IRQ & 40 & 4.91 & 33.36 & 3.14 & 0.83 & 0.86 & 0.77 & 1.04 & 0.74 & 0.68 & 0.62 & 0.66 & 0.50 & 0.52 & 0.64 & 0.64 & 0.61 \\ \hline
	SYR & 25 & 1.00 & 3.14 & 17.26 & 1.93 & 1.57 & 1.57 & 2.05 & 0.86 & 0.97 & 1.00 & 0.85 & 0.66 & 0.62 & 0.60 & 0.75 & 0.56 \\ \hline
	ASH & 34 & 0.75 & 0.83 & 1.93 & 11.62 & 3.09 & 2.15 & 2.95 & 1.01 & 1.10 & 1.01 & 1.15 & 0.74 & 0.91 & 0.58 & 0.78 & 0.58 \\ \hline
	ITJ & 37 & 0.61 & 0.86 & 1.57 & 3.09 & 28.45 & 2.48 & 2.41 & 0.98 & 0.85 & 0.95 & 0.86 & 0.75 & 0.82 & 0.66 & 0.67 & 0.57 \\ \hline
	GRK & 42 & 0.56 & 0.77 & 1.57 & 2.15 & 2.48 & 6.01 & 2.56 & 0.91 & 0.96 & 0.89 & 0.93 & 0.81 & 0.64 & 0.61 & 0.74 & 0.54 \\ \hline
	TUR & 34 & 0.75 & 1.04 & 2.05 & 2.95 & 2.41 & 2.56 & 4.46 & 0.90 & 0.95 & 0.94 & 0.90 & 0.70 & 0.81 & 0.71 & 0.75 & 0.59 \\ \hline
	N\_Italian & 21 & 0.68 & 0.74 & 0.86 & 1.01 & 0.98 & 0.91 & 0.90 & 2.37 & 1.39 & 1.36 & 1.43 & 0.84 & 1.24 & 0.51 & 0.66 & 0.61 \\ \hline
	Sardinian & 28 & 0.67 & 0.68 & 0.97 & 1.10 & 0.85 & 0.96 & 0.95 & 1.39 & 10.84 & 1.35 & 1.47 & 0.65 & 0.93 & 0.67 & 0.71 & 0.55 \\ \hline
	French & 29 & 0.50 & 0.62 & 1.00 & 1.01 & 0.95 & 0.89 & 0.94 & 1.36 & 1.35 & 1.63 & 2.08 & 0.83 & 1.46 & 0.55 & 0.59 & 0.53 \\ \hline
	Basque & 24 & 0.58 & 0.66 & 0.85 & 1.15 & 0.86 & 0.93 & 0.90 & 1.43 & 1.47 & 2.08 & 15.97 & 1.07 & 1.21 & 0.59 & 0.67 & 0.54 \\ \hline
	Adygei & 17 & 0.60 & 0.50 & 0.66 & 0.74 & 0.75 & 0.81 & 0.70 & 0.84 & 0.65 & 0.83 & 1.07 & 6.29 & 0.91 & 0.56 & 0.80 & 0.39 \\ \hline
	Russian & 25 & 0.47 & 0.52 & 0.62 & 0.91 & 0.82 & 0.64 & 0.81 & 1.24 & 0.93 & 1.46 & 1.21 & 0.91 & 5.80 & 0.48 & 0.57 & 0.37 \\ \hline
	Palestinian & 51 & 0.53 & 0.64 & 0.60 & 0.58 & 0.66 & 0.61 & 0.71 & 0.51 & 0.67 & 0.55 & 0.59 & 0.56 & 0.48 & 25.50 & 0.62 & 1.01 \\ \hline
	Druze & 47 & 0.66 & 0.64 & 0.75 & 0.78 & 0.67 & 0.74 & 0.75 & 0.66 & 0.71 & 0.59 & 0.67 & 0.80 & 0.57 & 0.62 & 49.59 & 0.65 \\ \hline
	Bedouin & 48 & 0.57 & 0.61 & 0.56 & 0.58 & 0.57 & 0.54 & 0.59 & 0.61 & 0.55 & 0.53 & 0.54 & 0.39 & 0.37 & 1.01 & 0.65 & 25.36 \\ \hline
\end{tabular}
\caption{IBD sharing within and across Jewish communities and other populations from the HGDP and PopRes datasets.}
\label{tbl:IBDdata:IBDJ}
\end{sidewaystable*}

\begin{figure*}
    \centering
    \begin{subfigure}[b]{1\textwidth}
		\vspace{-2mm}
            \centering
            \includegraphics[width=\textwidth]{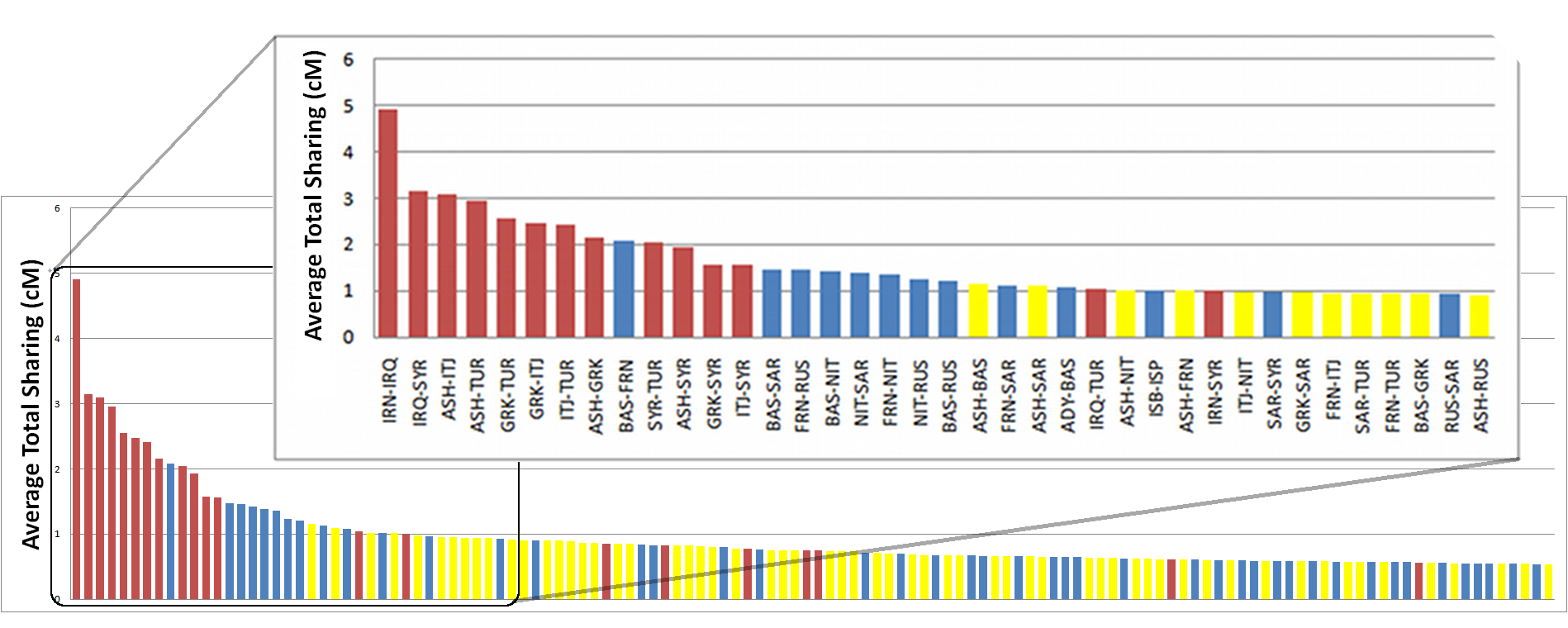}
            \caption{Cross-population average genome-wide IBD sharing per individual pair. Colors represent sharing between two Jewish communities (red), between a Jewish community and a non-Jewish community (yellow) and between non-Jewish communities (blue).}
		\label{fig:IBDdata:IBDJ:A}
    \end{subfigure}

    \begin{subfigure}[b]{0.408\textwidth}
            \centering
            \includegraphics[width=\textwidth]{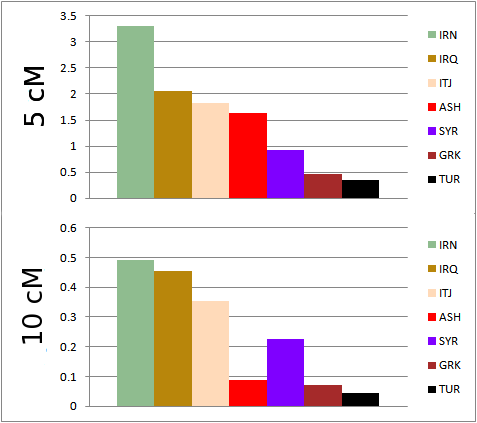}
            \caption{Decay of IBD sharing.}
		\label{fig:IBDdata:IBDJ:B}
    \end{subfigure}
    ~
    \begin{subfigure}[b]{0.6\textwidth}
            \centering
            \includegraphics[width=\textwidth]{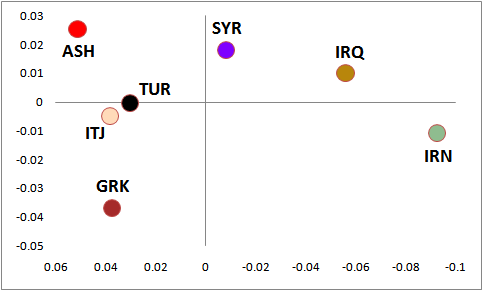}
            \caption{Isomap embedding of IBD sharing.}
		\label{fig:IBDdata:IBDJ:C}
    \end{subfigure}
    ~
	\caption{Summary of IBD sharing for Jewish communities.}
	\label{fig:IBDdata:IBDJ}
\end{figure*}

Within the different Jewish communities, three distinct patterns were observed. The Greek and Turkish Jews had relatively modest levels of IBD, similar to that observed in the French HGDP samples. The Italian, Syrian, Iranian, and Iraqi Jews demonstrated the high levels of IBD that would be expected for extremely inbred populations. Unlike the other populations, the Ashkenazi Jews exhibited increased sharing of segments at the shorter end of the range (i.e., $5$ cM length), but decreased sharing at the longer end (i.e., $10$ cM) (Figure \ref{fig:IBDdata:IBDJ:B}).

\begin{figure*}
    \centering
	\includegraphics[width=0.8\textwidth]{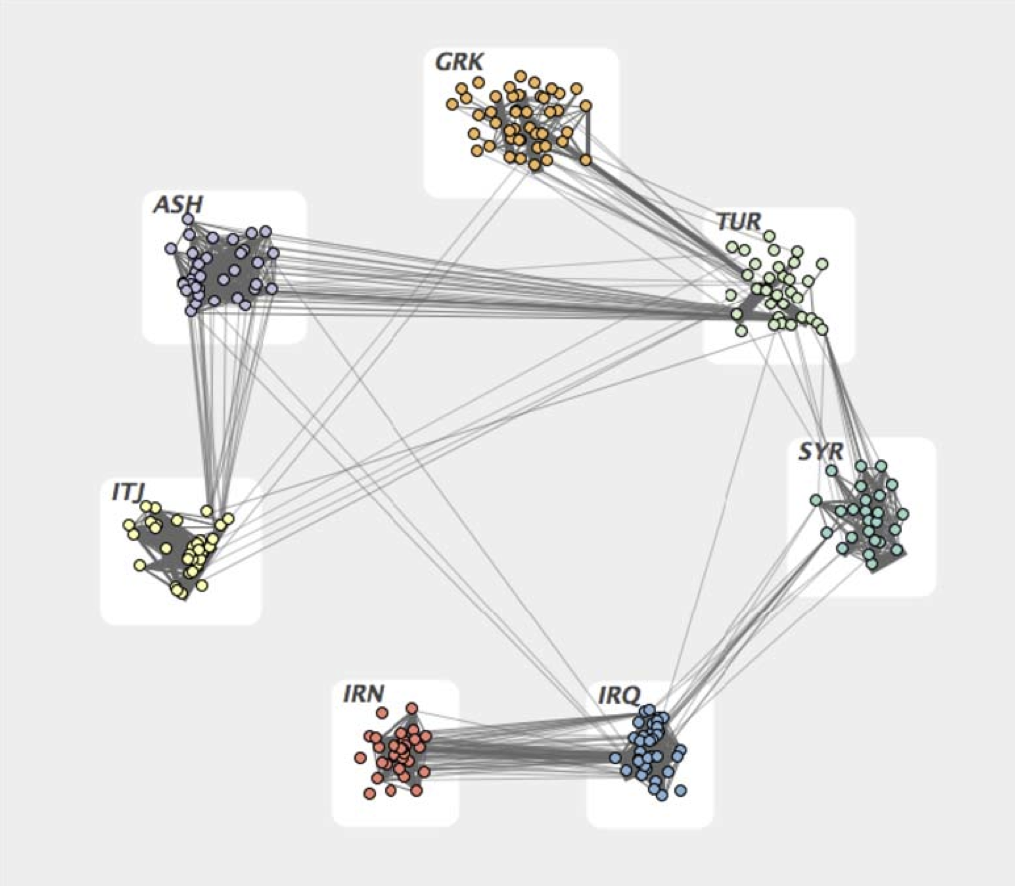}
	\caption{IBD sharing graph for the Jewish Hapmap groups.}
    \label{fig:IBDdata:IBD_graph_J}
\end{figure*}

As expected, the vast majority of long shared segments ($89\%$ of $15$ cM segments, $78\%$ of $10$ cM segments) were shared within communities. However, the genetic connections between the Jewish populations became evident from the frequent IBD across these Jewish groups ($63\%$ of all shared segments). The web of relatedness between the $27,966$ pairs of individuals in this study was intricate, even if restricted only to the $2,166$ pairs sharing a total $50$ cM or more, a level of sharing among third cousins (Figure \ref{fig:IBDdata:IBD_graph_J}). When population averages were examined, this network of IBD was consistent with the geographic distances between populations, with planar embedding representing $93\%$ of the initial information content (Figure \ref{fig:IBDdata:IBDJ:C}). The notable exception was that of Turkish and Italian Jews who were nearest neighbors in terms of IBD, but more distant on the geographical map, potentially reflecting their shared Sephardic ancestry. Jewish populations shared more and longer segments with one another than with non-Jewish populations, highlighting the commonality of Jewish origin. Among pairs of populations ordered by total sharing, $12$ out of the top $20$ were pairs of Jewish populations, and none of the top $30$ paired a Jewish population with a non-Jewish one (Figure  \ref{fig:IBDdata:IBDJ:A}).

\subsection{IBD sharing is enriched for Sephardic ancestry in modern Latino populations}

Modern day Latin America resulted from the encounter of Europeans with the indigenous peoples of the Americas in $1492$, followed by waves of migration from Europe and Africa. As a result, the genomic structure of present day Latin Americans was determined both by the genetic structure of the founding populations and the numbers of migrants from these different populations. In (\cite{velez2012impact}), we analyzed DNA collected from two well-established communities in Colorado (Hispanos, $33$ unrelated individuals) and Ecuador (Lojanos, $20$ unrelated individuals) with a measurable prevalence of the $BRCA1$ $c.185delAG$ and the $GHR$ $c.E180$ mutations, respectively, using Affymetrix Genome-wide Human SNP $6.0$ arrays to identify their ancestry. These mutations are found at relatively high frequency in Sephardic Jewish individuals, suggesting they may have been brought to these communities through Jewish migration from the realms that comprise modern Spain and Portugal during the Age of Discovery. In this work, several analyses identified enrichment for Sephardic Jewish ancestry. We here report a summary of IBD sharing analysis performed in this dataset.

For this analysis, the Hispano and Lojano datasets were combined with (1) $237$ samples from the Jewish HapMap Project (Affymetrix $6.0$), including Iranian, Iraqi, Syrian, Italian, Turkish, Greek and Ashkenazi Jews \cite{atzmon2010abraham}, described in the previous section; (2) $4$ US Hispanic/ Latino populations ($27$ Dominicans, $26$ Colombians, and $20$ Ecuadorians, as well as $27$ Puerto Ricans) from Illumina $610$ K arrays \cite{bryc2010genome}; (3) $50$ US Mexican samples from HapMap3 (Affymetrix $6.0$) \cite{altshuler2010integrating}.
We phased the genotype data for each group using the Beagle software package \cite{browning2007rapid}, then detected IBD segments using GERMLINE \cite{gusev2009whole} in Genotype Extension mode (preferred to the haplotype mode due to heterogeneous sample size and demographic background of the analyzed groups). The identified segments were used to exclude close relatives (sharing at least $800$ cM and at least ten segments of length $\geq 10$ cM ) from the analysis, obtain statistics on the average total IBD sharing within and across groups and identify cross-population regions of increased sharing. The total sharing between an average pair of individuals from two different populations was computed summing the length (in cM) of all IBD segments detected across the two populations and normalizing by the number of possible pairs of individuals (the product of the cardinality for the two groups). We normalized by $n \choose 2$ possible pairs when computing the average total sharing within a population of sample size $n$.

\begin{table*}
\begin{subtable}{1\linewidth}
	\centering
	\footnotesize
	\begin{tabular}{|c|c|c|c|c|c|c|c|c|c|c|c|c|c|c|c|c|c|}
	 \hline
	ASH & IRN & IRQ & SYR & ITL & GRK & TUR & HSP & LSN & MEX & TSI & CEU & CHB & YRI \\ \hline
	77.0 & 170.0 & 152.2 & 89.8 & 130.6 & 50.2 & 42.2 & 113.3 & 131.0 & 71.3 & 49.6 & 59.9 & 75.2 & 8.9 \\ \hline
	\end{tabular}
	\caption{IBD sharing within Jewish communities and other populations from the HapMap dataset.}
	\label{tbl:IBDdata:IBDJ_w}
\end{subtable}

\footnotesize
\begin{subtable}{1\linewidth}
\begin{tabular}{|c|c|c|c|c|c|c|c|c|c|c|c|c|c|c|c|c|c|}
	 \hline
	 & ASH & IRN & IRQ & SYR & ITL & GRK & TUR & HSP & LSN & MEX & TSI & CEU & CHB \\ \hline
	IRN & 11.92 & & & & & & & & & & & &  \\ \hline
	IRQ & 16.02 & 31.73 & & & & & & & & & & & \\ \hline
	SYR & 17.96 & 15.43 & 31.59 & & & & & & & & & & \\ \hline
	ITL & 24.51 & 15.20 & 24.59 & 25.74 & & & & & & & & & \\ \hline
	GRK & 21.35 & 15.05 & 24.99 & 27.00 & 33.68 & & & & & & & & \\ \hline
	TUR & 22.93 & 15.61 & 26.24 & 28.72 & 33.55 & 33.69 & & & & & & & \\ \hline
	HSP & 12.16 & 10.16 & 17.41 & 16.69 & 18.81 & 18.96 & 19.87 & & & & & & \\ \hline
	LSN & 8.36 & 7.73 & 12.41 & 11.62 & 13.01 & 12.74 & 13.57 & 43.75 & & & & & \\ \hline
	MEX & 10.34 & 9.25 & 14.80 & 14.74 & 15.95 & 15.95 & 17.15 & 52.07 & 53.69 & & & & \\ \hline
	TSI & 19.08 & 17.57 & 28.17 & 27.22 & 30.15 & 30.13 & 31.10 & 26.37 & 17.80 & 22.71 & & & \\ \hline
	CEU & 19.69 & 16.37 & 25.97 & 25.42 & 29.45 & 29.28 & 30.83 & 30.05 & 20.55 & 25.75 & 45.31 & & \\ \hline
	CHB & 1.88 & 1.87 & 3.18 & 2.42 & 2.52 & 2.65 & 2.69 & 10.43 & 10.03 & 12.15 & 3.56 & 3.88 & \\ \hline
	YRI & 0.01 & 0.01 & 0.03 & 0.02 & 0.03 & 0.02 & 0.02 & 0.05 & 0.02 & 0.08 & 0.03 & 0.02 & 0.02 \\ \hline
\end{tabular}
\caption{IBD sharing across Jewish communities and other populations from the HapMap dataset.}
\label{tbl:IBDdata:IBDJ_a}
\end{subtable}
\caption{IBD sharing within and across Jewish communities and other populations from the HapMap dataset.}
\end{table*}

Identity-by-descent showed elevated cross-population sharing between Hispano, Lojano and Mexican samples. The frequency of identity-by-descent (IBD) between unrelated individuals in a population is indicative of effective population size \cite{wright1931evolution}. We therefore analyzed the average genome-wide levels of IBD sharing within Latino ethnic groups. IBD sharing within Hispano and Lojano samples was higher than within other populations in this study, suggesting correspondingly higher levels of endogamy (Table \ref{tbl:IBDdata:IBDJ_w}).
We further analyzed rates of IBD sharing across different groups to investigate shared ancestry. Elevated cross-population sharing between Hispano, Lojano and Mexican samples (Table \ref{tbl:IBDdata:IBDJ_a}) was consistent with shared recent ancestry. When investigating potential shared ancestry between these groups and other populations, we observed that multiple populations shared segments IBD with Latinos (Table \ref{tbl:IBDdata:IBDJ_a}). More specifically, highest rates of such Latino-IBD sharing were observed in European and Tuscan samples followed by Sephardic and Mizrahi (Iranian, Iraqi and Syrian) Jewish communities. Lower rates of IBD were observed versus Ashkenazi samples in the Lojano samples, and to the Chinese group in Hispanos and Mexicans. Negligible IBD sharing with Yoruba samples was observed for all populations.

Besides detecting IBD sharing, we used the Xplorigin software package \cite{bonnen2009european} to investigate the proportion of European, Native American and Jewish ancestry of Hispano and Lojano samples in comparison to another Hispanic/Latino cohort from Mexico. Xplorigin builds a database of short haplotype frequencies for three reference populations, which are assumed to be the source of admixture for a studied group of samples. The haplotype frequencies are probabilistically used to assign locus-specific ancestry proportions to the analyzed individuals. Ancestry deconvolution was also applied to investigate the remote origin of regions shared IBD across populations.

We trained the Xplorigin software using $98$ randomly selected phased haplotypes from the following groups: European Basque and French from the HGDP dataset; Sephardic Italian, Greek and Turkish from the Jewish HapMap dataset; Native American Pima, Surui and Maya samples from the HGDP dataset. After pruning some markers during computational phasing, the number of makers used for this cross-platform analysis was $150,157$ SNPs. For each of the three reference groups we determined LD blocks and the frequency of haplotypes and transitions between haplotypes using Haploview \cite{barrett2005haploview}. The genome was then partitioned into short haplotype blocks, and Xplorigin's hidden Markov model was used to assign the most likely proportion of ancestry from the three reference populations to each observed individual.

We analyzed the proportions of ancestry in correspondence of IBD segments within and across populations. To overcome phase uncertainty for an IBD segment shared by two individuals, we considered the ancestry of both maternal and paternal chromosomes reported by Xplorigin in correspondence of the IBD region. The values reported in Table \ref{tbl:IBDdata:deconvIBD} were computed as follows: given a number of IBD segments between individuals of two populations $P1$ and $P2$, we report the average proportion of IBD ancestry of individuals from $P1$ in position $P1-P2$ of the table, and the average proportion of IBD ancestry of individuals from $P2$ in position $P2-P1$. The ancestry of IBD sharing within a population (table entries in positions $P1-P1$) was computed for both individuals of an IBD sharing pair. The reported mean ancestry proportion is computed as the genome-wide average ancestry proportion. To test for significance of the differences between genome-wide ancestry proportion and IBD ancestry proportions we performed random permutations of the IBD segments. We randomly shuffled IBD segments between populations $P1$ and $P2$, testing the ancestry proportions for the permuted set of IBD segments. The deviation from the genome-wide averages in correspondence of IBD segments was never observed for $1,000$ random permutations of each table entry.

\begin{table*}	
	\centering
	\begin{subtable}{.45\linewidth}
		\centering
	    \begin{tabular}{c|c|c|c|c}
			 & MEAN & MEX & HSP & LSN \\ \hline
			MEX & 0.309 & 0.206 & 0.239 & 0.205 \\ \hline
			HSP & 0.362 & 0.276 & 0.344 & 0.265 \\ \hline
			LSN & 0.312 & 0.22 & 0.239 & 0.291 \\
	    \end{tabular}
		\caption{European ancestry.}
	\end{subtable}
	
	\begin{subtable}{.45\linewidth}
		\centering
	    \begin{tabular}{c|c|c|c|c}
			 & MEAN & MEX & HSP & LSN \\ \hline
			MEX & 0.297 & 0.177 & 0.205 & 0.171 \\ \hline
			HSP & 0.342 & 0.226 & 0.327 & 0.211 \\ \hline
			LSN & \textcolor{blue}{0.305} & 0.203 & 0.232 & \textcolor{red}{0.342} \\
	    \end{tabular}
		\caption{Sephardic ancestry.}
	\end{subtable}

	\begin{subtable}{.45\linewidth}
		\centering
	    \begin{tabular}{c|c|c|c|c}
			 & MEAN & MEX & HSP & LSN \\ \hline
			MEX & \textcolor{blue}{0.394} & \textcolor{red}{0.617} & \textcolor{red}{0.557} & \textcolor{red}{0.624} \\ \hline
			HSP & \textcolor{blue}{0.296} & \textcolor{red}{0.498} & \textcolor{red}{0.328} & \textcolor{red}{0.524} \\ \hline
			LSN & \textcolor{blue}{0.382} & \textcolor{red}{0.576} & \textcolor{red}{0.53} & 0.367 \\
	    \end{tabular}
		\caption{Native American ancestry.}
\end{subtable}
\caption{Enrichment of ancestral components in IBD segments (red colors indicate statistically significant enrichment).}
\label{tbl:IBDdata:deconvIBD}
\end{table*}

Ancestry deconvolution showed sharing compatible with a history of Latino admixture with Europeans, Native Americans and Sephardic Jews. Many Latino populations are well known to include genetic ancestry components from Native Americans, Europeans and Africans, all admixed within the last $20$ generations. Comparing potential European, Sephardic Jewish and Native American ancestry, we observed proportions compatible with a history of Latino admixture from these three ethnicities (Table \ref{tbl:IBDdata:deconvIBD}). The Hispano samples showed increased European ancestry, whereas the Lojanos and Mexicans showed increased Native American ancestry. We further considered the IBD-shared segments among Latino samples, to explore correlation between the occurrence of such segments and admixture source population. Interestingly, these segments across all examined Latino populations were substantially enriched for Native American ancestry. As such segments indicate a recent common ancestor of the samples who share them. This indicates a small number of recent Native American founders, relative to other source populations. When considering IBD sharing in each Latino group separately, we further observed IBD sharing is also enriched for Sephardic ancestry ($p<0.001$) within the Lojano community. The relative enrichments in Sephardic versus European ancestry in IBD-shared segments proved robust to the choice of a source population, showing similar results when compared to Yoruba, who likely did not contribute significantly in terms of ancestry (see Table 2S in \cite{velez2012impact}).

\subsection{Jewish communities in North Africa}

Methods developed in the previous sections for the analysis of IBD sharing and ancestry deconvolution were adopted to analyze a dataset comprising several North African Jewish communities, together with samples from the previously described Jewish communities and several other North African populations, for a total of $509$ Jewish samples from $15$ populations and $114$ non-Jewish individuals from seven North African populations \cite{henn2012genomic} (samples listed in Table \ref{tbl:IBDdata:NAsamples}). Details of this and other analysis can be found in \cite{campbell2012north}.

\begin{table*}
\centering
\footnotesize
\begin{tabular}{|c|c|c|c|c|}
	 \hline
	Population ID & Female & Male & Total & Population \\ \hline
	ALGJ & 23 & 1 & 24 & Algerian Jewish \\ \hline
	ASHJ & 14 & 20 & 34 & Ashkenazi Jewish \\ \hline
	DJEJ & 0 & 17 & 17 & Djerban Jewish \\ \hline
	ETHJ & 13 & 3 & 16 & Ethiopian Jewish \\ \hline
	GEOJ & 4 & 9 & 13 & Georgian Jewish \\ \hline
	GRKJ & 25 & 29 & 54 & Greek Jewish \\ \hline
	IRNJ & 22 & 27 & 49 & Iranian Jewish \\ \hline
	IRQJ & 25 & 28 & 53 & Iraqi Jewish \\ \hline
	ITAJ & 20 & 19 & 39 & Italian Jewish \\ \hline
	LIBJ & 31 & 6 & 37 & Libyan Jewish \\ \hline
	MORJ & 32 & 6 & 38 & Moroccan Jewish \\ \hline
	SYRJ & 15 & 21 & 36 & Syrian Jewish \\ \hline
	TUNJ & 24 & 5 & 29 & Tunisian Jewish \\ \hline
	TURJ & 24 & 10 & 34 & Turkish Jewish \\ \hline
	YMNJ & 36 & 0 & 36 & Yemini Jewish \\ \hline
	ADYG & 10 & 7 & 17 & Adygei \\ \hline
	ALGE & 9 & 9 & 18 & Algerian \\ \hline
	BASQ & 8 & 16 & 24 & Basque \\ \hline
	BEDN & 20 & 27 & 47 & Bedouin \\ \hline
	DRUZ & 32 & 13 & 45 & Druze \\ \hline
	EGYP & 0 & 19 & 19 & Egyptian \\ \hline
	FREN & 17 & 12 & 29 & French \\ \hline
	LIBY & 1 & 16 & 17 & Libyan \\ \hline
	MORN & 0 & 18 & 18 & N Moroccan \\ \hline
	MORS & 5 & 5 & 10 & S Moroccan \\ \hline
	MOZA & 9 & 19 & 28 & Mozabite \\ \hline
	NITA & 7 & 14 & 21 & N Italian \\ \hline
	PALN & 34 & 17 & 51 & Palestinian \\ \hline
	RUSS & 9 & 16 & 25 & Russian \\ \hline
	SARD & 12 & 16 & 28 & Sardinian \\ \hline
	SOCC & 0 & 17 & 17 & Saharan \\ \hline
	TUNI & 0 & 15 & 15 & Tunisian \\ \hline
	AFRI & 1 & 24 & 25 & Sub-Saharan African \\ \hline
	ASIA & 10 & 15 & 25 & Asian \\ \hline
	\end{tabular}
\caption{Analyzed samples from Mediterranean and North African Jewish communities, and other non-Jewish populations.}
\label{tbl:IBDdata:NAsamples}
\end{table*}

IBD discovery was performed as previously described, although only Jewish samples and non-Jewish populations from the same geographic regions were analyzed to maintain high-density of SNP markers in the cross-platform analysis. Ancestry deconvolution using Xplorigin was run on a subset of the analyzed populations, with respect to their Maghrebi, Middle Eastern, and European ancestry, using $36$ non-Jewish Tunisian Berber, $48$ Palestinian, and $48$ Basque reference haplotypes, respectively.

\begin{figure*}
    \centering
	\includegraphics[width=1\textwidth]{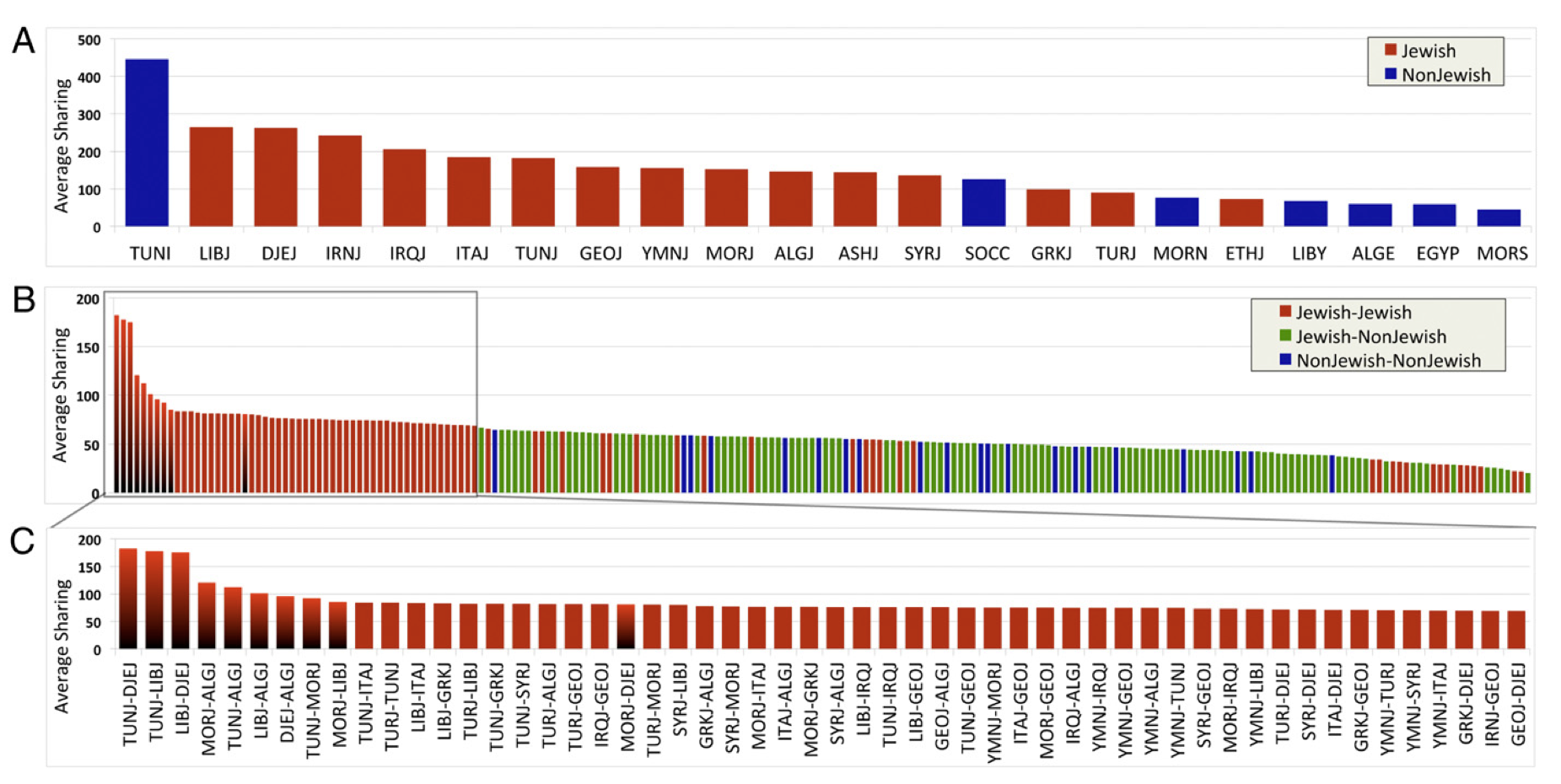}
	\caption{IBD sharing in North African communities. (A) Within groups (B) Across groups (C) Top sharing population pairs .}
    \label{fig:IBDdata:NAIBD}
\end{figure*}

As in the previously analyzed Jewish communities, North African Jewish populations showed a high degree of endogamy and IBD sharing between Jewish groups. We studied the frequency of IBD haplotypes shared by unrelated individuals within and across the analyzed groups. When IBD within populations was examined, the non-Jewish Tunisian Berbers exhibited the highest level of haplotype sharing, suggesting a small effective population size and high levels of endogamy (Figure \ref{fig:IBDdata:NAIBD}, panel A). With the exception of this Tunisian cohort, the Jewish populations generally showed higher IBD sharing than non-Jewish groups, indicating greater genetic isolation.

The relationships of the Jewish communities were outlined further by the IBD sharing across populations (Figure \ref{fig:IBDdata:NAIBD}, panels B and C), because the Jewish groups generally demonstrated closer relatedness with other Jewish communities than with geographically near non-Jewish populations. In particular, North African Jewish communities showed some of the highest levels of cross-population IBD sharing for the average pair of individuals. A strong degree of relatedness was observed across individuals from the Djerban, Tunisian, and Libyan Jewish communities. Noticeable proximity was also found between Jewish Algerian samples and other North African Jewish cohorts such as Moroccan, Tunisian, Libyan, and Djerban Jews, and across individuals from the Tunisian and Moroccan Jewish groups. Among non-Jewish North African groups, Algerians, South Moroccans, and West Saharan samples were found to share, on average, a smaller proportion of their genome IBD to other cohorts.

\begin{figure*}
    \centering
	\includegraphics[width=1\textwidth]{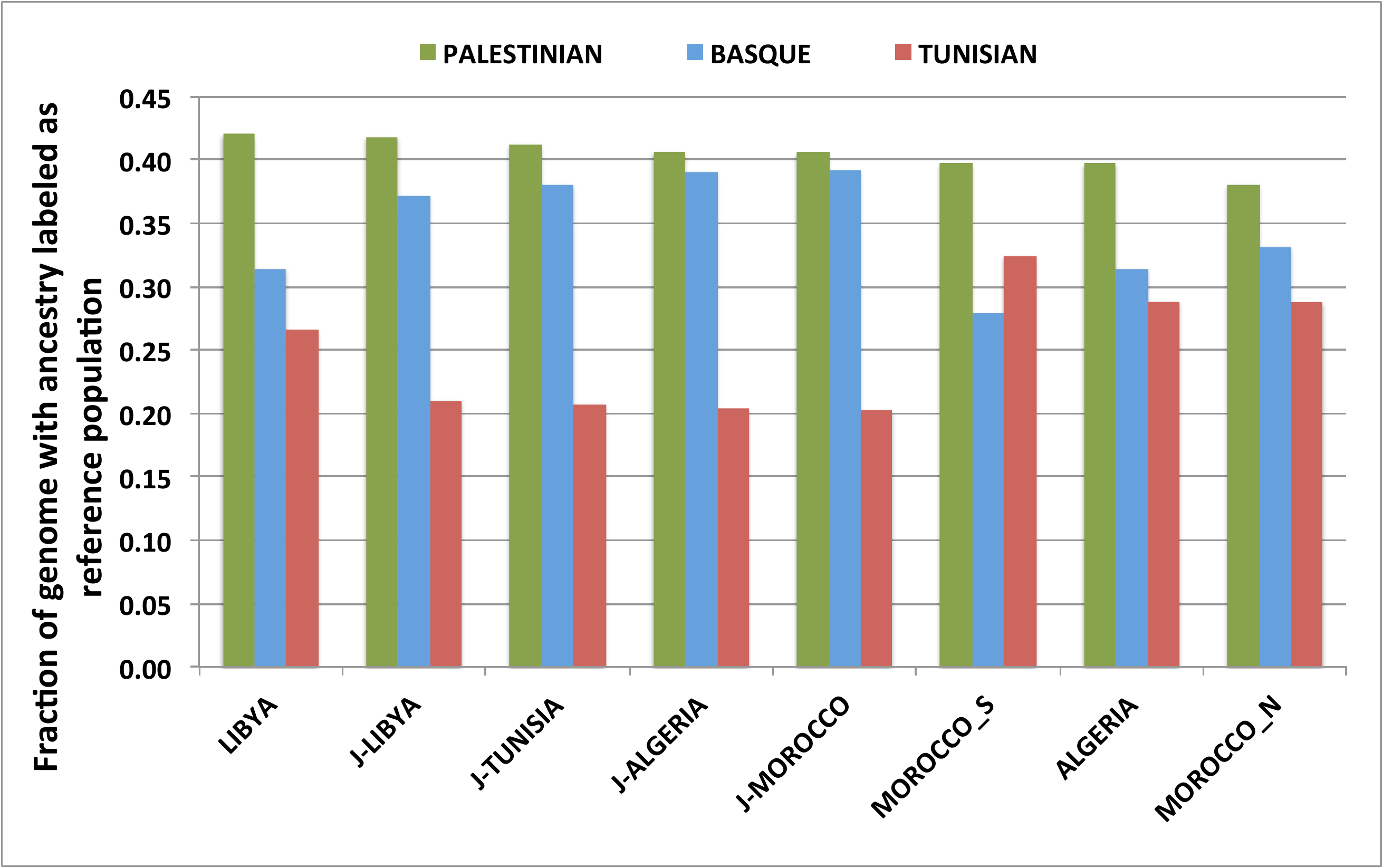}
	\caption{Ancestral deconvolution of Jewish and non-Jewish North African Communities.}
    \label{fig:IBDdata:NA-dec}
\end{figure*}

By using Xplorigin to perform ancestry deconvolution for a subset of the populations, the Maghrebi (Tunisian non-Jewish), European (Basque), and Middle-Eastern (Palestinian) ancestry components of North African Jewish communities were compared with the corresponding non-Jewish groups (Figure \ref{fig:IBDdata:NA-dec}). A stronger signal of European ancestry was found in the genomes of Jewish samples, with a decreased fraction of Maghrebi origins, whereas the Middle Eastern component was comparable across groups. In Jewish groups, geographical proximity to the Iberian Peninsula correlated with an increase in European ancestry and a decrease in Middle Eastern ancestry, whereas the Maghrebi component was only mildly reduced. Differences in ancestry proportions were found to be significant ($p<0.05$), except for the Maghrebi component of non-Jewish Northern Moroccan compared with non-Jewish Algerian samples, and the European component of Jewish Moroccan compared with Jewish Algerian samples.

\begin{table*}
	\begin{subtable}{.45\linewidth}
		\centering
		\small
		\begin{tabular}{|c|c c c c|c c c c|}
			 \hline
			& ALGE & LIBY & MORN & MORS & ALGJ & LIBJ & MORJ & TUNJ \\ \hline
			ALGE & 0.291 & \textcolor{OliveGreen}{0.291} & \textcolor{OliveGreen}{0.292} & \textcolor{red}{0.294} & \textcolor{OliveGreen}{0.259} & \textcolor{OliveGreen}{0.268} & \textcolor{OliveGreen}{0.253} & \textcolor{OliveGreen}{0.264} \\
			LIBY & \textcolor{OliveGreen}{0.274} & {0.249} & \textcolor{OliveGreen}{0.261} & \textcolor{red}{0.300} & \textcolor{OliveGreen}{0.241} & \textcolor{OliveGreen}{0.235} & \textcolor{OliveGreen}{0.249} & \textcolor{OliveGreen}{0.247} \\
			MORN & 0.312 & 0.295 & 0.297 & 0.317 & \textcolor{OliveGreen}{0.263} & \textcolor{OliveGreen}{0.259} & \textcolor{OliveGreen}{0.271} & \textcolor{OliveGreen}{0.275} \\
			MORS & \textcolor{OliveGreen}{0.323} & \textcolor{OliveGreen}{0.349} & \textcolor{red}{0.338} & \textcolor{red}{0.352} & \textcolor{OliveGreen}{0.345} & \textcolor{OliveGreen}{0.329} & \textcolor{OliveGreen}{0.348} & \textcolor{OliveGreen}{0.312} \\ \hline
			ALGJ & \textcolor{red}{0.205} & \textcolor{red}{0.203} & \textcolor{red}{0.207} & \textcolor{red}{0.221} & \textcolor{red}{0.192} & \textcolor{red}{0.194} & \textcolor{red}{0.196} & 0.199 \\
			LIBJ & \textcolor{red}{0.200} & 0.206 & \textcolor{red}{0.212} & \textcolor{red}{0.215} & \textcolor{red}{0.200} & 0.206 & \textcolor{red}{0.199} & 0.200 \\
			MORJ & \textcolor{red}{0.198} & 0.209 & \textcolor{red}{0.202} & 0.198 & 0.201 & 0.198 & \textcolor{OliveGreen}{0.202} & 0.201 \\
			TUNJ & \textcolor{red}{0.218} & 0.210 & \textcolor{red}{0.215} & \textcolor{red}{0.222} & \textcolor{red}{0.207} & \textcolor{red}{0.205} & 0.203 & \textcolor{red}{0.207} \\ \hline
		\end{tabular}
		\caption{Maghrebi ancestry.}
	\end{subtable}
	
	\begin{subtable}{.45\linewidth}
		\centering
		\small
		\begin{tabular}{|c|c c c c|c c c c|}
			 \hline
			& ALGE & LIBY & MORN & MORS & ALGJ & LIBJ & MORJ & TUNJ \\ \hline
			ALGE & 0.388 & \textcolor{red}{0.392} & 0.382 & 0.372 & 0.386 & 0.387 & 0.397 & 0.383 \\
			LIBY & 0.401 & 0.425 & 0.421 & 0.405 & 0.417 & 0.431 & 0.419 & 0.417 \\
			MORN & 0.368 & \textcolor{OliveGreen}{0.364} & 0.377 & \textcolor{OliveGreen}{0.361} & 0.377 & 0.395 & \textcolor{OliveGreen}{0.375} & 0.377 \\
			MORS & 0.371 & \textcolor{OliveGreen}{0.362} & \textcolor{OliveGreen}{0.362} & \textcolor{OliveGreen}{0.367} & 0.350 & \textcolor{OliveGreen}{0.355} & \textcolor{OliveGreen}{0.344} & 0.350 \\ \hline
			ALGJ & 0.380 & 0.406 & 0.385 & 0.398 & 0.405 & 0.393 & 0.398 & \textcolor{OliveGreen}{0.391} \\
			LIBJ & \textcolor{OliveGreen}{0.394} & 0.401 & \textcolor{OliveGreen}{0.395} & 0.395 & 0.398 & 0.410 & \textcolor{OliveGreen}{0.402} & 0.398 \\
			MORJ & \textcolor{OliveGreen}{0.368} & 0.372 & \textcolor{OliveGreen}{0.362} & \textcolor{OliveGreen}{0.395} & \textcolor{OliveGreen}{0.391} & \textcolor{OliveGreen}{0.368} & \textcolor{OliveGreen}{0.392} & \textcolor{OliveGreen}{0.391} \\
			TUNJ & \textcolor{OliveGreen}{0.396} & 0.411 & \textcolor{OliveGreen}{0.407} & 0.414 & \textcolor{OliveGreen}{0.409} & 0.420 & \textcolor{OliveGreen}{0.420} & 0.409 \\ \hline
		\end{tabular}
		\caption{Middle Eastern ancestry.}
	\end{subtable}

	\begin{subtable}{.45\linewidth}
		\centering
		\small
		\begin{tabular}{|c|c c c c|c c c c|}
			 \hline
			& ALGE & LIBY & MORN & MORS & ALGJ & LIBJ & MORJ & TUNJ \\ \hline
			ALGE & 0.291 & 0.335 & 0.326 & \textcolor{OliveGreen}{0.334} & \textcolor{red}{0.355} & \textcolor{red}{0.345} & \textcolor{red}{0.350} & \textcolor{red}{0.352} \\
			LIBY & \textcolor{red}{0.325} & 0.326 & \textcolor{red}{0.318} & 0.294 & \textcolor{red}{0.342} & \textcolor{red}{0.334} & \textcolor{red}{0.332} & \textcolor{red}{0.342} \\
			MORN & 0.321 & \textcolor{red}{0.341} & 0.326 & 0.322 & \textcolor{red}{0.360} & \textcolor{red}{0.346} & \textcolor{red}{0.354} & \textcolor{red}{0.360} \\
			MORS & \textcolor{red}{0.306} & \textcolor{red}{0.289} & \textcolor{red}{0.299} & \textcolor{red}{0.281} & \textcolor{red}{0.305} & \textcolor{red}{0.315} & \textcolor{red}{0.308} & \textcolor{red}{0.305} \\ \hline
			ALGJ & \textcolor{OliveGreen}{0.415} & 0.391 & 0.409 & 0.381 & 0.403 & 0.414 & 0.406 & 0.410 \\
			LIBJ & 0.406 & 0.393 & 0.393 & \textcolor{OliveGreen}{0.390} & 0.402 & 0.384 & \textcolor{red}{0.399} & 0.402 \\
			MORJ & 0.435 & 0.418 & \textcolor{red}{0.436} & \textcolor{red}{0.407} & \textcolor{red}{0.408} & \textcolor{red}{0.435} & \textcolor{red}{0.406} & \textcolor{red}{0.408} \\
			TUNJ & 0.386 & 0.380 & 0.378 & 0.365 & \textcolor{red}{0.384} & 0.375 & \textcolor{red}{0.378} & \textcolor{OliveGreen}{0.384} \\ \hline
		\end{tabular}
		\caption{European ancestry.}
	\end{subtable}
	\caption{Enrichment of ancestral components in IBD segments (green color indicates statistically significant depletion, red indicates enrichment).}
	\label{tbl:IBDdata:deconvIBDNA}
\end{table*}

In addition to genome-wide proportions, this ancestry painting analysis was intersected with regions that harbor long-range IBD haplotypes, as done in the analysis of Sephardic ancestry in Latin American populations. In Jewish populations, the ancestry proportions in corresponding IBD regions highlighted mild, but in some cases significant, deviations from genome-wide averages (Table \ref{tbl:IBDdata:deconvIBDNA}), whereas stronger differences were observed in the recent ancestry for the corresponding non-Jewish communities. In these groups, recently co-inherited regions exhibited significantly increased European ancestry, with significantly decreased Maghrebi ancestry, compared with genome-wide averages. This phenomenon was generally stronger for loci shared IBD with individuals from Jewish communities. This increase in European ancestry and corresponding decrease in Maghrebi ancestry may be interpreted in several ways: (1) This increase may be due to the inherently higher European ancestry of Jewish segments planted into the genomes of non-Jewish populations. (2) Alternatively, the difference in genome-wide ancestries between Jewish and non-Jewish groups alone could explain this observation in the case of recent symmetric gene flow in both directions. However, this second scenario alone is inconsistent with the data, because it would imply a comparable decrease of European ancestry in regions IBD to non-Jewish populations to be observed in Jewish genomes. (3) The observed increase of European ancestry could be similarly explained by European segments newly planted in both populations. This explanation is also unlikely, because it would result in a comparable increase of European ancestry in Jewish genomes, which is instead observed to only mildly increase compared with genome-wide averages. The increase in European ancestry is stronger in IBD regions of length between $3$ and $4$ cM, compared with regions at least $4$ cM long (see Supplementary Table 7 in \cite{campbell2012north}), which is compatible with European admixture occurring several generations before present, through ancestors that resided in the Iberian Peninsula.

\chapter{IBD sharing and demography}
\label{chap:IBDmodel}
The results described in Chapter \ref{chap:IBD_in_data} outlined that IBD sharing in purportedly unrelated individuals does carry information about population-scale features such as demographic history, population stratification and natural selection. This motivated developing a theoretical framework that combines coalescent theory and modeling of IBD sharing in pedigree structures, which enables quantitative approaches to studying hidden relatedness in relation to these population features. In this chapter, we describe this framework and show its application in the inference of recent demographic events in two real populations: Ashkenazi Jewish and Kenyan Maasai, which were both shown to harbor substantial IBD sharing across pairs of unrelated individuals, although such sharing seems to be emerging from distinct demographic backgrounds. The remainder of the text will occasionally refer to supplementary figures and tables. These can be consulted online\footnote{\url{http://www.sciencedirect.com/science/MiamiMultiMediaURL/1-s2.0-S0002929712004727/1-s2.0-S0002929712004727-mmc1.pdf/276895/FULL/S0002929712004727/524f0ffb44ca9b5087aa3a4c41eb0202/mmc1.pdf}} as supplementary materials of the article \cite{palamara2012length}, where additional details of the presented analysis can also be found.

\section{A coalescent-based model for the relationship between demographic events and shared IBD haplotypes}
\label{sec:model:methods}

As shown in Chapter \ref{chap:intro}, coalescent theory \cite{kingman1982coalescent} indicates that, at a specific locus of their genome, two haploid gametes from a Wright-Fisher population of constant (haploid) effective population size $N_e$ have a probability of $1/N_e$ of finding a common ancestor at each generation. The time (in generations before present [gbp]) for these two individual gametes to reach a most recent common ancestor (MRCA) when their lineages are traced back into the past is geometrically distributed and has an expected value of $N_e$. More generally, if a population is composed of $N(g)$ haploid individuals at generation $g$, then the chance of finding a common ancestor at that generation is ${N(g)}^{-1}$, and the time distribution to a common ancestor assumes a more complex form. The relationship between the probability of finding common ancestors and the size of a population is appealing for demographic reconstruction. One can in fact study the distribution of time to a common ancestor at the average genomic locus for many pairs of individuals and can therefore gain information on a population's size across different time scales.
In the proposed methodology, we rely on haplotype sharing to obtain a probabilistic estimate of the time to coalescence at any genomic site for any pair of individuals in the population at hand. The extent of a co-inherited IBD haplotype is probabilistically determined by the generation of the MRCA for the two individuals at the considered locus. Unfortunately, individual segments carry little information about specific sites unless the common ancestor is extremely recent (e.g., less than $10$ gbp \cite{huff2011maximum}). However, because we are interested in genome-wide, population-wide summary statistics, significant information can be gathered from a large number of segments co-inherited by different pairs of individuals from the analyzed population sample. In fact, the number of considered pairs grows quadratically with the sample size, and the number of expected IBD segments increases as shorter segment lengths are considered. Leveraging these principles, we derive analytical results for the distribution of IBD sharing across purportedly unrelated individuals. As detailed below, we express these quantities as a function of historical demography in the population.

\subsection{IBD and demographic history in Wright-Fisher populations}
Formally, consider a random pair of haploid individuals sampled from the studied population and a specific locus along their genome. Note that although we present this analysis in the context of haploid individuals, the following results are easily adapted to the case of diploid individuals by the appropriate multiplication or division by a factor of two. We are interested in modeling the probability that the chosen locus is spanned by a non-recombinant IBD segment of a specific genetic length. We abstract this length as a continuous random variable $L$ and denote its probability density function by $p(l|\theta)$, where $\theta$ encodes a parameterization of the population's demographic history. In the simplest case of a constant population size, $\theta$ is only parameterized by the constant population size $N_e$. We assume neutrality throughout; therefore, this is a Wright-Fisher population \cite{wright1931evolution}, and we employ the notation $\theta=\theta_{WF}=\langle N_e\rangle$. For more complex scenarios, such as an exponentially expanding population, this parameterization might include the sizes of the ancestral and current populations, $N_a$ and $N_c$, respectively, and the duration of the exponential expansion $G$. In such a case, we write $\theta=\theta_{EXP}=\langle N_a,N_c,G\rangle$. In the remainder of this work, we refer to the effective population size in a coalescent model simply as population size. For practical purposes, we focus on closed intervals $R=[u,v]$ of possible values for $L$ and derive a closed-form expression for $p_R(l|\theta)=\int_{u}^{v}p(l|\theta)dl$.

We denote time in generations before the present throughout. The time $g_{mrca}$ of the individuals' MRCA at the considered locus is generally unknown. We therefore marginalize it as

\begin{equation}
\begin{split}
\int_{u}^{v}p(l|\theta)dl=\int_{u}^{v} \sum_{g=1}^{\infty}p(l,g_{mrca}=g|\theta)dl.
\end{split}
\label{eq:model:1}
\end{equation}

When the time to the MRCA is known, the length of the resulting shared segment is only dependent on the number of generations separating the two individuals (i.e., $l\independent \theta|g_{mrca}$). Manipulating this expression, we therefore obtain

\begin{equation}
\begin{split}
\int_{u}^{v}p(l|\theta)dl= \sum_{g=1}^{\infty}p(g_{mrca}=g|\theta)\int_{u}^{v}p(l|g_{mrca}=g)dl.
\end{split}
\label{eq:model:2}
\end{equation}

The distribution of the distance to the first recombination event encountered as we move either upstream or downstream of a chosen genomic site is exponentially distributed (it has a mean of $g/50$ cM) because this is a haplotype shared by two individuals separated by $2g$ generations. The total length of the shared segment is therefore distributed as the sum of two independent exponential random variables parameterized by their mean of $g/50$ cM, resulting in an Erlang-2 distribution with the same parameter. We therefore have

\begin{equation}
\begin{split}
\int_{u}^{v} p(l|\theta)dl= \int_{0}^{\infty} \left[p(t_{mcra}=t|\theta_{WF})\int_{u}^{v}Erl_2\left(l;\frac{t}{50}\right)dl\right]dt,
\end{split}
\label{eq:model:3}
\end{equation}

where we also standardly switch to a continuous time axis \cite{hudson1983properties} by replacing the discrete $g_{mrca}$ with a continuous $t_{mrca}$, still measured in generations. Note that we are not measuring time in units of $N_e$ generations as it is often done in the coalescent literature \cite{griffiths1991two}. To complete the above formulation, we substitute the distribution of the time to MRCA for a specific demographic setting $\theta$. In the coalescent framework, for the simple case of a population of constant size $N_e$ and non-overlapping generations, the probability of finding a common ancestor at $g_{mrca}=g$ is geometric with parameter $p(g_{mrca}=g|\theta)=N_e^{-1}$ (or exponential at the continuous limit). Substituting this expression into Equation \ref{eq:model:3}, we obtain the desired relationship between sharing of IBD haplotypes and population size:

\begin{equation}
\begin{split}
p_R(l|\theta_{WF}) &= \int_{0}^{\infty} \left[\frac{e^{−t/N_e}}{N_e} \int_{u}^{v} Erl_2(l;\frac{t}{50})dl\right]dt \\
~\\
&=\frac{100 N_e^2 (v−u)[25(u+v)+u v N_e]}{(50+u N_e)^2(50+v N_e)^2}.
\end{split}
\label{eq:model:4}
\end{equation}

\subsection{Varying population size}
When more complex population dynamics are considered, the probability of coalescence cannot be modeled through a simple geometric distribution. In general, for a population with demographic history $\theta$, we can define a function $N(g,\theta)$ to express the population size at generation $g$. We can then express the chance of coalescence as

\begin{equation}
\begin{split}
p(g_{mrca}=g|\theta) = \frac{1}{N(g,\theta)} \prod_{j=1}^{g-1} \left( 1 - \frac{1}{N(j,\theta)}\right).
\end{split}
\label{eq:model:5}
\end{equation}

Equation \ref{eq:model:5} is very general and might lead to more complex instantiations for Equation \ref{eq:model:3}. However, we consider a special and useful case in which the population history converges to $N_a=\lim_{g\to +\infty}N(g,\theta)$. By definition, there exists a finite time $G$ before which $N(g,\theta)=N_a, \forall \{g > G\}$. In practice, we consider $G$ to be the time before the period in history we aim to describe in detail, and we also note that demographic events preceding a sufficiently ancient generation $G$ are unlikely to affect the probability of sharing IBD haplotypes longer than a chosen threshold. We observe that for any such converging history $\theta$, we can always obtain a closed-form expression regardless of the specific form of $N(g,\theta)$ for $g\leq G$. For a population size of $N(g,\theta)$, such that $N(g,\theta)=N_a$ for all $g>G$, Equation \ref{eq:model:5} can in fact be rewritten as

\begin{equation}
\begin{split}
\int_{u}^{v} p(l|\theta)dl=\phi_1(l,\theta,u,v,1\dots G) + \phi_2(l,\theta,u,v,G+1\dots \infty),
\end{split}
\label{eq:model:6}
\end{equation}

where

\begin{equation}
\begin{split}
\phi_1 (l,\theta,u,v,1\dots G)=\sum_{g=1}^{G} \left[ \prod_{j=1}^{g-1} \left( 1 - \frac{1}{N(j,\theta)}\right) \frac{1}{N(g,\theta)} \int_{u}^{v} Erl_2\left(l;\frac{g}{50}\right)dl\right],
\end{split}
\end{equation}

and

\begin{equation}
\begin{split}
\phi_2(l,\theta,u,v,G+1\dots \infty) &= \frac{1}{N_a} \prod_{j=1}^{G}\left( 1 - \frac{1}{N(j,\theta)}\right) \\
& \times \sum_{g=G+1}^{\infty}\left(1 - \frac{1}{N_a}\right)^{g - G - 1}\int_{u}^{v} Erl_2\left(l;\frac{g}{50}\right)dl.
\end{split}
\end{equation}

Continuous time allows a closed-form expression for $\phi_2$, as

\begin{equation}
\begin{split}
\sum_{g=G+1}^{\infty}\left( 1 - \frac{1}{N_a}\right)^{g-G-1}\int_{u}^{v} Erl_2\left(l;\frac{g}{50}\right)dl \approx \\
\approx e^{-C(G+1)} \times[f(u,G,C)-f(v,G,C)],
\end{split}
\label{eq:model:CF1}
\end{equation}

where $C=\log(1-\frac{1}{N_a})$ and

\begin{equation}
\begin{split}
f(x,G,C)=\frac{e^{(G+1)(C-x/50)}[x(100+x+Gx)-50C(50+x+Gx)]}{(x-50C)^2}.
\end{split}
\label{eq:model:CF2}
\end{equation}
whereas $\phi_2$ adds up to a finite number of summands. The function $N(g,\theta)$ can thus be arbitrarily defined to describe different demographic scenarios. Consider, for instance, the case of an ancestral population of size $N_a$: it exponentially expands during $G$ generations to reach the current size $N_c$, parameterized by $\theta_{EXP}=\langle N_a,N_c,G\rangle$ as discussed above. The population size can be modeled (under the assumption of continuous time) as

\begin{equation}
\begin{split}
N(t,\theta_{EXP}\langle N_a,N_c,G\rangle)=
	\begin{cases}
	N_c e^{-rt} & if~t \leq T,\\
	N_a & \mbox{otherwise}. \\
	\end{cases}
\end{split}
\label{eq:model:7}
\end{equation}
where $r=[\log(N_c)-\log(N_a)]/T$ is the population expansion rate.
\subsection{Sharing distribution}

In the following section, we present explicit expressions for the case of Wright-Fisher populations (i.e., $\theta = \langle N_e \rangle$). Note, however, that these results are general, and analogous calculations can be performed for other demographic models.

Consider a specific site $\varsigma$ and a length range $R=[u,v]$. We are interested in IBD segments whose length lies within that interval, spanning the site $\varsigma$. We consider the event of such a segment being shared between a randomly chosen pair of individuals from a studied population, and we define an indicator random variable for such an event as

\begin{equation}
\begin{split}
\operatorname{I}(\varsigma,R=[u,v])=
	\begin{cases}
	1 & \mbox{if } \varsigma \mbox{ is traversed by a segmetns of length } u \leq l \leq v, \\
	0 & \mbox{otherwise}. \\
	\end{cases}
\end{split}
\end{equation}

where we omit the dependence on the demographic model $\theta$ to simplify the notation. We now use these indicator variables to derive the expected fraction of genome spanned by IBD segments whose length is in this interval. Consider a dense set of sites $\Gamma$ along the genome. Assume all sites are at equal genetic distance from adjacent sites. We have that

\begin{equation}
\begin{split}
\operatorname{E}_R[f|\theta]=\operatorname{E}\left[\frac{1}{\lvert \Gamma \rvert} \sum_{\varsigma \in \Gamma}\operatorname{I}(\varsigma,R)\right] &=\frac{1}{\lvert \Gamma \rvert} \sum_{\varsigma \in \Gamma} \operatorname{E}[\operatorname{I}(\varsigma,R)] \\
&=\frac{1}{\lvert \Gamma \rvert} \sum_{\varsigma \in \Gamma} \int_{u}^{v} p(l|\theta)dl \\
&=\int_{u}^{v}p(l|\theta)dl.
\end{split}
\end{equation}

For given values of the demographic parameters $\theta$, this predicts the fraction $f$ of the genome shared through segments of length within specific intervals. To obtain the proportion of segments of a given length $l$, we divide $p(l|\theta)$ by $l$ and multiply by a normalizing constant:

\begin{equation}
\begin{split}
p(s=l|\theta)=\frac{p(l|\theta)}{l} \times \frac{1}{\int_{0}^{\infty} p(l|\theta)/l ~ dl} = \frac{2\times 50^2 N_e}{(50+lN_e)^3}.
\end{split}
\label{eq:model:10}
\end{equation}

The probability of finding a segment within the length range $R=[u,v]$ is thus

\begin{equation}
\begin{split}
p(s\in R|\theta) = \int_{u}^{v} p(s=l|\theta)dl=50^2\left[(50+N_e u)^{-2}-(50+N_e v)^{-2}\right].
\end{split}
\label{eq:model:11}
\end{equation}

Equations \ref{eq:model:10} and \ref{eq:model:11} allow computing the length distribution of a segment in the range $R$,

\begin{equation}
\begin{split}
p_R(s=l|\theta)=
	\begin{cases}
	\frac{p(s=l|\theta)}{p(s\in R|\theta)} & \mbox{if } s \in R, \\
	0 & \mbox{otherwise}. \\
	\end{cases}
\end{split}
\end{equation}

and the expected length of such a segment,

\begin{equation}
\begin{split}
\operatorname{E}_R[s|\theta] = \frac{\int_{u}^{v} l \times p(s=l∣\theta)dl} {p(s\in R|\theta)}=\frac{50v+2u(25+N_e v)}{100+N_e(u+v)}.
\end{split}
\label{eq:model:13}
\end{equation}

We note that for a typical pair of sharing individuals, the number and length of IBD segments are approximately independent \cite{huff2011maximum}. This allows us to express the expected genome-wide sharing between two individuals as the product of the expected number of IBD segments, $\lambda_R$, and the expected length of a shared segment in the considered length range, $\operatorname{E}_R[s|\theta]$. For a genome of size $\gamma$ cM, $\gamma \times \operatorname{E}_R[f|\theta]\approx \operatorname{E}_R[s|\theta]\times \lambda_R$. We can thus compute the expected number of segments found in the considered length range as

\begin{equation}
\begin{split}
\lambda_R \approx \gamma \times \frac{\operatorname{E}_R[f|\theta]}{\operatorname{E}_R[s|\theta]} = \gamma \times \frac{50N_e^2(v−u)[100+N_e(u+v)]}{(50+uN_e)^2(50+vN_e)^2}.
\end{split}
\label{eq:model:14}
\end{equation}

We model the number of shared segments as a Poisson random variable, $p_R(s=n|\theta)\approx Poiss(n,\lambda_R)$; thus, the standard deviation for the segment distribution is $\sigma_R[s|\theta]=\sqrt{\lambda_R}$. If the considered length range is not too wide, the variance of the segment lengths can be neglected, and we can obtain a simple approximation for the standard deviation of the fraction of genome shared through segments in the length range $R$ by scaling $\sigma_R[s|\theta]$ by the expected length of a segment and by dividing it by the genome size:

\begin{equation}
\begin{split}
\sigma_R[f|\theta] &\approx \frac{\operatorname{E}_R[s|\theta] \sqrt{\lambda_R}}{\gamma} \\
&= \sqrt{\frac{\operatorname{E}_R[f|\theta] \operatorname{E}_R[s|\theta]}{\gamma}} \\
&= \frac{10N_e [25v + u(25+N_e v)]}{(50+N_e u)(50+N_e v)} \times \sqrt{\frac{2(v-u)}{\gamma[100+N_e(u+v)]}}
\end{split}
\label{eq:model:15}
\end{equation}

Finally, the obtained quantities can be used for expressing the full distribution of the portion $\tau$ of the genome shared through segments of a desired length again under the assumption of independence between number and length of shared segments. Define $l_n$ to be the sum of $n$ segments of length in the range $R$:

\begin{equation}
\begin{split}
	p_R(l_n=x|\theta) =
		\begin{cases}
		\delta(x) & \mbox{if } n=0, \\
		\operatorname{conv}[p_R(s=l|\theta),n] & \mbox{otherwise}. \\
		\end{cases}\end{split}
\end{equation}

where $\delta(\cdot)$ is the Dirac delta function and $\operatorname{conv}[p_R(s=l|\theta),n]$ is the nth convolution of $p_R(s=l|\theta)$ (e.g., $\operatorname{conv}[p_R(s=l|\theta),3]=p_R(s=l|\theta) \ast p_R(s=l|\theta) \ast p_R(s=l|\theta))$. The probability of sharing a total of $x$ cM through segments of the desired length is then

\begin{equation}
\begin{split}
p_R(\tau=x|\theta) &= \sum_{n=0}^{\infty} p_R(s=n,l_n=x|\theta) \\
&= \sum_{n \mid p_R(s=n|\theta)\neq 0}[p_R(s=n|\theta)p_R(l_n=x|\theta)].
\end{split}
\label{eq:model:17}
\end{equation}

Note that although we have considered the general length range $R=[u,v]$, the interval $R=[u,\infty)$ represents a particular and useful case in which all segments longer than a detectable threshold u are considered. When $v\longrightarrow \infty$, the length distribution simplifies to

\begin{equation}
\begin{split}
p_R(s=l|\theta)=\frac{2N_e(50+N_e u)^2}{(50+lN_e)^3}, \mbox{ ~ for } l\in R,
\end{split}
\end{equation}

and the expected length becomes

\begin{equation}
\begin{split}
E[s|\theta]=50N_e+2u.
\end{split}
\end{equation}

The expected number of segments is therefore

\begin{equation}
\begin{split}
\lambda_R \approx \gamma \times 50N_e (50+N_e u)^2.
\end{split}
\label{eq:model:numSeg_u}
\end{equation}

The approximation for the standard deviation of the genome fraction shared through segments in a specified length range provided in Equation \ref{eq:model:15} becomes inaccurate when long length intervals are considered. When $v\longrightarrow \infty$, we obtain an improved approximation by multiplying by a numerically computed factor of $75/(50+u)$:

\begin{equation}
\begin{split}
\sigma_R[f|\theta] \approx \frac{75}{50+u} \times \frac{25 + N_e u}{50 + N_e u} \times \frac{200} {\gamma N_e}.
\end{split}
\end{equation}

Additional calculations for higher moments of IBD sharing in the Wright Fisher model can also be found in \cite{carmi2013variance}.

\subsection{Inference}
In the case of Wright-Fisher populations, we can obtain an estimate of the population size $N_e$ by comparing the sharing observed in a specific length range to Equation \ref{eq:model:4} and by solving for $N_e$. The observed sharing in the length range $R=[u,v]$ can be computed from the analyzed data as

\begin{equation}
\begin{split}
\hat{p}_R = \frac{\sum_{i\mid u \leq l_i \leq v}l_i}{\gamma {n \choose 2}},
\end{split}
\label{eq:model:18}
\end{equation}

where $l$ is the length of a detected IBD segment and $n$ represents the number of haploid individuals (see above for discussion of the diploid case). A closed-form solution for $N_e$ can be computed for a given observed value of $\hat{p}_R$. In the particular case of $v\longrightarrow \infty$, where we consider all segments longer than a detectable threshold $u$, such a solution assumes a simpler form. Equation \ref{eq:model:4} becomes

\begin{equation}
\begin{split}
\int_{u}^{v} p(l|\theta_{WF})dl = \frac{100(25+N_e u)}{(50+N_e u)^2},
\end{split}
\label{eq:model:19}
\end{equation}

and an estimate of $N_e$ can be computed as

\begin{equation}
\begin{split}
\hat{N}_e = \frac{50(1-\hat{p}_R+\sqrt{1-\hat{p}_R})}{u\hat{p}_R}.
\end{split}
\label{eq:model:20}
\end{equation}

The number of IBD segments can also be used to derive a similar, improved estimator. Assuming all individual pairs are independent, and identically distributed, a Poisson likelihood for the segment counts can be obtained using the expectation of Equation \ref{eq:model:numSeg_u}. Setting the derivative of the logarithm of such likelihood to zero, we obtain

\begin{equation}
\begin{split}
\hat{N}_e = \frac{50 \left(1-\eta+\sqrt{1-2\eta}\right)}{u\eta},
\end{split}
\label{eq:model:segEst}
\end{equation}
where $\eta=\frac{2\hat{s}_R u}{\gamma {n \choose 2}}$ and $\hat{s}_R$ are the total number of segments longer than $u$ cM observed for all $n \choose 2$ pairs of genomes $\gamma$ cM long.

For much of the analysis reported in this paper, we minimized the squared deviation between the observed IBD sharing and the theoretical expectation (Equation \ref{eq:model:6}) for a tested demographic model. The performance of this approach is comparable to that of estimation based on segment counts. To compute a distance between observed and predicted sharing, we thus evaluate

\begin{equation}
\begin{split}
\delta_R = \left[\log(\hat{p}_R)-\log(\operatorname{E}_R[f|\theta])\right]^2
\end{split}
\label{eq:model:21}
\end{equation}

and average this quantity across a collection of intervals $\Pi = \{R_j\}_{1 \leq j \leq \lvert \Pi \rvert}$:

\begin{equation}
\begin{split}
\delta_\Pi = \sqrt{\frac{1}{\lvert \Pi \rvert} \sum_{j=1}^{\lvert \Pi \rvert} \delta_{R_j}}.
\end{split}
\label{eq:model:22}
\end{equation}

The transformation to log space in Equation \ref{eq:model:21} has the effect of making the error contributions along the dynamic range of length intervals more uniform than in linear space. Grid-search minimization of Equation \ref{eq:model:22} can therefore be employed for exploring a large portion of the parameter space. Upon convergence to a grid point of least deviation from the theoretical expectation, a full likelihood-based approach can be used for retrieving the most likely values for the demographic-model parameters in a smaller portion of the parameter space and can thus allow substantial computational savings. An alternative to the minimization of the squared deviation is maximizing a composite likelihood based on Poisson counts of the observed segments. We have observed comparable performance for either approach.

\subsection{Evaluation on synthetic data}

To evaluate the accuracy of the proposed model and of the inference procedure, we simulated a large number of synthetic populations by using the GENOME coalescent simulator \cite{liang2007genome}. We extracted ground-truth information on shared segments to eliminate the noise introduced by methods for IBD discovery. To this extent, the coalescent simulator was modified to output shared segments descending from the same ancestor as observed in the synthetic genealogy (according to definition (b) for IBD segments in Section\ref{intro:subsec:IBD_def}). For all the simulations, we generated a total of $500$ diploid samples for a single chromosome made of $27,800$ non-recombining blocks with an inter-block recombination rate of $10^{−4}$, mimicking the genetic length of chromosome $1$ (${\sim}278$ cM). We verified that the use of non-recombining blocks of $0.01$ cM did not introduce significant biases in our analysis (see Supplementary Figure $1$). We simulated $900$ synthetic populations that underwent exponential contraction and expansion (see Supplementary Table $1$ for the range of demographic parameters). We applied a gradient-driven local-minimization procedure to retrieve the parameter values that minimize Equation \ref{eq:model:22}. In order to avoid local minima, we initially performed a grid search in a predefined box volume of the parameter space (see Supplementary Table $1$ for the parameters list). We then refined the least-squares solution by using a gradient-based optimization from the best point on the grid.

The accuracy of our inference procedure depends on the length of the analyzed genomic region and on the number of samples for which IBD segments are observed. In particular, it follows from Equation \ref{eq:model:18} that upon fixing $\hat{p}_R$ and $\sum_{i\mid u \leq l_i \leq v}l_i$, the result is unchanged for several values of $\gamma$ and $n$. In terms of accuracy of the proposed evaluation, an equivalent configuration would have been the use of ${\sim}140$ diploid individuals for the entire genetic length of the autosomal genome (${\sim}3,500$ cM for the HapMap 3 genetic map; see Supplementary Figure $2$). The choice of length intervals $R_j=[u_j,v_j]$ also affects the inference results: segments of length between $1$ and $2$ cM, for instance, might have originated from a wide span of generations in the past, whereas segments of length $10$-$11$ cM tend to have a more deterministic (and more recent) origin. Frequency bins of different sizes can be used for focusing on specific time periods. For all the analyses reported in this paper, we adopted a combination of bins of uniform length and bins of length intervals corresponding to specific percentiles of the Erlang-$2$ distribution. In particular, we used length values between the $21.4$th and the $31.4$th percentiles of the Erlang-$2$ distributions with parameter $\lambda = \frac{k}{50}$ (the maximum likelihood estimate occurs at the $26.4$th percentile) for several consecutive integral values of $k$ (i.e., $k = 2, 3, \dots 43$).

\subsection{Real data analysis}
We applied the proposed inference procedure to genotype samples of $500$ AJ individuals from Jerusalem (Israel) and $143$ MKK individuals from Kinyawa (Kenya), already analyzed in Chapter \ref{chap:IBD_in_data}. The AJ individuals were typed on the Illumina $1$M platform and are self-reported unrelated individuals. After quality control, a total of $745,811$ autosomal SNPs were used for the analysis. The MKK samples comprise $56$ unrelated trio-phased individuals and $87$ unrelated individuals from the HapMap 3 data set previously introduced. As a result of the availability of haplotype phase information, we focused our analysis on the $56$ trio-phased samples and used $1,387,466$ markers for the analysis.

The AJ samples were phased with the Beagle software package \cite{browning2007rapid}, whereas trio-phased MKK individuals were downloaded from the HapMap website\footnote{\url{http://hapmap.ncbi.nlm.nih.gov}}. IBD sharing was estimated with the GERMLINE software package \cite{gusev2009whole}. We tweaked the parameters of the GERMLINE algorithm to improve the quality of IBD detection for the specific data set by using the following procedure. Using GERMLINE's default \emph{haplotype extension} parameters, we extracted IBD segments from the real data and then used the analytical inference procedure to retrieve demographic parameters. We simulated a synthetic population by using the inferred demography and extracted ground-truth IBD segments. We ran GERMLINE on the synthetic genotypes several times and changed the \emph{err\_hom}, \emph{err\_het}, \emph{bits} to find a set of parameters that minimized the deviation of the genotype-inferred IBD sharing density from that obtained from ground-truth data. We then used these parameters to extract IBD segments from the real data again and iterated the procedure until convergence. The GERMLINE parameters to which we converged were \emph{-min\_m 1 -err\_hom 0 -err\_het 2 -bits 25 -h\_extend} for the Beagle-phased AJ data and \emph{-min\_m 1 -err\_hom 2 -err\_het 2 -bits 60 -h\_extend} for the trio-phased MKK data.

\subsubsection{Demographic Model Selection in the AJ Population}
We tested increasingly flexible models to infer the demographic history of the AJ population. In order to control for potential over fitting, we evaluated the parameters obtained for different models by using a likelihood approach. To this extent, after optimizing the model parameters by using the least-squares approach, we used rejection sampling to retrieve parameters corresponding to a local maximum likelihood for each model. We then used the Akaike information criterion (AIC, \cite{akaike1974new}) to compare models while controlling for their different degrees of freedom (see the algorithm reported in Supplementary Table $2$).

\begin{figure*}
    \centering
	\includegraphics[width=0.9\textwidth]{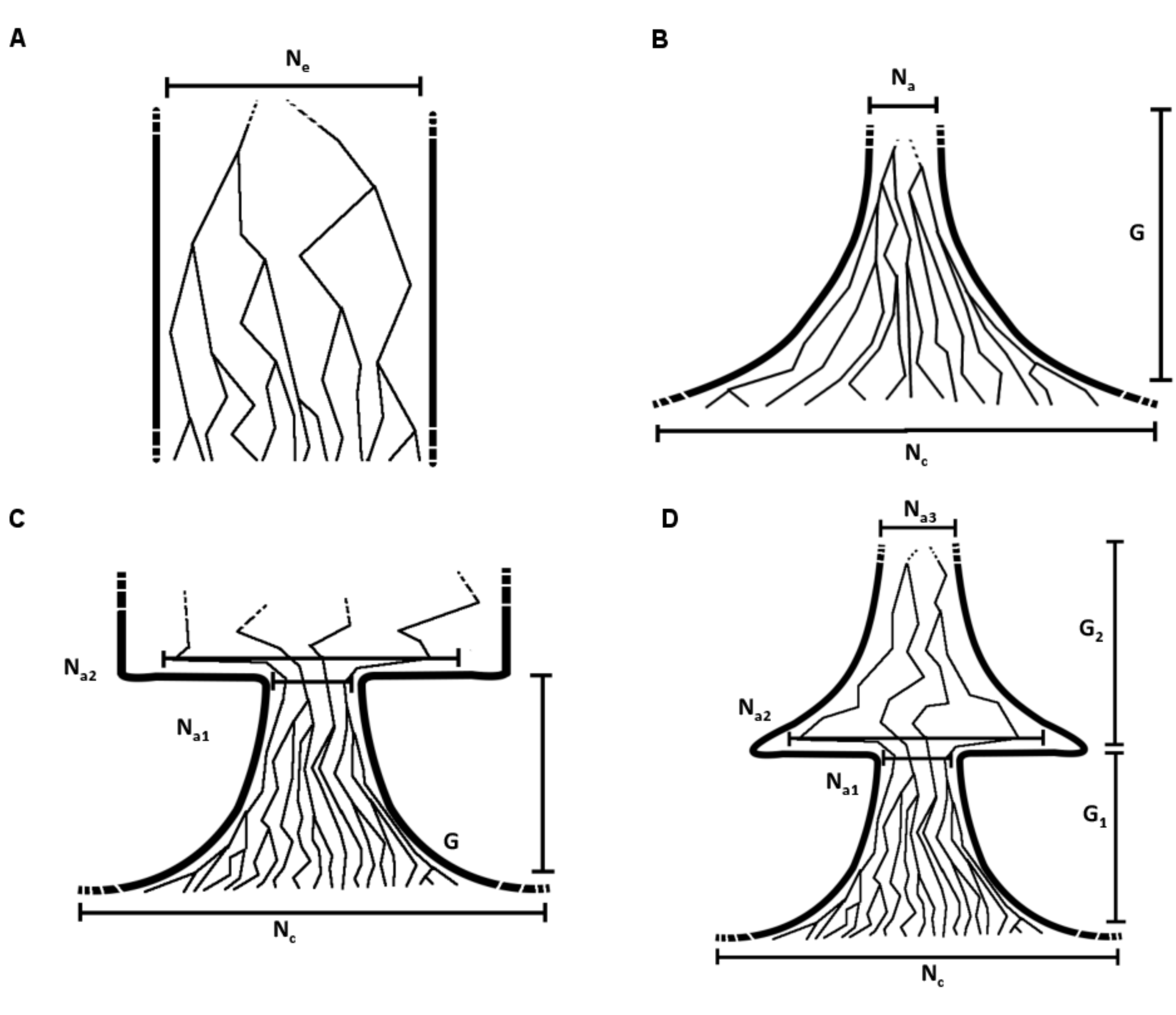}
	\caption{Demographic Models. (A) Population of constant size. (B) Exponential expansion (contraction for $N_a>N_c$). (C) A founder event followed by exponential expansion. (D) Two subsequent exponential expansions separated by a founder event.}
    \label{fig:model:models}
\end{figure*}

Three models were used for the inference in the AJ population (see Figure \ref{fig:model:models} and an additional description in the Results): (1) a model of exponential expansion ($\mathcal{M}_E$), (2) a model including a founder event followed by exponential expansion ($\mathcal{M}_{FE}$), and (3) a model of two exponential-expansion periods separated by a founder event ($\mathcal{M}_{EFE}$). The $\mathcal{M}_{E}$ model did not provide enough flexibility to fit the IBD-sharing summary extracted for the AJ population, resulting in a poor fit (particularly for shorter segments) and unrealistically large values for the recent population size. We therefore excluded this model from further analysis. For models $\mathcal{M}_{FE}$ and $\mathcal{M}_{EFE}$, we used the following rejection-sampling approach to maximize the model likelihood around the least-squares solution obtained in the previous step. (1) For each model, for each model parameter, we generated a list of neighboring points by allowing each parameter to vary by  $\pm 3\%$ of its current value. (2) For each point on such a local grid, we sampled several random data sets of sharing individuals by using the corresponding demographic parameters (details in Supplementary Table $3$). We created each data set by sampling random sharing values for independent individual pairs from the distribution of Equation \ref{eq:model:17}. (3) For each analyzed set of parameter values, we computed a likelihood as the fraction of data points for which the deviation between AJ and sampled sharing was smaller than a tolerance threshold $\delta$ ($\delta\approx 0.089$ for $\mathcal{M}_{FE}$ and $\delta\approx 0.037$ for $\mathcal{M}_{EFE}$). (4) We updated the current point to the most likely point in the analyzed neighborhood, if any, and iterated steps $1$-$3$ until no point with a higher likelihood was found. (5) We applied the AIC to compare models.

For both models, only one iteration of the above local maximization was required. The most likely parameter values in the grid matched those obtained with the least-squares approach, except for the current population size, which increased by $3\%$ for model $\mathcal{M}_{FE}$  and decreased by $3\%$ for model $\mathcal{M}_{EFE}$. When comparing the two models, we used a tolerance threshold of $\delta\approx 0.037$ and obtained an AIC value of $19.21$ for the $\mathcal{M}_{EFE}$ model, which allows five parameters to vary (such $\delta$ results in a likelihood of $0.01$ for the $\mathcal{M}_{EFE}$ model). Using the same acceptance threshold, we thus required a log likelihood of at least $-5.6$ (a likelihood of ${\sim}3.7 \times 10^{−3}$) for model $\mathcal{M}_{FE}$ , which has four parameters, to be selected. None of the 105 sampled points were accepted with such a threshold, leading us to choose the $\mathcal{M}_{EFE}$ model. The likelihoods of additional parameter values estimated for the $\mathcal{M}_{EFE}$  model with the use of a wider grid are reported in Supplementary Table $4$.

Note that when sampling from Equation \ref{eq:model:17}, we assumed independence of the analyzed sharing length intervals $R_i$ and of the pairs within a data set, potentially underestimating the variance of randomly sampled summaries of IBD. To account for the presence of small correlations, we thus performed full coalescent simulations according to the most likely set of parameters of each model by only sampling a synthetic chromosome $1$ for $500$ diploid individuals. We repeated the rejection-based comparison by using $104$ such points for each model and obtained an equivalent result.

\subsubsection{Accounting for Phase Errors}
The inference procedure described in the previous sections assumes that high-quality IBD information is available. When real data sets are analyzed, several sources of noise, such as computational phasing errors, might distort summary statistics of haplotype sharing. In the absence of reliable probabilistic measures for the quality of shared segments, modeling this potential bias is complicated. To account for this additional noise, we refined the inferred AJ demographic model by using simulations that mimic SNP ascertainment, inaccurate phasing, and IBD discovery in the analyzed data sets. We expected the distortion of IBD summary statistics in the AJ data set to not be substantial (Supplementary Figure $3$). The preliminary inference based on the assumption of high-quality IBD information therefore provides an efficient means for exploring large portions of the parameter space and for performing model comparison. This can be followed by such simulation-based refinement, which requires considerable computation.

After finding the most likely parameters and selecting model $\mathcal{M}_{EFE}$ for the AJ data as previously described, we refined the obtained solution by using a local-search approach. We iteratively varied one demographic parameter at a time and kept a tested value if it resulted in a decreased deviation from the AJ data summary. Note that in order to account for the stochastic variation observed across multiple independent simulations of the same demographic history, we would need to generate several synthetic data sets for each tested set of demographic parameters. However, we did not repeat such simulations multiple times as a result of computational constraints.

For all coalescent simulations in real-data inference, we used the GENOME software package. The simulated chromosomes have the same genetic length as their real-data equivalent and a mutation rate of $1.1 \times 10^{−8}$ per site per generation \cite{roach2010analysis}. To reduce the computational burden, we used non-recombining block units of $10$ kb for MKK simulations and $20$ kb units for AJ simulations, resulting in an IBD length resolution of $0.01$ and $0.02$ cM, respectively. Synthetic markers were randomly ascertained to match the same density of the real data. We matched the spectrum of the real data sets by randomly selecting the same proportion of variants for each frequency bin and used a bin size of $2\%$. No missing genotypes were allowed in simulated data because occasional missing genotypes in the real data were imputed during Beagle phasing or excluded from the analysis if not reliably imputed. All simulations were carried out for the entire autosomal genome.

\section{Results of evaluation and real data analysis}
\subsection{Simulated data}

\begin{figure*}
    \centering
	\includegraphics[width=\textwidth]{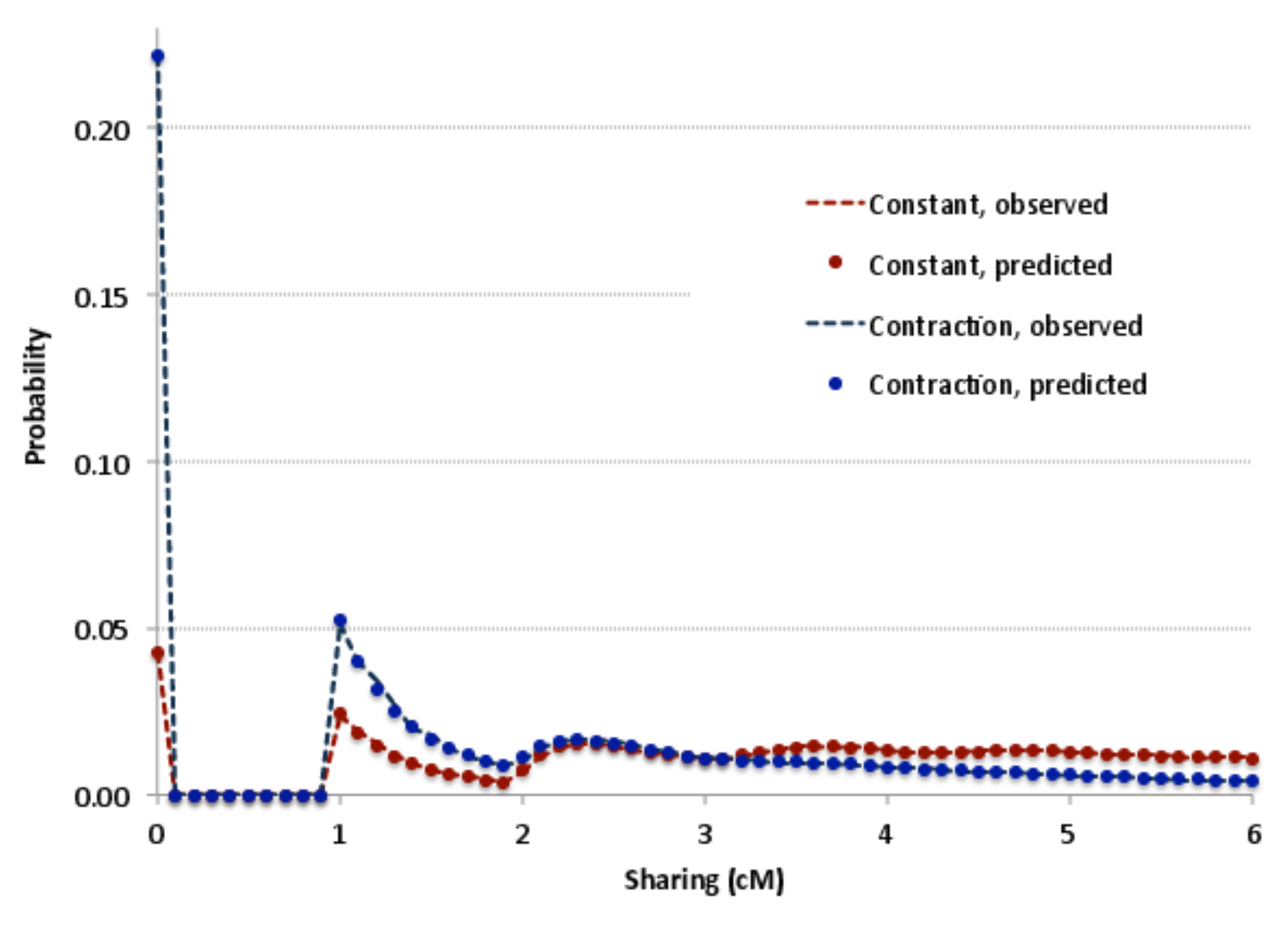}
	\caption{Analytical (dots) and empirical (dashed lines) distribution for total IBD (Equation \ref{eq:model:17}) for a constant population of $2,000$ diploid individuals (red, $R=[1,4]$) and an exponentially contracting population ($A=50,000$, $C=500$, $G=20$, $R=[1,\infty)$, in blue).}
    \label{fig:model:total}
\end{figure*}

\begin{figure*}
    \centering
	\includegraphics[width=\textwidth]{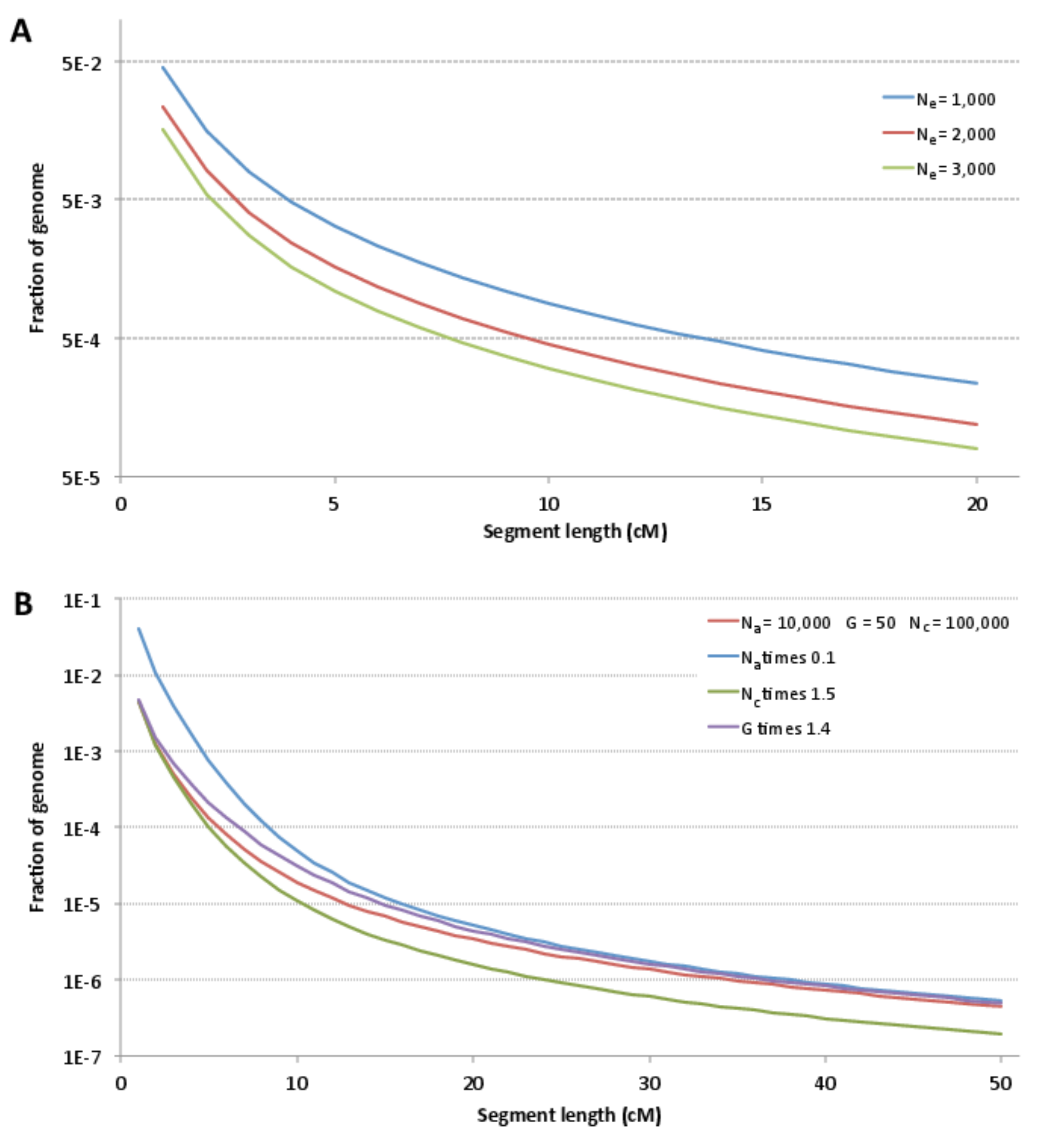}
	\caption{Effects of demographic parameters on IBD length distributions. Wright-Fisher models (A) and exponential population expansion (B).}
    \label{fig:model:distr}
\end{figure*}

The described methods were implemented in \emph{DoRIS}, a freely available software tool\footnote{\url{http://www.cs.columbia.edu/~pier/doris/}}. We tested the accuracy of the proposed model through extensive simulation of synthetic populations with known demographic history. For each simulated population, we analyzed a region of length equivalent to chromosome $1$ for $500$ diploid samples (see Section \ref{sec:model:methods}). All the derived theoretical quantities were found in good agreement with the values obtained from simulation (see Supplementary Figure $4$ for an evaluation summary and Figure \ref{fig:model:total} for examples of total haplotype-sharing distributions). We noted that for populations of constant size, as expected, a smaller population size causes a larger fraction of the genome to be shared through IBD segments for the average pair in the population (Figure \ref{fig:model:distr}). Furthermore, the frequency of segments at different length intervals is informative of population size at different time scales. Consider the case of an exponential expansion (Figure \ref{fig:model:models}.A) with the following parameterization: $N_a$ is the size of the ancestral population when exponential expansion began, $N_c$ denotes the population size at the current generation, and $G$ represents the number of generations during which the exponential expansion took place. A small ancestral population size $N_a$ causes a higher rate of remote coalescent events and a consequently larger fraction of the genome to be spanned by short segments of IBD. Similarly, a small value of $N_c$ increases the chance of coalescence in the more recent generations, causing a larger fraction of the genome to be spanned by long segments. For fixed Na and $N_c$, variations of the duration of expansion $G$ affect the expansion rate and have a noticeable effect on the slope of the sharing distribution, i.e., the genome fraction spanned by mid-length segments.

\begin{figure*}
    \centering
	\includegraphics[width=\textwidth]{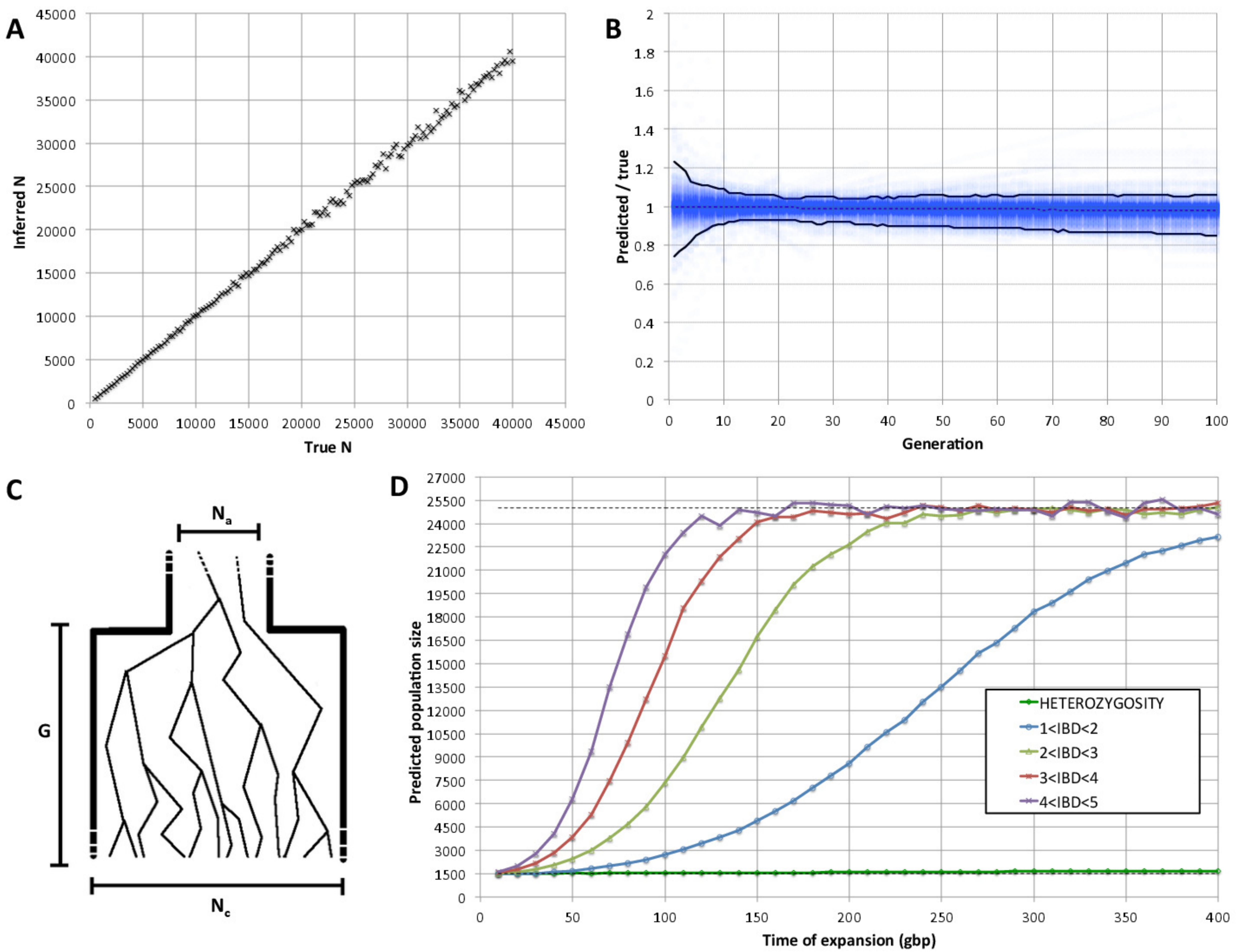}
	\caption{Performance for constant-size populations (A), expanding and contracting populations (B), and a suddenly expanding population (C) studied with a constant-size model (D).}
    \label{fig:model:eval}
\end{figure*}

We used the relationship of Equation 4 to infer the size of a Wright-Fisher population by using a realistic chromosome 1 simulated for several populations, each with its own constant size $N_e$ ranging from $500$-$40,000$ individuals. In the analysis of IBD information for 500 diploid samples in each such synthetic population, the predicted value was highly correlated with the true size of the synthetic populations $(r = 0.9994$; Figure \ref{fig:model:eval}). Across all tested values of $N_e$, the ratio between true and estimated population size had a median of $1.00$ and a $95\%$ confidence interval (CI) of $0.97$-$1.03$.

\subsection{IBD and heterozygosity in an expanding population}
To outline IBD's particular sensitivity to recent demographic variation, we examined the effects of variable population size on demographic inference conducted either through the proposed approach based on IBD haplotypes or through a classical approach based on heterozygosity. We focused on the scenario in which a population of $3,000$ ancestral individuals suddenly expands to a size of $25,000$ individuals $G$ generations before the present (Figure \ref{fig:model:eval}.C). We varied $G$ from $10$-$400$ generations and simulated the ascertainment of IBD haplotypes by extracting information on shared haplotypes along a realistic chromosome $1$ for $500$ diploid samples. For both IBD-based and heterozygosity-based reconstructions, we assumed and inferred a constant population size $N_e$. We used the relationship of Equation \ref{eq:model:4} for the IBD model and Watterson's estimator of Equation \ref{eq:intro:watterson} for the heterozygosity-based approach (the heterozygosity $\theta$ was estimated from the synthetic sequences, and $\mu$ matched the simulated mutation rate). An estimate of $N_e$ was obtained for each data set across all simulated times of expansion (Figure \ref{fig:model:eval}.D). As expected, the obtained estimate of $N_e$ tended to lie in the range between the ancestral and the current size of the population. Long, recently originated segments provide a better prediction of the current population size, especially for remote expansions. In contrast, the high frequency of shorter segments of more remote origins biases the inference toward a smaller population size when these segments are taken into account. For example, the effects of a small ancestral population size can be observed on segments between $4$ and $5$ cM in length only for expansions that occurred fewer than $120$ generations ago; in contrast, when segments between $1$ and $2$ cM in length are analyzed, traces of a smaller ancestral population are still notable, even for expansions that occurred as far back as $400$ generations ago. When comparing these results to population-size estimates obtained with heterozygosity from full synthetic genomic sequence, we observed the heterozygosity-based estimates of $N_e$ to be strongly biased toward the small size of the ancestral population. Although they present less instability than do the IBD-based estimates, the inferred values approached the ancestral population size, even for expansions that occurred $400$ generations before the present. This analysis outlines the unique sensitivity of long-range IBD sharing to recent demographic variation.

\subsection{Evaluation of the inference in populations of varying size}
We tested the accuracy of our inference procedure for the cases of either an exponential increase or decrease in population size (expansion or contraction, respectively). We simulated $450$ synthetic populations that underwent an exponential expansion and $450$ that underwent exponential contraction. We analyzed the IBD sharing of $500$ diploid samples from each simulated population along a $278$ cM chromosome. We evaluated the accuracy of the inferred demography by using the ratio between true and predicted sizes of each analyzed population (Figure \ref{fig:model:eval}.B) for all generations between $1$ and $100$. We found our inferred population size to be within $10\%$ of the true value $95\%$ of the time. The population size of recent generations was harder to infer because of the scarcity of long IBD segments in very large populations (this scarcity is due to a low chance of recent coalescent events).

Note that the reconstruction accuracy is influenced by sample size and length of the analyzed region (see Section \ref{sec:model:methods}). The rates of expansion and contraction also substantially affect the ability to recover the correct population size; faster expansion and contraction rates incur more noisy estimates (the testing reported in Figure \ref{fig:model:eval}.B included extreme and possibly unrealistically large rates of expansion and contraction). This was evident when we classified the synthetic populations as either strong or mild contraction or expansion events and separately assessed the inference accuracy for each of these classes (see Supplementary Figure $5$).

\subsection{Two periods of expansion in the Ashkenazi Jewish population}

\begin{figure*}
    \centering
	\includegraphics[width=\textwidth]{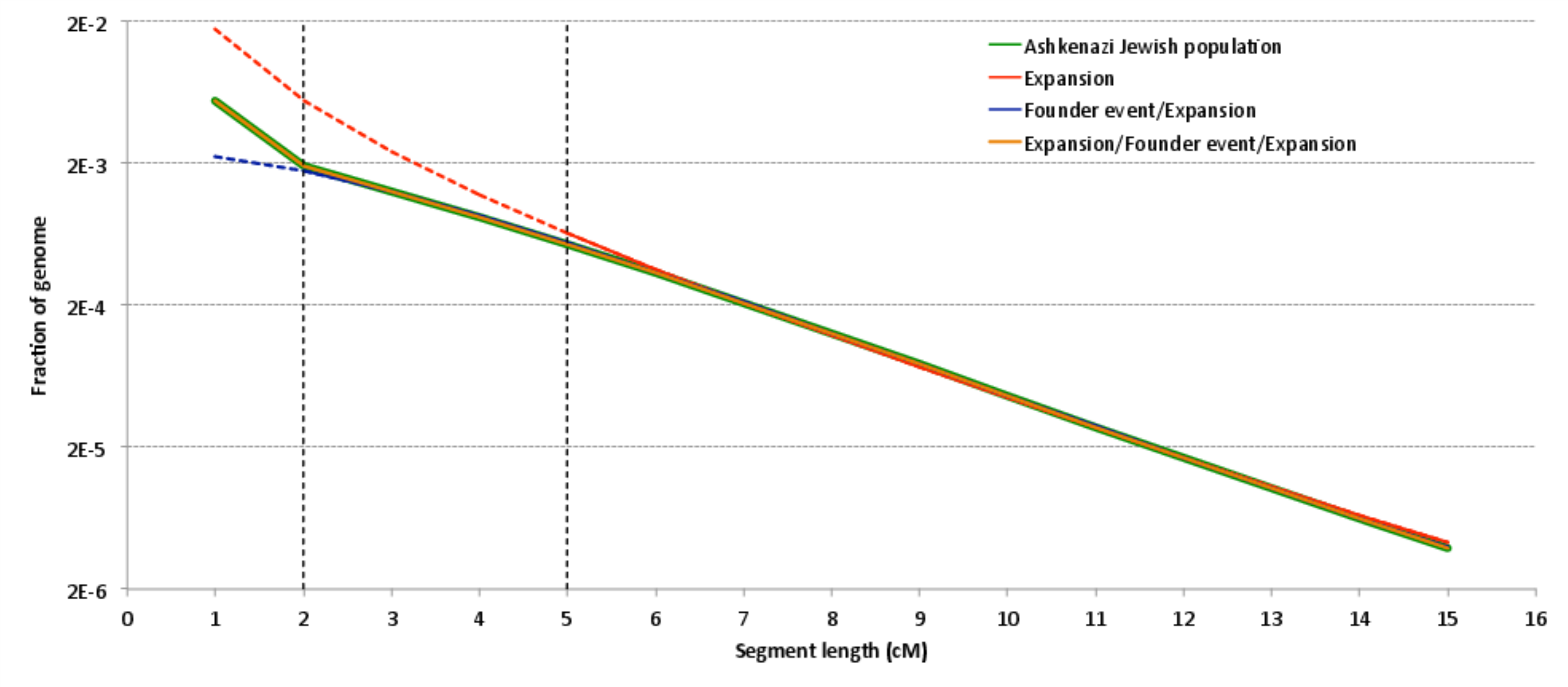}
	\caption{AJ inference. Observed distribution of haplotype sharing (green line); exponential expansion for only long ($>5$cM) segments (red line, best fit: $N_c=97,700,000, G=26, N_a=1,300$); founder event-expansion (purple line, best-fit: $N_c=12,800,000; G=35; N_{a1}=230; N_{a2}=70,600$); exponential expansion-founder event-exponential expansion (orange line, best-fit: $N_c=42,000,000; G_1=33; N_{a1}=230; N_{a2}=37,800; N_{a3}=1,800; G_2=167$).}
    \label{fig:model:AJ}
\end{figure*}

We analyzed the demographic history of the AJ population by applying our method to a real data set of $500$ individuals (segment-length distributions in Figure \ref{fig:model:AJ}). We initially tested several models by using the proposed procedure. After inferring the most likely parameters for the chosen model, we used simulations to refine the analytical solution and account for potential errors in IBD detection (see Supplementary Table 2 for an algorithmic summary of the analysis).

As a first step, we fitted a simple model of exponential growth (Figure \ref{fig:model:models}.B). If only long ($\geq 5$ cM) segments are considered, the parameters of this model can be optimized to provide a good match for the observed sharing. This supports the occurrence of an expansion event in the recent history of this population, as reported in our previous analysis using a simpler simulation-based approach \cite{gusev2012architecture}. However, exponential growth alone is unable to provide a good fit for the observed frequency of shorter segments, suggesting additional demographic dynamics during more ancient AJ history. The decay in the frequency of medium-length segments, between $2$ and $5$ cM, was weaker than that observed for longer ones, suggesting a founder event---a reduction of the ancestral population size and subsequent rapid expansion. Indeed, a refined model that allows such an event to predate exponential expansion (Figure \ref{fig:model:models}.C) provides a good fit for the frequency of all segments of length $\geq 2$ cM. We note that such a severe founder event was also reported in a previous analysis based on lower throughput data \cite{slatkin2004population,atzmon2010abraham} and is consistent with historical reports of this population \cite{finkelstein1960jews}. However, this model does not adequately explain why a further change in the slope of the sharing spectrum was observed for short segments between $1$ and $2$ cM of length. Such a steep increase in the frequency of short segments can again support the occurrence of an exponential growth preceding the observed founder event. We therefore optimized parameters for a model that allows two subsequent exponential-expansion periods separated by a founder event (Figure \ref{fig:model:models}.D). We focused our analysis on generations $1-200$ (i.e., setting $G_1+G_2=200$ in Figure \ref{fig:model:models}.D). The considered model allows $N_{a3}$ founders to exponentially expand to a population of $N_{a2}$ individuals during $G_2$ generations. After a founder event, $N_{a1}$ individuals are randomly selected and exponentially expand to reach a current population of $N_c$ individuals during the remaining $G_1$ generations. Using this model, we were able to obtain a good fit for the entire IBD frequency spectrum, corresponding to the parameter values $N_{a3} \sim 1,800$, $N_{a2} \sim 37,800$, $N_{a1} \sim 230$ ,and $G_1=33$ (therefore, $G_2 = 167$) and $N_c \sim 42,000,000$. Model comparison based on the AIC supports this model over simpler demographic scenarios. We note that the most recent expansion period was inferred to have a considerably high rate ($r \sim 0.37$, defined in Equation \ref{eq:model:7}). More complex models (e.g., inferring the value of $G_2$ and allowing for a founder event predating the remote expansion) did not significantly improve on the reported demography.

When real data is analyzed, the quality of computational phasing and IBD detection might affect the reconstruction accuracy. Inaccuracies in the recovery of long-range IBD haplotypes are reflected in the inferred current size of the AJ population, which is extremely large. This is most likely due to long IBD segments being shortened to smaller segments because of switch errors during computational phasing, in addition to greater uncertainty associated with the inference of recent large population sizes (Figure \ref{fig:model:eval}.B and Supplementary Figure $5$). We therefore refined inferred parameters to take into account such potential bias by using realistic coalescent simulations that also reproduce noise due to computational phasing and IBD discovery. We obtained an improved fit for a population composed of ${\sim}2,300$ ancestors $200$ generations before the present; this population exponentially expanded to reach ${\sim}45,000$ individuals $34$ generations ago. After a severe founder event, the population was reduced to ${\sim}270$ individuals, which then expanded rapidly during $33$ generations (rate $r \sim 0.29$) and reached a modern population of ${\sim}4,300,000$ individuals.

\subsection{IBD in the Maasai: the village model}

\begin{figure*}
    \centering
	\includegraphics[width=\textwidth]{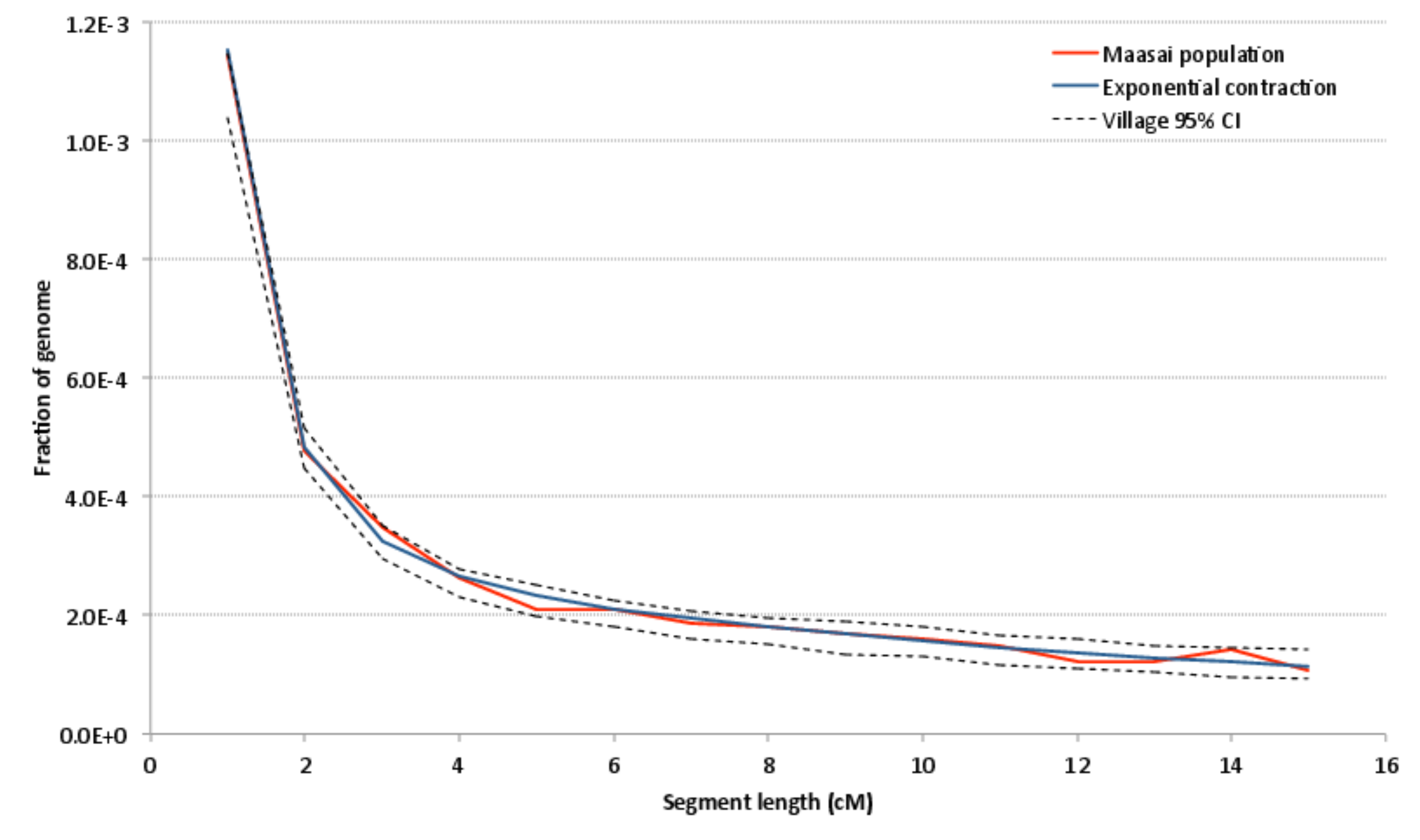}
	\caption{Demographic analysis of the Maasai. Observed distribution of haplotype sharing (red); single-population expansion model (blue); several small demes that interact through high migration rates (dashed CI obtained through random resampling of $200$ synthetic data sets).}
    \label{fig:model:MKK}
\end{figure*}

We additionally investigated the demographic profile of $56$ samples of self-reported unrelated MKK individuals from the HapMap 3 data set. We detected high levels of segmental sharing across individuals, consistent with recent analysis of hidden relatedness in this sample (see \cite{pemberton2010inference,gusev2012architecture}, Chapter \ref{chap:IBD_in_data}). Genome-wide IBD sharing was elevated among all individual pairs, suggesting high rates of recent common ancestry across the entire group rather than the presence of occasional cryptic relatives due to errors during sample collection (Supplementary Figure $6$). Optimizing a model of exponential expansion and contraction (Figure \ref{fig:model:models}.B), we obtained a good fit to the observed IBD frequency spectrum (Figure \ref{fig:model:MKK}), suggesting that an ancestral population of ${\sim}23,500$ individuals decreased to ${\sim}500$ current individuals during the course of $23$ generations ($r \sim -0.17$). We note that this result might not be driven by an actual gradual population contraction in the MKK individuals, but it most likely reflects the societal structure of this semi-nomadic population. Although little demographic evidence has been reported, the MKK population is in fact believed to have a slow but steady annual population growth \cite{coast2001maasai}. We hypothesized that a high level of migration across small-sized MKK villages (Manyatta) provides a potential explanation for the observed IBD patterns in this population. In such a model, a small genetic pool for recent generations gradually becomes larger as a result of migration across villages as one moves back into the past. To validate the plausibility of this hypothesis, we simulated a demographic scenario in which multiple small villages interact through high migration rates. This setting is similar to Wright's island model \cite{wright1943isolation}, and we shall refer to it as the village model in this case (Supplementary Figure $7$). We extracted IBD information for one of the simulated villages and attempted to infer its demographic history by using a single-population model of exponential expansion and contraction (Figure \ref{fig:model:models}.B). Indeed, the single-population model provides a good fit for this synthetic sample, and the severity of the gradual contraction of the population was observed to be proportional to the simulated migration rate. We thus used the village model to analyze the MKK demography and relied on coalescent simulations to retrieve its parameters: migration rate, size, and number of villages that provide a good fit for the empirical distribution of IBD segments. We observed a compatible fit for this model, in which $44$ villages of $485$ individuals each intermix with a migration rate of $0.13$ individuals per generation (Figure \ref{fig:model:MKK}).

Note that, although our simulations involved several villages of constant size, adequate choices of migration rates would result in the signature of a drastic contraction even among expanding villages (and, therefore, overall expanding population). From a methodological point of view, we further note that LD might also provide information for inferring such a ``village effect''. However, although current strategies for IBD detection allow finding shared haplotypes in the presence of computational phasing errors, LD analysis over long genomic intervals is substantially affected by noisy phase information (Supplementary Figure $8$).

\chapter{Reconstructing recent migration events}
\label{chap:migration}
In Chapter \ref{chap:IBDmodel} we introduced a model that allows expressing several relevant quantities of IBD sharing across purportedly unrelated individuals from a single population. We here extend this analysis to the case of individuals sampled from a number of different demes, of which we want to investigate recent demographic events such as population size fluctuation and migration. Details of this analysis can be found in \cite{palamara2013inference}. Note that in the remainder we measure genetic length in Morgans (M) rather than centimorgans (cM), used in the past chapter. This will often allow providing more compact expressions for the described quantities.

\section{IBD distributions in the presence of migration}
\label{IBDMig}

\begin{figure*}
	\vspace{-2mm}
        \centering
        \begin{subfigure}[b]{0.6\textwidth}
			\vspace{-2mm}
                \centering
                \includegraphics[width=\textwidth]{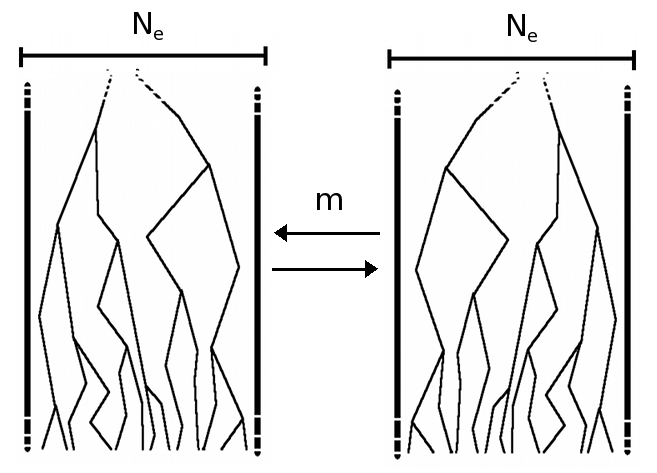}
                \caption{Constant, symmetric population size and migration rate}
                \label{fig:02a}
        \end{subfigure}
        
        \begin{subfigure}[b]{0.7\textwidth}
			\vspace{-2mm}
                \centering
                \includegraphics[width=0.75\textwidth]{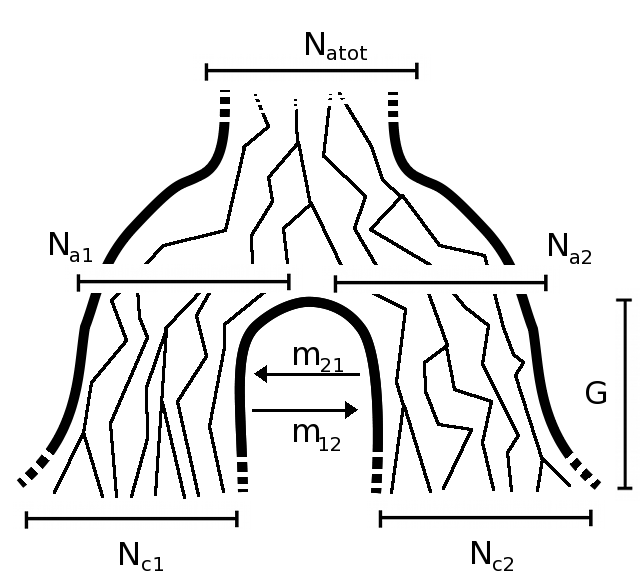}
                \caption{Population split followed by size changes and migration}
                \label{fig:02b}
        \end{subfigure}
        ~
        \caption{Two demographic models that involve two populations and migration between them. In model \ref{fig:02a} the populations have the same constant size $N_e$, and exchange individuals at the same rate $m$. In model \ref{fig:02b}, a population of constant ancestral size $N_{atot}$ splits $G$ generations in the past, resulting in two populations whose sizes independently fluctuate from $N_{a1}$ and $N_{a2}$ individuals to $N_{c1}$ and $N_{c2}$ individuals during $G$ generations. During this period, the populations interact with asymmetric migration rates $m_{12}$ and $m_{21}$.}
		\label{fig:02}
\end{figure*}

We begin discussing the case of multiple populations referring to a simple scenario, where two populations of constant size $N_e$ exchange individuals at a fixed rate $m$ per individual, per generation (see model in Figure \ref{fig:02a}). We encode this migration rate using the matrix
\begin{equation}
\mathbf{Q} =
\begin{pmatrix}
-m & m  \\
m & -m
\nonumber
\end{pmatrix}
\label{eq:mig:04}
\end{equation}
We consider two individuals, $i$ and $j$, each sampled from either population. We trace the ancestors of these individuals at one genomic site, and encode their state (in terms of population their ancestors belong to), using a vector of dimensionality $2$. If individual $i$ is sampled from population $1$ and individual $j$ from population $2$, for example, the state at generation $0$ is known and we write it as $\mathbf{v}_i(0)=(1,0)$, $\mathbf{v}_j(0)=(0,1)$. If both are sampled from population $1$, $\mathbf{v}_i(0)=(1,0)$, $\mathbf{v}_j(0)=(1,0)$. After $t$ generations (measured in continuous time), the probability that the ancestor of individual $i$ at this genomic location belongs to either population is given by
\begin{equation}
\mathbf{v}_i(t)=(1,0)e^{t\mathbf{Q}}=\left(\frac{e^{-2 m t}}{2} (1 + e^{2 m t}), \frac{e^{-2 m t}}{2} (e^{2 m t}-1)\right)
\label{eq:mig:05}
\end{equation}
if individual $i$ was sampled from population 1, or, symmetrically
\begin{equation}
\mathbf{v}_i(t)=(0,1)e^{t\mathbf{Q}}=\left(\frac{e^{-2 m t}}{2} (e^{2 m t}-1), \frac{e^{-2 m t}}{2} (1 + e^{2 m t})\right)
\label{eq:06}
\end{equation}
if it was sampled from population 2. We are interested in expressing the probability that individuals $i$ and $j$ coalesce at time $t$. This requires both individuals to be in the same population, in which case coalescence happens at rate $1/N_e$. Formally $p(t|m,N_e)=\mathbf{v}_i(t)\mathbf{v}_j(t)^\intercal/N_e$, which in this setting becomes
\begin{equation}
p(t|m,N_e)\approx\frac{1+e^{-4 m t}}{2N_e}
\label{eq:07}
\end{equation}
if $\mathbf{v}_i(0)=\mathbf{v}_j(0)$, and
\begin{equation}
p(t|m,N_e)\approx\frac{1-e^{-4 m t}}{2N_e}
\label{eq:08}
\end{equation}
otherwise. Note that a Taylor approximation was made in equations \ref{eq:07} and \ref{eq:08}. A more detailed derivation is reported in the Appendix of this chapter. To compute $\int_u^v p(l|\boldsymbol{\theta})\, \mathrm{d} l$, we plug the coalescence probability in Equation \ref{eq:model:2} (or its continuous version). Also, for simplicity we take $R=[u,\infty)$, obtaining
\begin{equation}
\int_u^\infty p(l|\boldsymbol{\theta})\, \mathrm{d} l=\frac{1}{2N_eu} + \frac{m + u}{2N_e(2 m + u)^2}
\label{eq:09}
\end{equation}
if $\mathbf{v}_i(0)=\mathbf{v}_j(0)$, and
\begin{equation}
\int_u^\infty p(l|\boldsymbol{\theta})\, \mathrm{d} l=\frac{m (4 m + 3 u)}{2 N_e u (2 m + u)^2}
\label{eq:10}
\end{equation}
otherwise. Recall that $\int_u^v p(l|\boldsymbol{\theta})\, \mathrm{d} l = f_R$, which is the expected fraction of genome shared through segments of length between $u$ and $v$ by an individual pair. To infer $\hat{N}$ and $\hat{m}$, we therefore consider the observed average fraction of genome shared through IBD segments longer than a threshold $u$, for all pairs of individuals sampled from the same population or from different populations (which we call $\hat{f}_{s}$ and $\hat{f}_{d}$, respectively, now omitting the dependence on the length range). We then solve the system obtained by equating $\hat{f}_{s}$ and $\hat{f}_{d}$ to the quantities in (\ref{eq:09}) and (\ref{eq:10}), to obtain the estimators
\begin{equation}
\begin{split}
\hat{N}_e = \frac{1}{(\hat{f}_{d}+\hat{f}_{s})u}\\
\hat{m} = \frac{u\left(3\hat{f}_{s}-5\hat{f}_{d}-\sqrt{2\hat{f}_{d}\hat{f}_{s}-7\hat{f}_{d}^2+9\hat{f}_{s}^2}\right)}{8(\hat{f}_{d}-\hat{f}_{s})}
\end{split}
\label{eq:11}
\end{equation}

A simple generalization of the above scenario consists in allowing the two considered populations to differ in their effective population sizes, $N_{e1}$ and $N_{e2}$. In this scenario it is still possible to obtain a closed form expression for $\int_u^v p(l|\boldsymbol{\theta})\, \mathrm{d} l$, and a closed form estimator for $\hat{N}_{e1}$, $\hat{N}_{e2}$, $\hat{m}$, which are reported in the Appendix.

\subsection{The general case}
\label{sec:general}

Although the previously discussed case of constant population sizes and migration rates has a simple formulation and can be used to gain initial insight into the recent demography of a study cohort, such population dynamics are oversimplified and generally unrealistic. Luckily, given a few reasonable assumptions, population sizes and migration rates can be allowed to arbitrarily fluctuate in time, still permitting a closed form computation of $\int_u^v p(l|\boldsymbol{\theta})\, \mathrm{d} l$.

Consider two populations whose sizes at generation $g$ are expressed as $N_1(g)$ and $N_2(g)$. The rate at which these two populations exchange individuals can be encoded in a discrete migration matrix 
\begin{equation}
\mathbf{M}(g) =
\begin{pmatrix}
1-m_{12}(g) & m_{12}(g)  \\
m_{21}(g) & 1-m_{21}(g)
\end{pmatrix}
\label{eq:13}
\end{equation}
where $m_{12}(g)$ represents the probability of an individual migrating from population $1$ to population $2$ at generation $g$ (backwards in time). After $g$ generations, the probability that the ancestor of individual $i$ at a genomic location belongs to either population is given by the vector $\mathbf{v_i}(0)\prod_{k=0}^g \mathbf{M}(k)$. Define the matrix $\mathbf{N}(g)$ to be diagonal with $1/N_1(g)$ and $1/N_2(g)$ as its diagonal elements. The probability of coalescence from generation $g-1$ to generation $g$ is then
\begin{equation}
c_g = \left[\mathbf{v_i}(0)\prod_{k=0}^g \mathbf{M}(k)\right] \mathbf{N}(g) \left[ \mathbf{v_j}(0)\prod_{k=0}^g \mathbf{M}(k) \right]^\intercal
\label{eq:14}
\end{equation}
And the probability of the two individuals to coalesce $g$ generations before present is
\begin{equation}
p(g|\mathbf{M}(g),\mathbf{N}(g)) = c_g\prod_{k=1}^{g-1}(1-c_k)
\label{eq:15}
\end{equation}
Equation \ref{eq:15} can be used to compute
\begin{equation}
\begin{split}
\int_u^v p(l|\mathbf{M}(g),\mathbf{N}(g))\, \mathrm{d} l = \sum_{g=1}^\infty \left[ c_g\prod_{k=1}^{g-1}(1-c_k) \int_u^v p(l|g)\, \mathrm{d} l \right] \\
\end{split}
\label{eq:16}
\end{equation}

Note that Equation \ref{eq:16} is very general, and we can allow additional demographic changes to take place. For instance, by setting $N_2(g)=0$, $m_{12}(g)=0$ and $m_{21}(g)=1$ for all $g>G$, we encode a population split that occurred $G$ generations ago. In practice, a pair of populations will have split a number of generations back in time, and it is therefore convenient to consider models of the kind depicted in Figure \ref{fig:02b}. In this model a population of constant size $N_{atot}$ splits $G$ generations in the past, forming two populations of size $N_{a1}$ and $N_{a2}$. The size of these two groups then fluctuates in time, to reach a present size of $N_{c1}$ and $N_{c2}$. During their separation, the populations exchange individuals at a rate of $m_{12}$ and $m_{21}$ per generation, per individual. Of course, other models can be defined, allowing variable migration rates, and different population size dynamics.

For mathematical convenience, it is safe to assume the ancestral population size becomes constant a number of generations in the past. Models where the ancestral population size ($N_{atot}$ in Figure \ref{fig:02b}) is constant from generation $G$ to infinity allow for a closed form computation of Equation \ref{eq:16}, no matter which demographic dynamics take place from generation 0 to $G$ (Equations \ref{eq:model:CF1} and \ref{eq:model:CF2}). Furthermore extremely remote demographic events have negligible impact on shared haplotypes of currently detectable lengths (e.g. $>1$ cM).

\subsection{Simulations, ancestry deconvolution and real data}

We tested our framework using extensive simulation of realistic chromosomes under several demographic models, using the GENOME coalescent simulator (\cite{liang2007genome}). For computational convenience, we set the size of the simulator's non-recombinant segments between $0.01$ and $0.025$ cM, always using a recombination rate of $1$cM/Mb. A modified version of the simulator was used to extract ground truth IBD haplotypes from the simulated genealogies, defined as segments co-inherited by pairs of individuals from their most recent common ancestor (see definition (b) in Section \ref{intro:subsec:IBD_def}). For some of the simulations we inferred shared haplotypes using the GERMLINE software package (\cite{gusev2009whole}) on phased genotype data, which was obtained setting GENOME's mutation rate to $1.1\times10^{-8}$ per base pair (\cite{roach2010analysis}). Genotypes were post-processed to mimic the information content of array data. To this extent, we computed the allele frequency spectrum of European individuals from the HapMap 3 dataset (\cite{frazer2007second}), using frequency bins of $2\%$. We then randomly selected the same proportion of alleles from the simulated genotypes. We obtained an average density of ${\sim}230$ single nucleotide polymorphisms/Mb.

In order to compare the proposed IBD-based approach for migration inference to the approach of \cite{gravel2012population}, which is based on ancestry deconvolution, we simulated synthetic datasets under several demographic models, and extracted genotype data as previously described. We then ran the PCAdmix software (\cite{brisbin2012pcadmix}) with windows of size $0.3$cM and the genetic map used in the simulations. The output of PCAdmix was used to infer migration rates via the Tracts software package (\cite{gravel2012population}). IBD information was computed in the same datasets running the GERMLINE software, and the output was used to infer migration rates using the DoRIS software package, which implements the proposed framework. Perfectly phased haplotypes were used in input for both PCAdmix and GERMLINE. Only migration rates were inferred, while all other demographic parameters were set to the true simulated values for both Tracts and DoRIS.

To demonstrate the use of the DoRIS framework on real data, we analyzed 56 trio-phased samples from the HapMap 3 dataset. Phased genotypes were downloaded from the HapMap 3 web page at http://hapmap.ncbi.nlm.nih.gov. IBD haplotypes were extracted using GERMLINE, as previously described in \cite{palamara2012length}.

\section{Results of evaluation and real data analysis}

\begin{figure*}[!ht]
	\vspace{-2mm}
        \centering
        \begin{subfigure}[b]{0.5\textwidth}
			\vspace{-2mm}
                \centering
                \includegraphics[width=\textwidth]{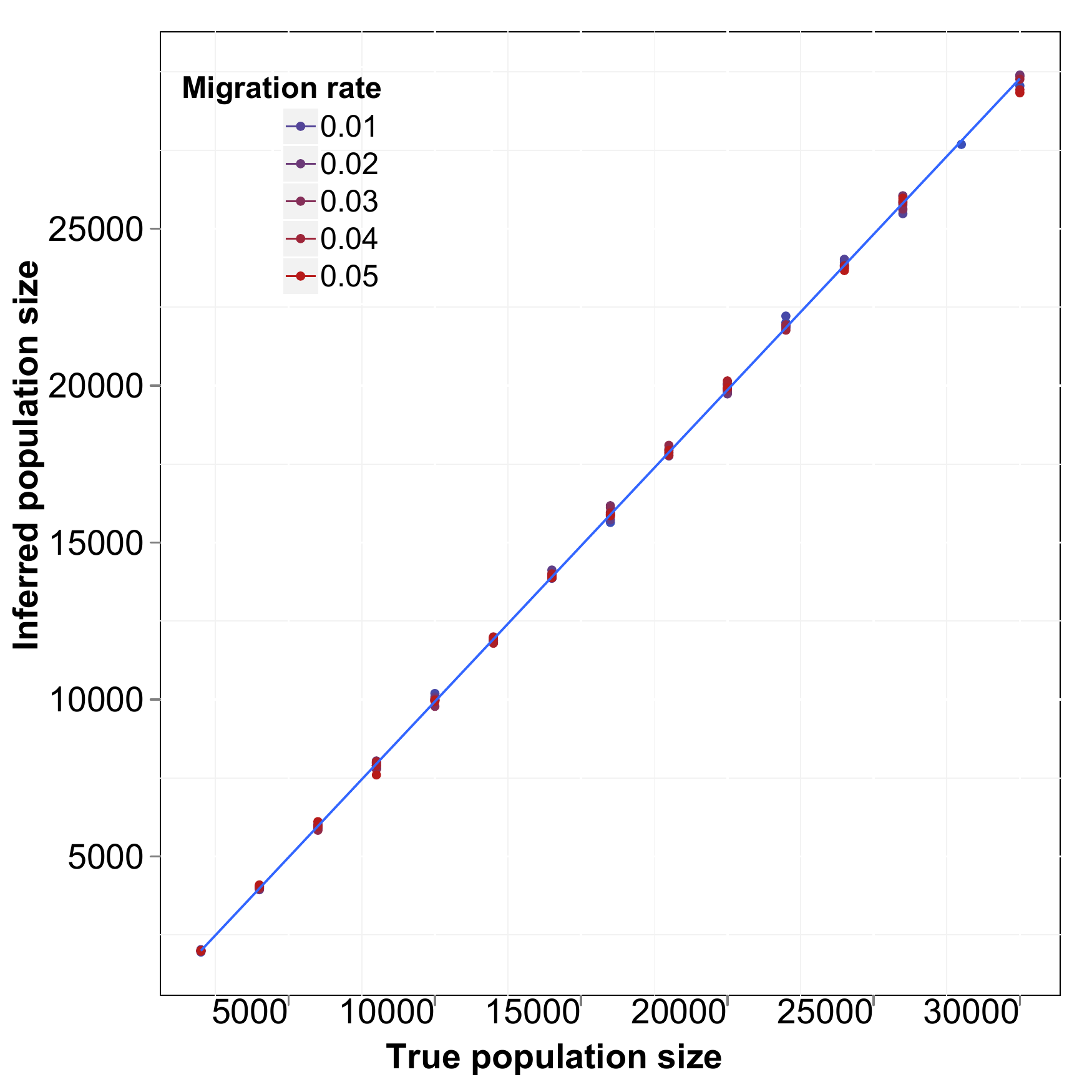}
                \caption{True vs. inferred effective population size}
                \label{fig:03a}
        \end{subfigure}
        ~
        \begin{subfigure}[b]{0.5\textwidth}
			\vspace{-2mm}
                \centering
                \includegraphics[width=\textwidth]{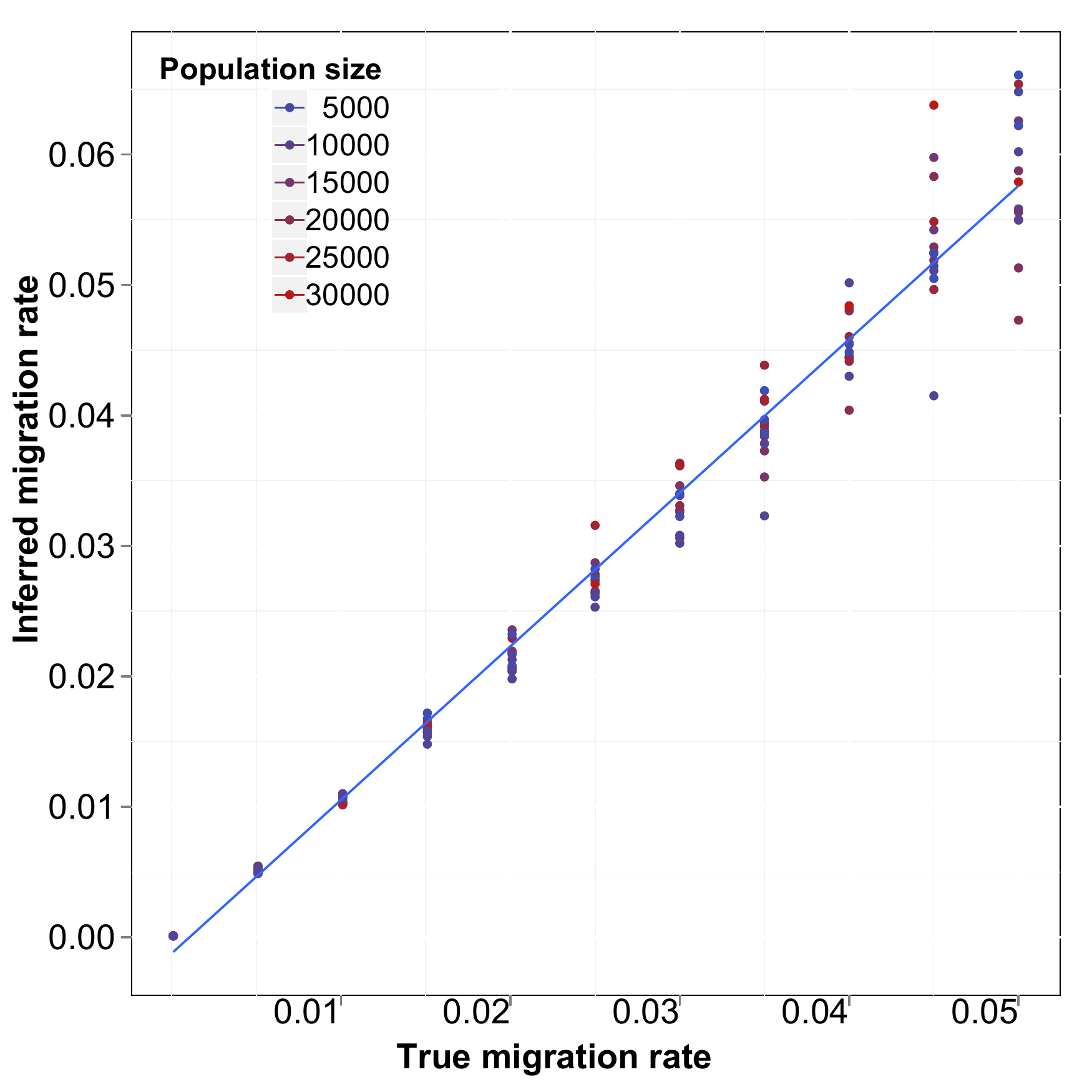}
                \caption{True vs. inferred migration rate}
                \label{fig:03b}
        \end{subfigure}
        ~
        \caption{True vs. inferred parameters for the model in Figure \ref{fig:02a}.}
		\label{fig:03}
\end{figure*}

\subsection{Constant size and symmetric migration rates}

In order to test the accuracy of demographic inference based on the proposed model, we initially simulated a number of populations of constant size $N_e$, which exchange individuals at a constant, symmetric migration rate $m$, as depicted in the model of Figure \ref{fig:02a}. We simulated 15 possible sizes of synthetic populations, ranging from $2,000$ to $30,000$ haploid individuals, with increments of $2,000$. For each population size, we simulated 11 possible migration values, uniformly chosen between $10^{-4}$ and $5\times10^{-2}$. For a total of $165$ datasets, we simulated a chromosome of $300$ centimorgans for $500$ haploid individuals from each subpopulation, and computed IBD sharing within and across populations. The simulations used non-recombining blocks of $0.02$ cM. This resolution may introduce small biases in the analysis, which we found to be negligible in our previous work. We then used Equation \ref{eq:11} to estimate $\hat{m}$ and $\hat{N_e}$, with results shown in Figure \ref{fig:03}. To test the model's accuracy, for this analysis we only considered ground-truth IBD segments extracted from the synthetic genealogies.

\begin{figure}[!ht]
	\vspace{-2mm}
\centering
\includegraphics[width=0.7\textwidth]{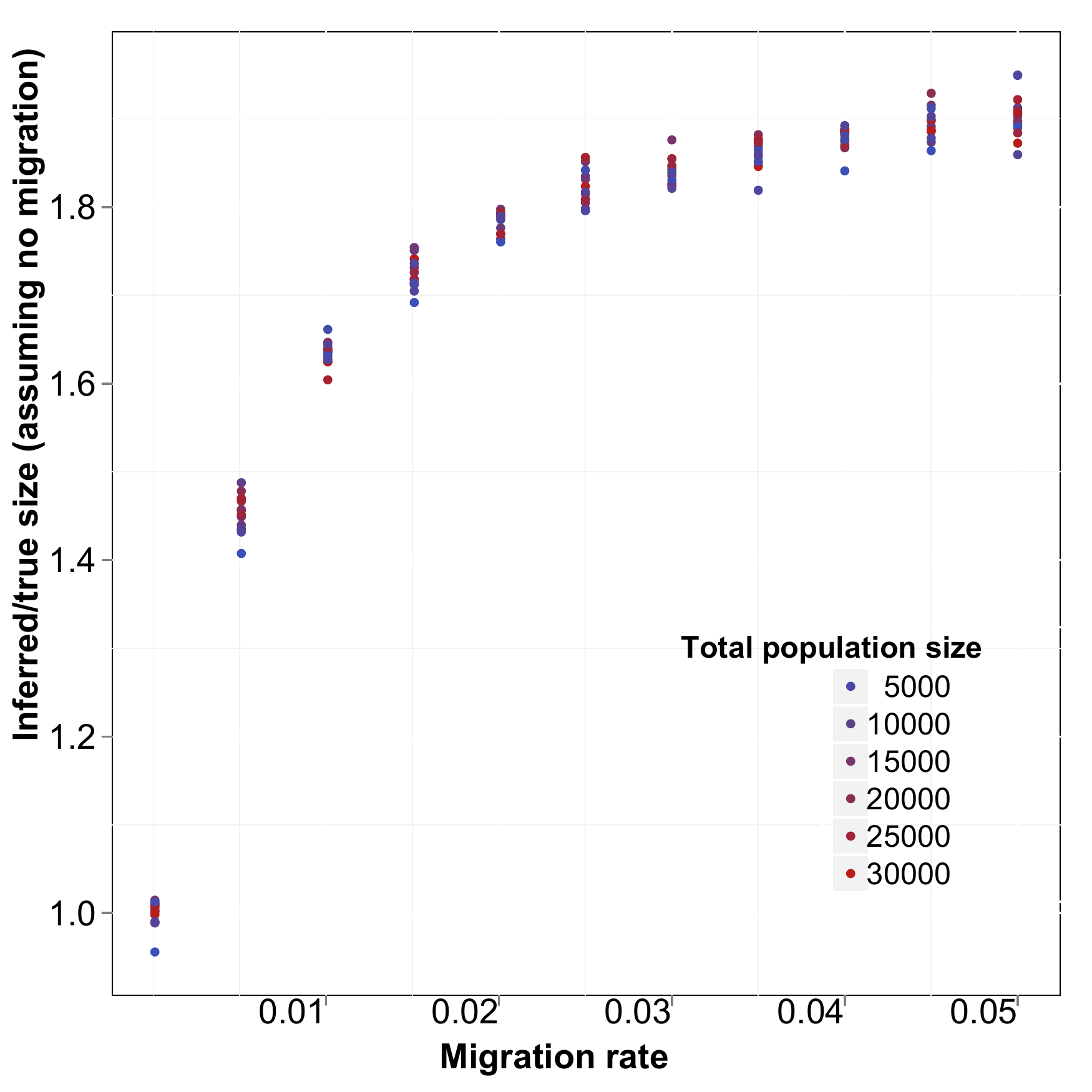}
\caption{Inference of recent effective population size using Equation \ref{eq:model:20}, which neglects migration. The ratio between inferred and true population size (y axis) increases as the migration rate (x axis) is increased, approaching the sum of population sizes for both populations (twice the true size).}
\label{fig:04}
\end{figure}

We obtained a good correspondence between the true population size and the size inferred via the estimator of Equation \ref{eq:11}, with almost perfect correlation shown in Figure \ref{fig:03a}. Inferred migration rates were also very close to the simulated rates, although a moderate upward bias and higher estimation variance for large migration rates was observed in this case (Figure \ref{fig:03b}). In addition to using the effective population size estimator of Equation \ref{eq:11}, we used the estimator of Equation \ref{eq:model:20}, which assumes constant population sizes and no migration. As expected, the inferred recent effective population size was in this case inflated by the presence of migration, as shown in Figure \ref{fig:04}. When migration rates are increased, the inferred population size quickly approaches the total population size (in this case $2N_e$).

\subsection{Dynamic size and asymmetric migration rates}

\begin{figure*}
	\vspace{-2mm}
    \centering
    \begin{subfigure}[b]{0.5\textwidth}
		\vspace{-2mm}
            \centering
            \includegraphics[width=\textwidth]{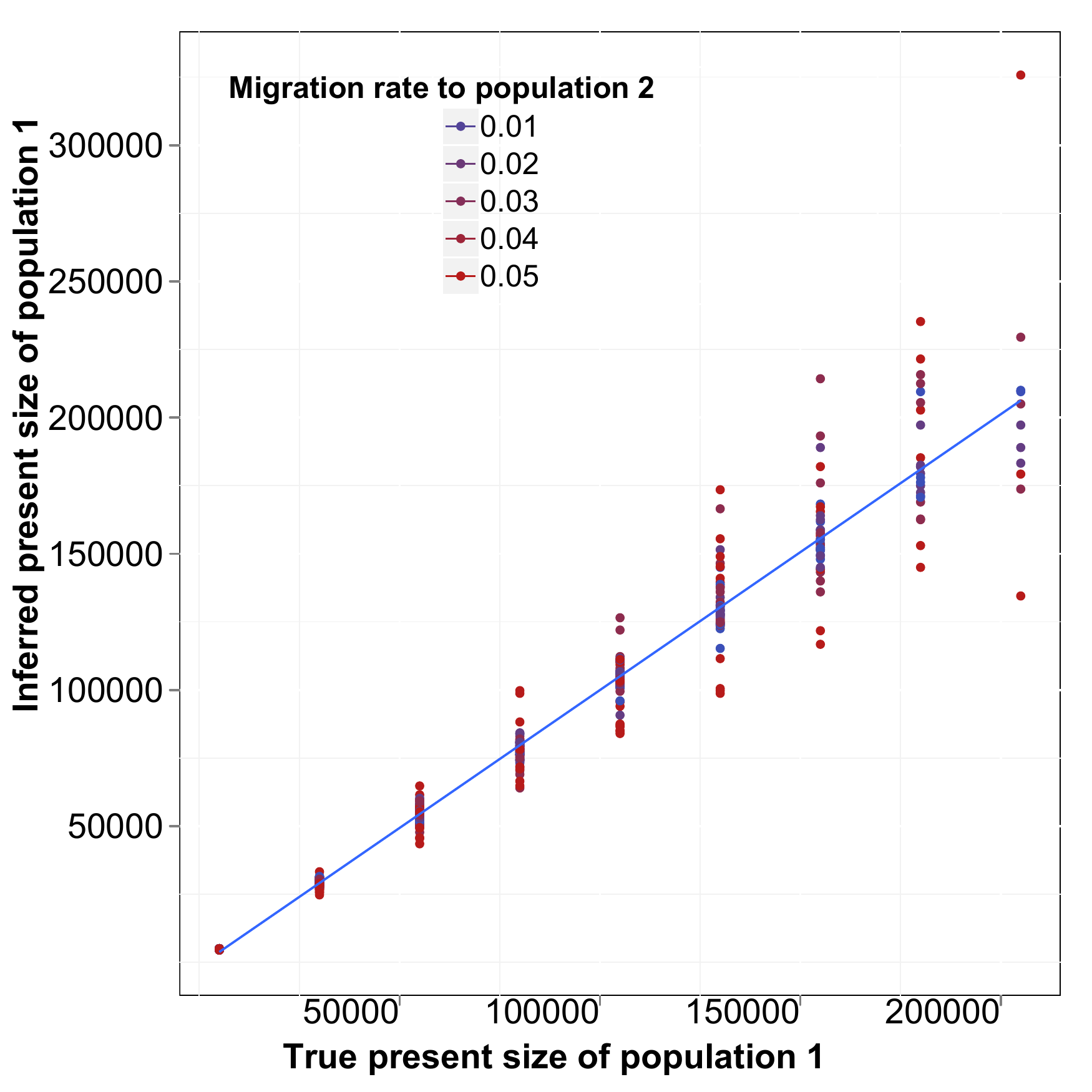}
            \caption{True vs. inferred value of $N_{c1}$}
            \label{fig:05a}
    \end{subfigure}
    ~
    \begin{subfigure}[b]{0.5\textwidth}
		\vspace{-2mm}
            \centering
            \includegraphics[width=\textwidth]{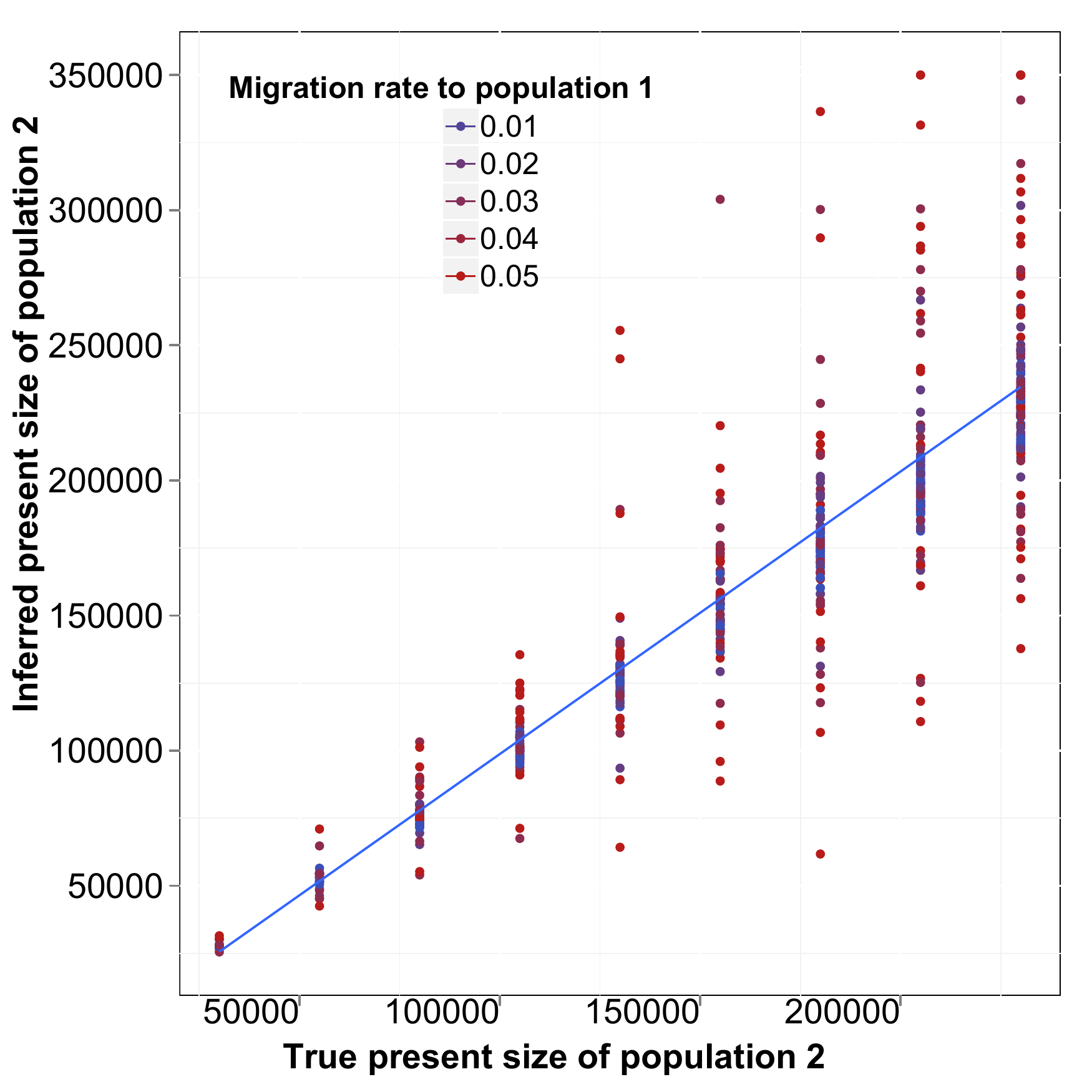}
            \caption{True vs. inferred value of $N_{c2}$}
            \label{fig:05b}
    \end{subfigure}
    ~
      \centering
      \begin{subfigure}[b]{0.5\textwidth}
              \centering
              \includegraphics[width=\textwidth]{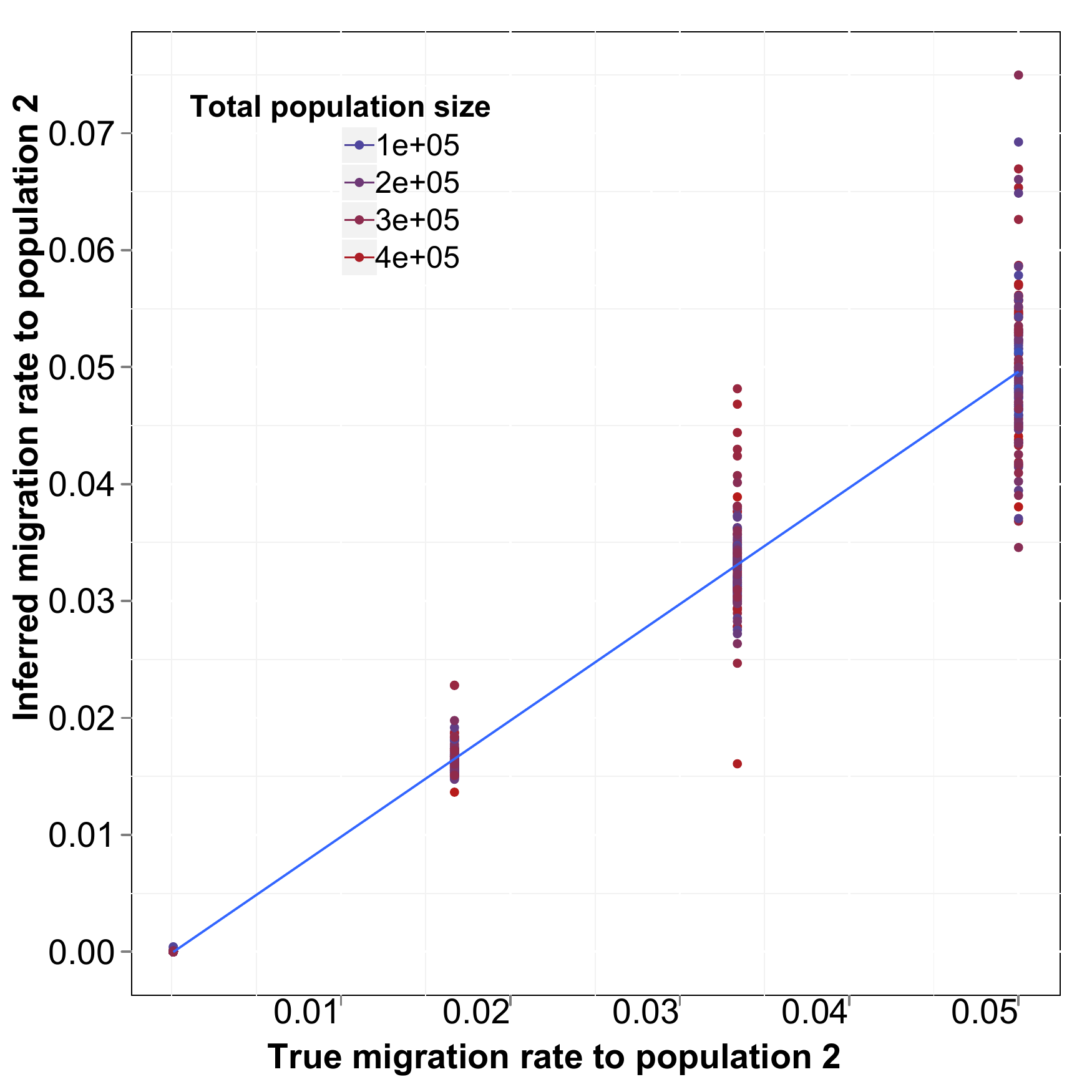}
              \caption{True vs. inferred value of $m_{12}$}
              \label{fig:05c}
      \end{subfigure}
      ~
      \begin{subfigure}[b]{0.5\textwidth}
              \centering
              \includegraphics[width=\textwidth]{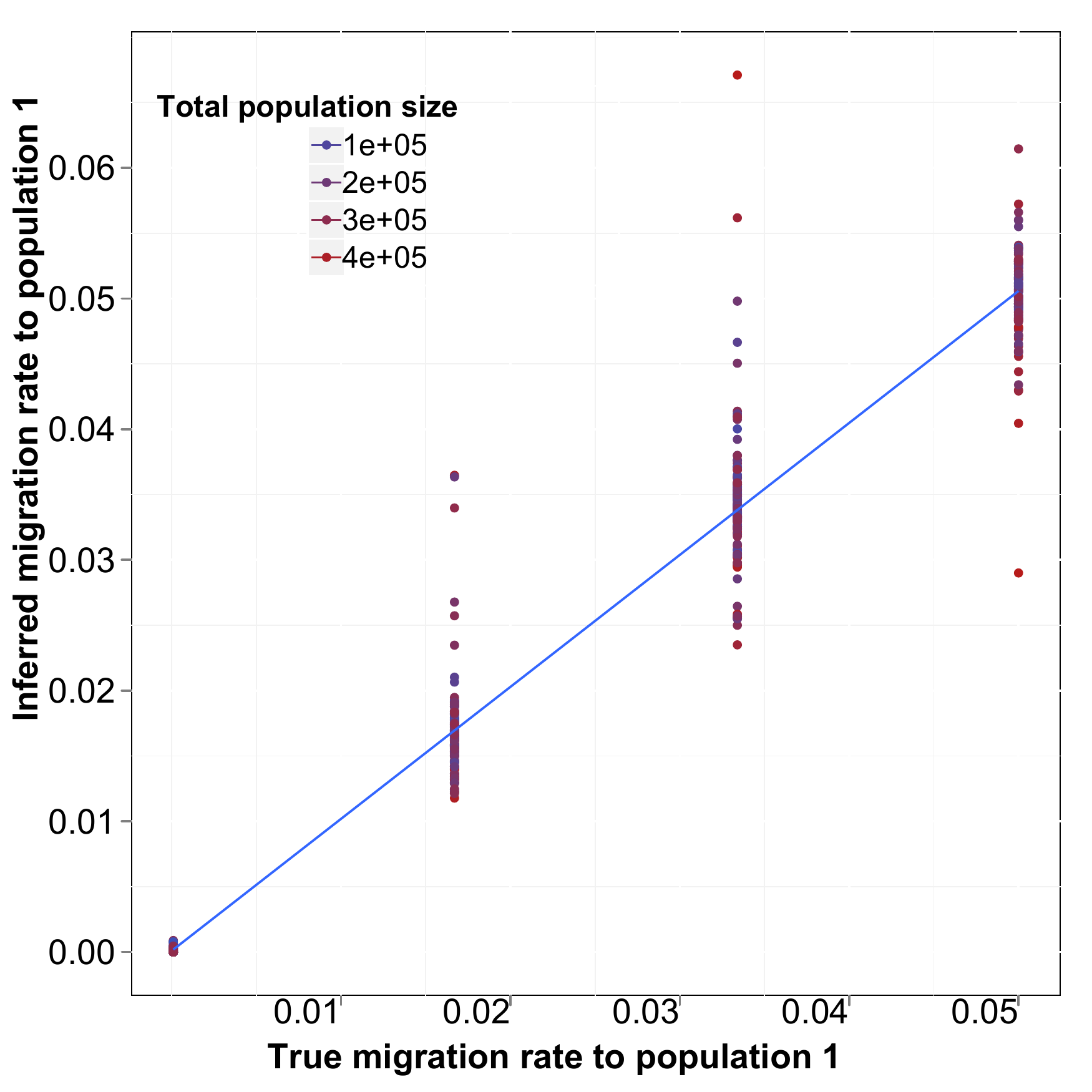}
              \caption{True vs. inferred value of $m_{21}$}
              \label{fig:05d}
      \end{subfigure}
      ~
        \caption{Results of the evaluation of our method on synthetic populations with demographic history depicted in the model of Figure \ref{fig:02b}. Higher variance in the method's accuracy is observed due to limited sample sizes and increased population sizes. Higher migration rates further decrease the rate of coalescent events in the recent generations (Figure \ref{fig:05b}), resulting in additional uncertainty. However no significant bias is observed in the inference.}
		\label{fig:05}
\end{figure*}
We then tested our model's performance in the more complex demographic scenario depicted in Figure \ref{fig:02b}, where a population splits into two subpopulations which grow at different exponential rates, interacting with asymmetric migration rates. We simulated a chromosome of ${\sim}275$ cM for $500$ haploid individuals per subpopulation. Simulated non-recombinant blocks had size $0.025$ cM. In all simulated scenarios, we kept $N_{atot}$ fixed to $10,000$ haploid individuals, while $N_{a1}$ and $N_{a2}$ were kept fixed at $5,000$ individuals. For $N_{c1}$ and $N_{c2}$ we simulated all possible combinations of sizes between $5,000$ and $205,000$ haploid individuals, with increments of $15,000$ (excluding cases where $N_{c1} = N_{c2}$). Note that on average the simulated values of $N_{c1}$ were smaller, resulting in higher inference accuracy compared to $N_{c2}$. For each pair of population sizes we simulated values of $m_{12}$ and $m_{21}$ using all combinations of the migration rates $0.0001$, $0.0167$, $0.0334$, and $0.5$.

A total of 540 synthetic populations were tested. For each synthetic population we extracted the average fraction of genome shared through haplotypes of different length intervals by pairs of individuals within each population or across populations. As in our previous work, we used a combination of intervals of uniform length and length intervals corresponding to quantiles of the Erlang-2 distribution, which is used in $p(l|t)$. Inference performance was tested via minimization of the root mean squared deviation between observed and predicted average fraction of shared genome. Note that a likelihood based approach (e.g. considering the number of shared segments) could be used based on the quantities derived in the previous chapter. We scanned several possible values for one parameter at a time, performing a line search while fixing the remaining model parameters to the true simulated value. The results of this analysis are reported in Figure \ref{fig:05}.

As expected, due to the large recent effective population sizes we simulated, the variance of the inference accuracy was higher in this scenario, suggesting that more than a single chromosome for $500$ diploid individuals may be required for the analysis of these demographies. A single chromosome of ${\sim}250$ centimorgans sampled in $500$ diploid individuals is in fact equivalent for the purpose of this inference to the analysis of all the autosomal chromosomes for ${\sim}150$ diploid samples (see \cite{palamara2012length}). Larger population sizes result in lower signal to noise ratio for the estimation of the expected fraction of genome shared via IBD segments, and increasing sample size or analyzing additional chromosomes is expected to reduce the variance in the inference performance. Lower accuracy was observed in the inference of $N_{c2}$ since, as previously mentioned, this simulated subpopulation was on average larger. Inferred population sizes were more accurate in the presence of low migration rates (represented by colors in figures \ref{fig:05a} and \ref{fig:05b}), as high migration further reduces the chance of early coalescent events, exacerbating the effects of large population sizes. Overall, no significant bias was observed in the recovered parameter values, suggesting our model provides a good match for the empirical distributions.

\subsection{Applicability of the model to genotype data}

While the previous analysis was mainly concerned with testing the model's accuracy, and relied on ground-truth IBD sharing extracted from the simulated genealogies, it is interesting to ask whether this approach can be used on genotype data. To this extent, we simulated genotypes for the demographic model of Figure \ref{fig:02a}. We set the population sizes to $4,000$ or $12,000$ diploid individuals per population, and extracted 300 diploid sampled from each group. The migration rate was symmetric, and set to $0.04$ per individual, per generation. Chromosomes of $150$ cM were simulated using non-recombinant blocks of size $0.01$ cM, and the synthetic genotypes were post-processed to reproduce the density and allele frequency spectrum of realistic SNP array data. In addition to extracting the ground truth IBD information as previously described, we inferred IBD haplotypes from the simulated genotypes using the GERMLINE software. The results suggest that when accurate phase information is available (e.g. for the X Chromosome, or for trio-phased samples), GERMLINE is able to recover the IBD sharing distribution across any pair of samples with high fidelity (Figure \ref{fig:06}). However, when the samples were computationally phased using the Beagle software (\cite{browning2007rapid}), GERMLINE had an inconsistent performance, accurately recovering the IBD sharing in the case of $N=4,000$, while poorly inferring long haplotypes in the case of $N=12,000$. This suggests that additional care must be taken when analyzing computationally phased data, particularly when analyzing cross-population IBD spectra, were the quality of the inferred IBD haplotypes will likely vary from population to population, as a result of different underlying demographic histories.

\subsection{MKK analysis, revisited}

To demonstrate the applicability of our method to real data, we analyzed the HapMap 3 Masai dataset, which was already studied in our previous work using a simulation-based approach. We here revisit this analysis, using the described analytical framework.

Cryptic relatedness across individuals in this dataset is extremely common, and does not appear to be due to the presence of occasional outliers among the samples. Demographic reports are not supportive of recent population bottlenecks in this group, which is though to be slowly but steadily expanding (\cite{coast2001maasai}). The Masai are a semi-nomadic people, and individuals often reside in small communities (\emph{Manyatta}) of tens to few hundreds of members. To study their demography, we therefore use a model where $V$ villages of constant size $N$ exchange individuals at a constant and symmetric rate $m$. This model is similar to the one depicted in Figure \ref{fig:02a}, with symmetric migration rates across several populations. We assumed that all samples were extracted from the same village, and used the model described in Section \ref{sec:general} for the analysis. We performed a grid search testing migration rates from $0.01$ to $0.4$, with intervals of $0.01$, village sizes from $50$ to $4,000$ with steps of $10$, and number of villages from $3$ to $150$ with increments of $1$. We also obtained $95\%$ confidence intervals for the inferred values using a bootstrap approach, by creating $400$ resamples randomly selecting individuals with replacement, then recomputing the optimal parameters using a gradient-driven procedure, which was initialized using the parameters inferred using the original samples (note, however, that small correlations exist for IBD sharing across individual pairs, and this method may provide optimistic intervals). Using this approach, we obtained the following estimates: $V=58$ ($95\%$ CI: $46$ to $75$), $N=400$ ($95\%$ CI $360$ to $470$), and $m=0.1$ ($95\%$ CI $0.09$ to $0.12$).

\subsection{Comparison with existing methods}

The structure of long-range haplotypes is known to carry relevant information about recent population dynamics, but this genomic feature has only recently become observable thanks to the development of modern high-throughput genomic technologies. As a consequence, methods that rely on a population's haplotypic structure to reconstruct demographic events have only recently arose. A model proposed in \cite{pool2009inference}, and recently expanded in \cite{gravel2012population}, provides a way to analyze the distribution of migrant tracts and infer the timing and intensity of very recent migration events. In order to analyze the distribution of migrant haplotypes, however, ancestry deconvolution needs to be accurately performed. This typically requires the availability of two suitable reference populations, which are required to be sufficiently diverged from each other. The amount of required divergence depends on the specific method used for the deconvolution, but in general this poses significant constraints in terms of the demographic scenarios that can be analyzed using these methods. 

\begin{figure}[!ht]
	\vspace{-2mm}
\centering
\includegraphics[width=0.65\textwidth]{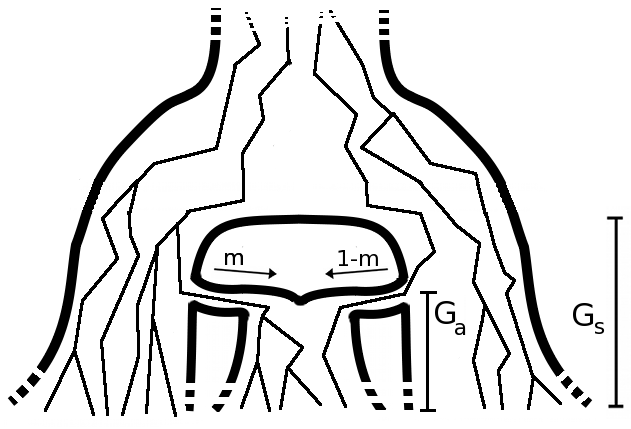}
\caption{The model used to simulate admixed populations.}
\label{fig:07}
\end{figure}

\begin{figure}[!ht]
	\vspace{-2mm}
\centering
\includegraphics[width=0.8\textwidth]{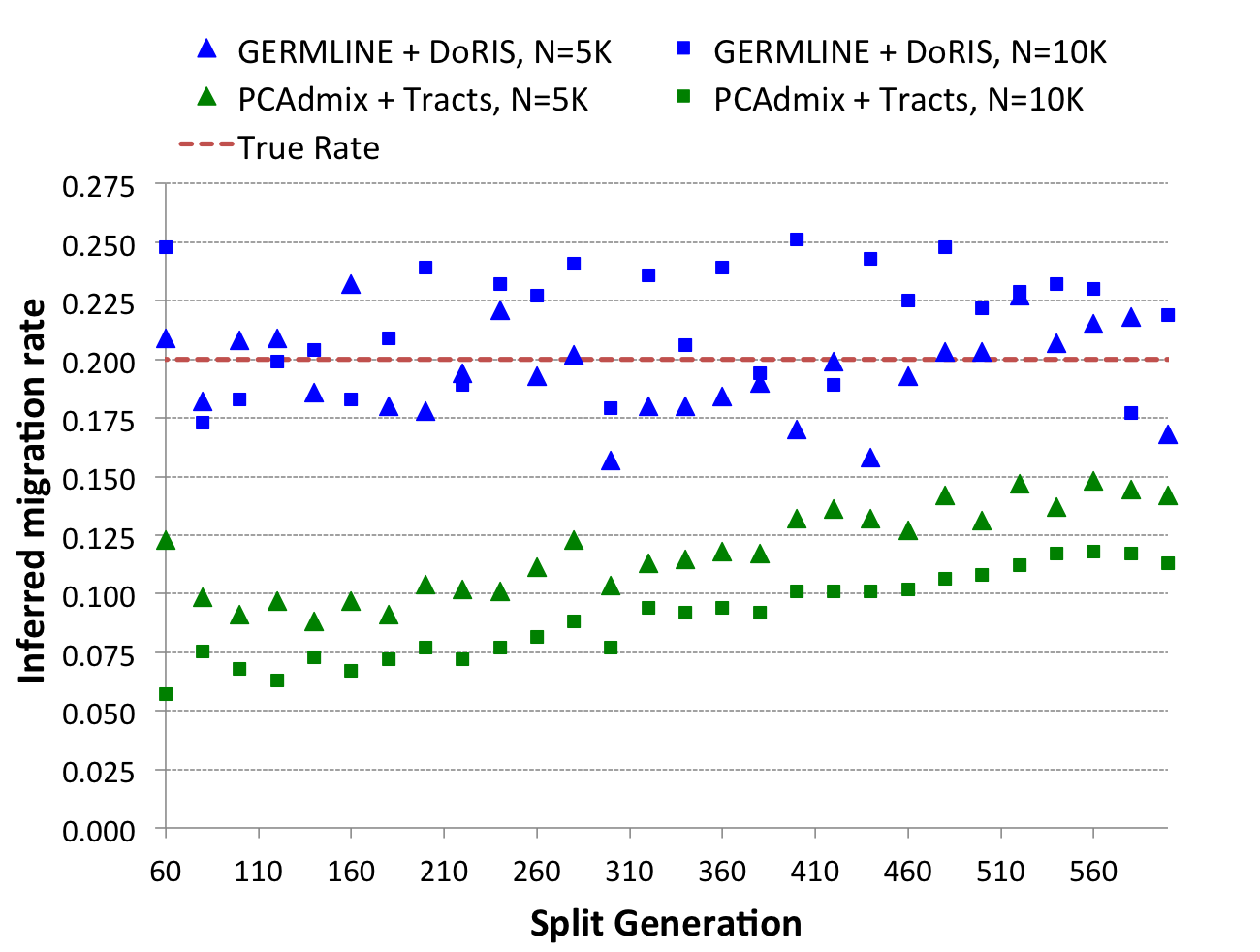}
\caption{We created several simulation genotype datasets using the model in Figure \ref{fig:07}, varying $G_s$ while keeping $m=0.2$, $G_a=25$, and using constant populations of size $5,000$ or $10,000$ diploid individuals. We inferred the value of $m$ using PCAdmix+Tracts, or GERMLINE+DoRIS, here reported as a function of $G_s$.}
\label{fig:08}
\end{figure}

To compare our IBD-based approach to methods based on ancestry deconvolution, we simulated the demographic scenario of Figure \ref{fig:07}, where two populations split $G_s$ generations in the past, and $G_a$ generations in the past contribute a fraction of genomes to the creation of a group of admixed individuals, with probability $m$ and $1-m$, through a unique pulse of migration. All three population sizes were fixed to either $N=5,000$ or $N=10,000$, $m$ was set to $0.2$, and $G_a$ was $25$ in all simulations. We varied $G_s$ from $40$ to $600$, with increments of 20, and extracted genotype data on a single 400 cM chromosome for 250 diploid samples in each of the three extant populations. We used the output of the PCAdmix software as input for the Tracts program (\cite{gravel2012population}), and the IBD segments retrieved by GERMLINE as input for the DoRIS software. Note that for the IBD analysis we only used the 250 admixed samples and the 250 samples from the population contributing ${\sim}m$ haplotypes at generation $G_a$, while the samples from the third population were ignored. In both cases we inferred the value of $m$, setting all other parameters to the true simulated values, with results shown in Figure \ref{fig:08}.

\begin{figure}[!ht]
	\vspace{-2mm}
\centering
\includegraphics[width=0.8\textwidth]{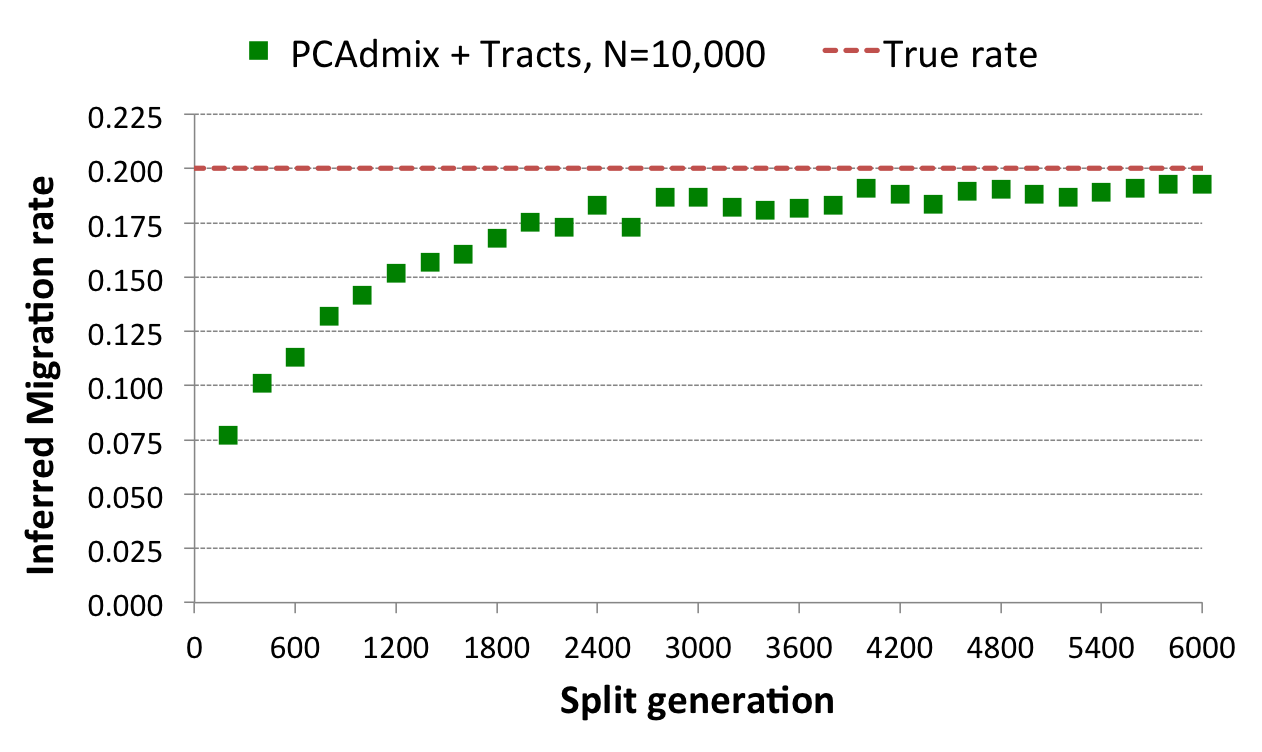}
\caption{We created several datasets using the model in Figure \ref{fig:07}, varying $G_s$ from $200$ to $6,000$, and using $m=0.2$, $G_a=25$ with population sizes of $10,000$ diploid individuals. We inferred the value of $m$ using PCAdmix+Tracts from phased genotype data.}
\label{fig:09}
\end{figure}

DoRIS performed better on average (mean inferred $m=0.205$, std $0.025$), while providing slightly noisy results, suggesting the need for a larger sample size and/or the analysis of additional chromosomes. The migration rate inferred by Tracts (mean $m=0.104$, std $0.0233$) was strongly biased. We note that in this setting Tracts is essentially used to only report the proportion of ancestry inferred by the deconvolution method, which is the actual source of inaccuracy. Even for populations that diverged 600 generations in the past (${\sim}15,000$ years before present assuming a generation of 25 years), the recovered rate was substantially lower than the simulated rate. The case of $N=5,000$ yielded better estimates, due to the higher drift found in smaller populations, which improved the power of PCAdmix to call migrant tracts. We additionally run the PCAdmix+Tracts analysis on longer time scales, simulating values of $G_s$ from $200$ to $6,000$, with intervals of $200$ generations, using $N=10,000$. Even for several thousand generations since the split of the reference populations, a small bias was observed (Figure \ref{fig:09}).

\begin{figure}[!ht]
	\vspace{-2mm}
\centering
\includegraphics[width=0.8\textwidth]{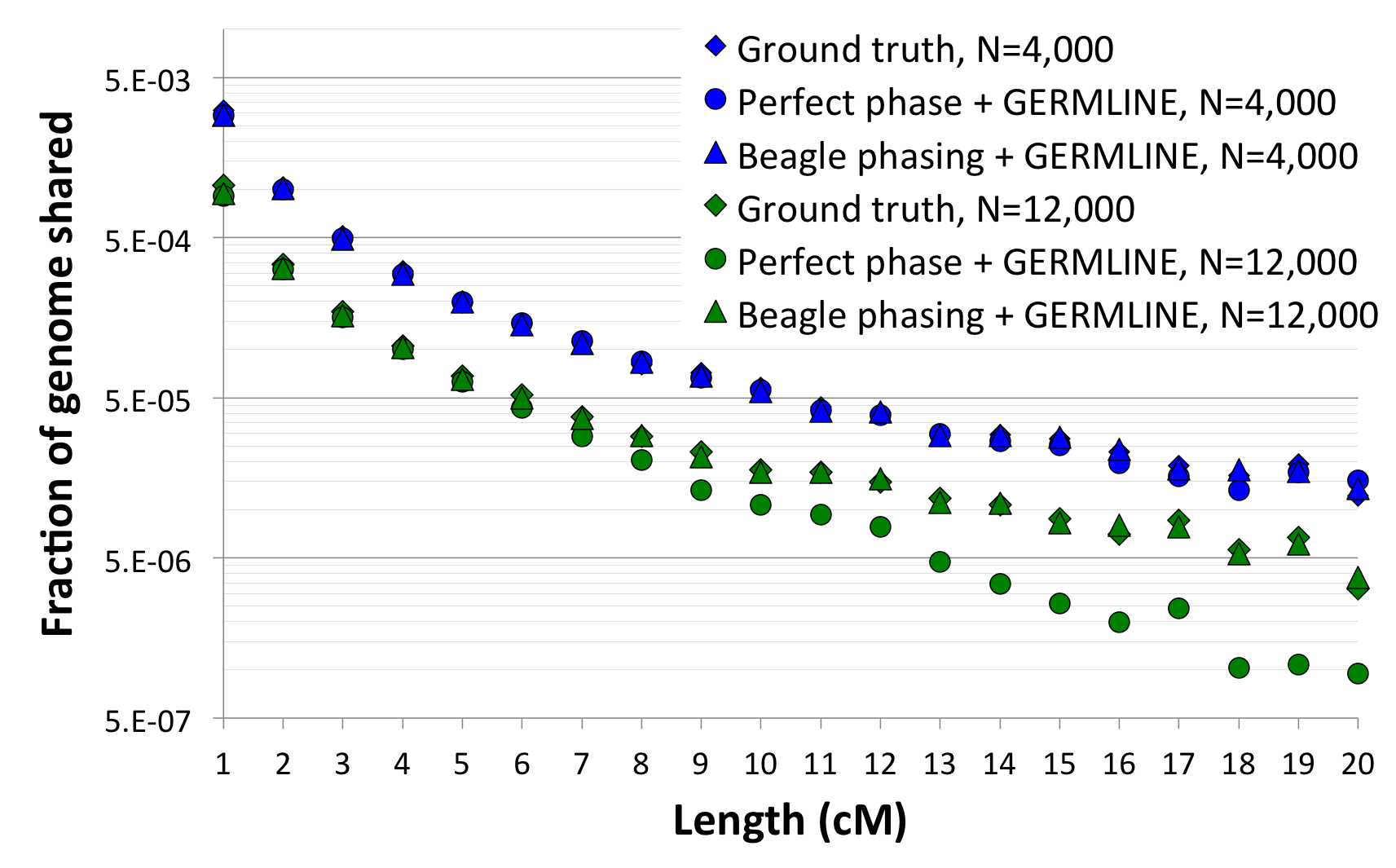}
\caption{We simulated a chromosome of 150 cM for 600 individuals using the model in Figure \ref{fig:02a}, setting population sizes to $4,000$ and $12,000$ diploid individuals, with a migration rate of $0.04$. IBD sharing was extracted directly from the simulated genealogy (diamonds), or inferred trough GERMLINE using perfectly phased (circles) or computationally phased (triangles) chromosomes.}
\label{fig:06}
\end{figure}

This analysis suggests that while the methods that rely on ancestry deconvolution are a useful tool for the specific case of recently admixed groups arising from strongly diverged populations, they may not be suitable for the analysis of fine-scale migration events, such as those that occurred across populations that split few tens to hundreds of generations in the past. It is however possible that adjusting some of the parameters used for the GENOME simulations and for the PCAdmix software, or using other deconvolution methods, the obtained accuracy may be increased. Furthermore, the development of methods for ancestry deconvolution in sequence data, where rare variants are observable, is expected to substantially increase the power of this analysis, although the effects of limited population divergence are likely to still affect the accuracy of methods that do not explicitly take this aspect into account. An additional difference to be noted between the two considered approaches is that Tracts does not model population size changes in the populations, focusing on relative migration rates, while DoRIS allows recovering both population size fluctuations and migration rates, thus providing insights into the magnitude of migration events. This increased flexibility, however, may complicate the inference, also in light of our observation that large sample sizes are required for the IBD analysis.

\section{IBD sharing outlines regional-scale demographic history: the Genome of the Netherlands}

\begin{figure}[!ht]
	\vspace{-2mm}
\centering
\includegraphics[width=\textwidth]{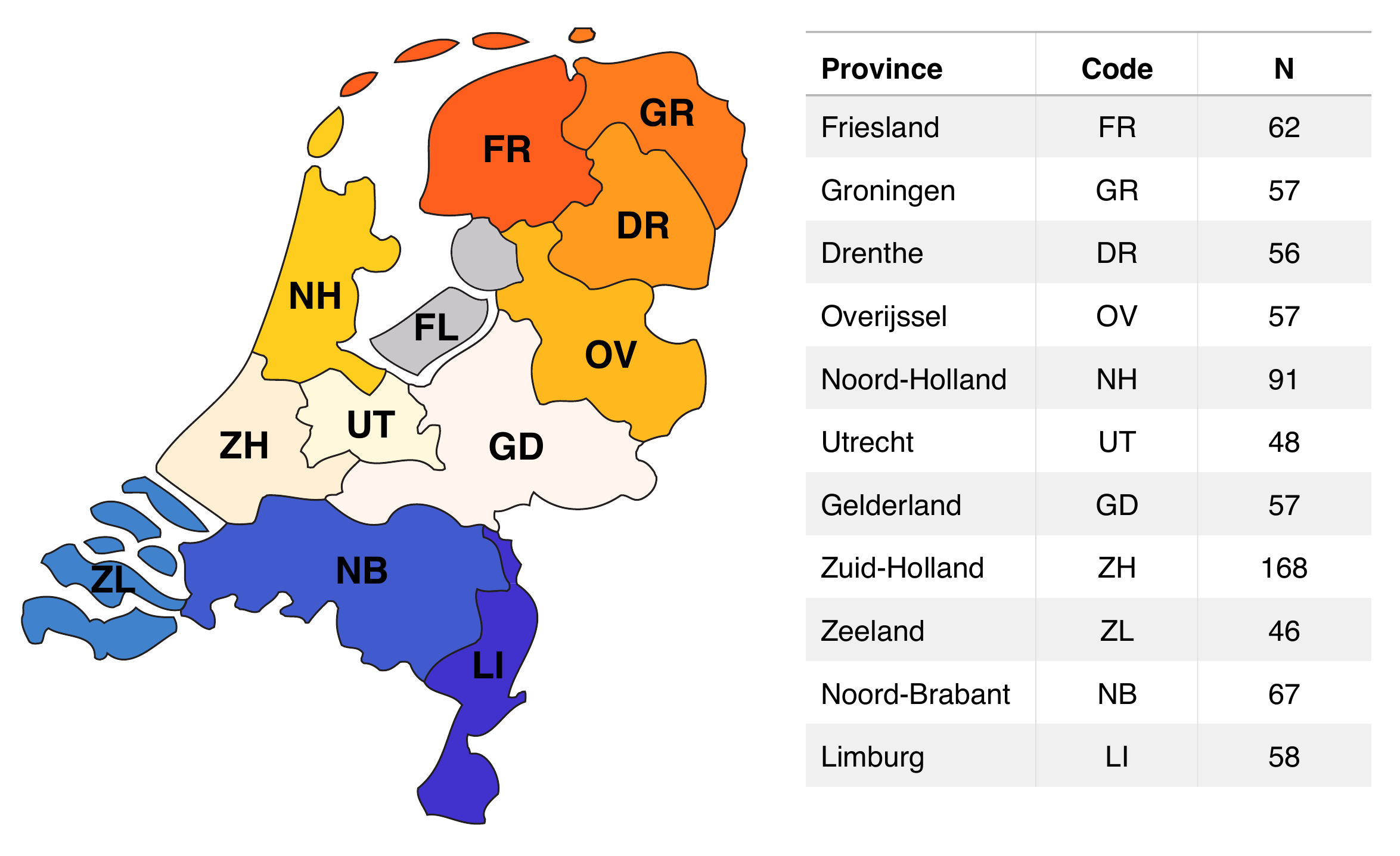}
\caption{A map of the analyzed provinces and the number of collected samples.}
\label{fig:mig:GoNLmap}
\end{figure}

The Genome of the Netherlands (GoNL) Project was established with the goal of characterizing genomic variation in the Dutch population \cite{boomsma2013genome}. To this extent, $250$ trio-families from $11$ provinces of the Netherlands were sequenced with an average coverage of $14$-$15$x. Sequencing data provides a very large number of informative (high frequency) variants, which added to the trio design of the GoNL project results in reliable inference of haplotype phase, therefore enabling accurate detection of IBD sharing. Furthermore, fine-grained information about genomic variation across $11$ provinces from a single country provides the opportunity to demonstrate the methods developed in this and the previous chapter for the inference of fine-scale demographic history, using the DoRIS software package. We here present preliminary results from ongoing analysis of these samples.

We initially removed all variants with a minimum allele frequency below $1\%$, and also excluded from downstream analysis all markers with trio-phasing posterior probability \cite{menelaou2013genotype} below $1.0$, obtaining $3,525,142$ SNPs. We used the genetic map provided for the Phase I integrated variant set release (v3) of the $1,000$ Genomes Project\footnote{\url{http://mathgen.stats.ox.ac.uk/impute/data_download_1000G_phase1 _integrated.html}} (build $37$, hg19 coordinates). For all markers that were not found in the genetic map, we inferred the genetic distance by linear interpolation assuming uniform recombination rate between the two closest markers found upstream and downstream in the available map. We run GERMLINE using the parameters \emph{-min\_m 1 -err\_hom 2 -err\_het 0 -bits 75 -haploid}, i.e. requiring a minimum IBD segment length of $1$ cM, allowing at most $2$ mismatches in windows of $75$ markers, for perfectly phased haploid chromosomes. We excluded regions with unusual density of IBD sharing, which may be caused by false positive/negative segments due to low density of markers, deviations from neutrality or presence of common structural variations that affect recombination. To this extent, we restricted our analysis to regions with IBD density within $5$ standard deviations from the mean genome-wide sharing. We further required the analyzed regions to be at least $45$ cM long, obtaining a total of $26$ regions spanning $2,160.26$ cM (Table \ref{tbl:mig:GoNLregions}), IBD sharing density per site per pair $3.07 \times 10^{-3}$, std $1.35 \times 10^{-3}$).

\begin{table*}
\centering
\footnotesize
\begin{tabular}{|c|c|c|c|c|}
	 \hline
	From Chromosome & From (genetic) & To (genetic) & From (physical) & To (physical) \\ \hline
	1 & 97.5 & 150.5 & 66,874,699 & 118,837,888 \\ \hline
	2 & 36.5 & 115 & 17,246,473 & 85,384,179 \\ \hline
	2 & 209 & 257.5 & 193,010,478 & 235,351,139 \\ \hline
	3 & 1 & 190 & 678347 & 176030190 \\ \hline
	4 & 101 & 217.5 & 85315581 & 189,657,996 \\ \hline
	5 & 38 & 148 & 22,657,926 & 141,420,437 \\ \hline
	6 & 54 & 111 & 33,954,192 & 103,983,460 \\ \hline
	6 & 150.5 & 197.3 & 139,903,959 & 170,245,872 \\ \hline
	7 & 1 & 63 & 962,247 & 38,722,532 \\ \hline
	7 & 66.5 & 172 & 41,688,961 & 152,254,508 \\ \hline
	8 & 78 & 166 & 55,170,178 & 139,553,601 \\ \hline
	9 & 83 & 158 & 72,512,292 & 132,515,730 \\ \hline
	10 & 44 & 181.5 & 19,570,732 & 134,866,854 \\ \hline
	11 & 5 & 160.9 & 2,047,054 & 134,587,122 \\ \hline
	12 & 16.5 & 90 & 6,476,123 & 75,656,510 \\ \hline
	12 & 98.5 & 158.5 & 82,586,486 & 128,401,829 \\ \hline
	13 & 1.4 & 128.8 & 20,518,406 & 114,094,544 \\ \hline
	14 & 1.1 & 54 & 20,545,390 & 59,184,876 \\ \hline
	14 & 58 & 113.5 & 63,846,103 & 104,808,535 \\ \hline
	15 & 70 & 149.9 & 50,284,344 & 101,969,749 \\ \hline
	17 & 1.1 & 84.5 & 163,278 & 55,936,970 \\ \hline
	18 & 37.5 & 84 & 11,962,813 & 59,189,703 \\ \hline
	19 & 26 & 105.9 & 7,857,579 & 158,513,172 \\ \hline
	20 & 19 & 83 & 5,649,902 & 52,818,462 \\ \hline
	21 & 1.9 & 63.7 & 15,636,220 & 47,031,048 \\ \hline
	22 & 21 & 74.1 & 23,874,416 & 50,493,062 \\ \hline
\end{tabular}
\caption{Regions that passed quality control and were analyzed in the GoNL dataset.}
\label{tbl:mig:GoNLregions}
\end{table*}

\begin{figure}[!ht]
	\vspace{-2mm}
\centering
\includegraphics[width=\textwidth]{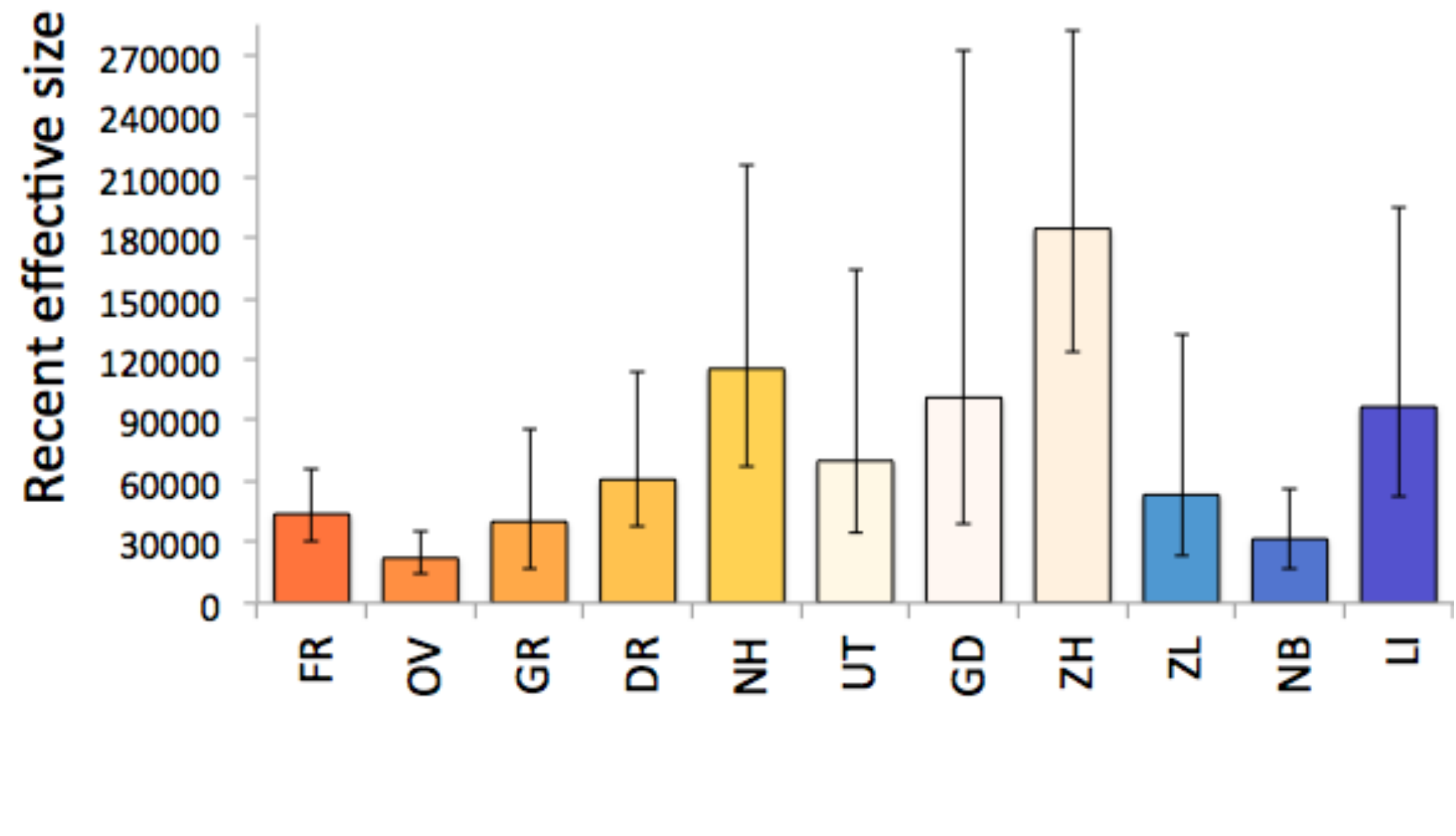}
\caption{Reconstructed recent population sizes.}
\label{fig:mig:GoNLrecent}
\end{figure}

\begin{figure}[!ht]
	\vspace{-2mm}
\centering
\includegraphics[width=\textwidth]{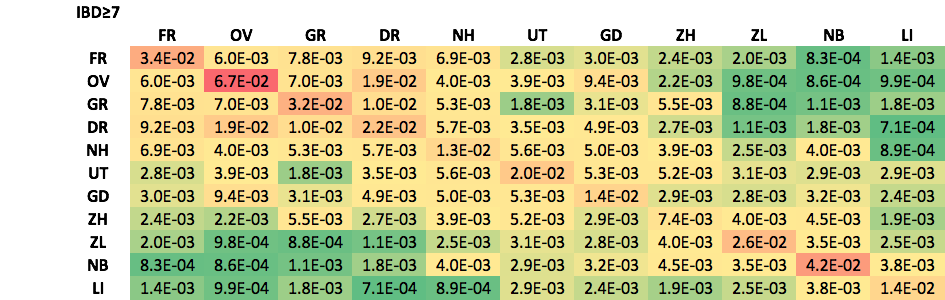}
\caption{Sharing of segments of at least $7$ cM within and across provinces.}
\label{fig:mig:GoNL7}
\end{figure}

When we analyzed long IBD segments (at least $7$ cM), recently co-inherited within each province, differences in the inferred effective population sizes (obtained using the methods of Chapter \ref{chap:IBDmodel}) were substantial (Figure \ref{fig:mig:GoNLrecent}), reflecting recent differentiation as a result of heterogeneous growth and migration events that occurred in the past $20$-$25$ generations (expected time to recent common ancestor based on segment length in a model of exponential expansion: ${\sim}500$ years ago). In particular, Zuid-Holland exhibits an effective population size of $> 100,000$ individuals (eight times that of Overijssel), suggestive of recent expansion and possibly gene flow from other provinces. The sharing of such long IBD segments (also across provinces) supports localized recent common ancestry, with all provinces sharing, on average, the largest number of long IBD segments with other individuals from the same geographic region (Figure \ref{fig:mig:GoNL7}).


\begin{figure}[!ht]
\vspace{-2mm}
\centering
\includegraphics[width=\textwidth]{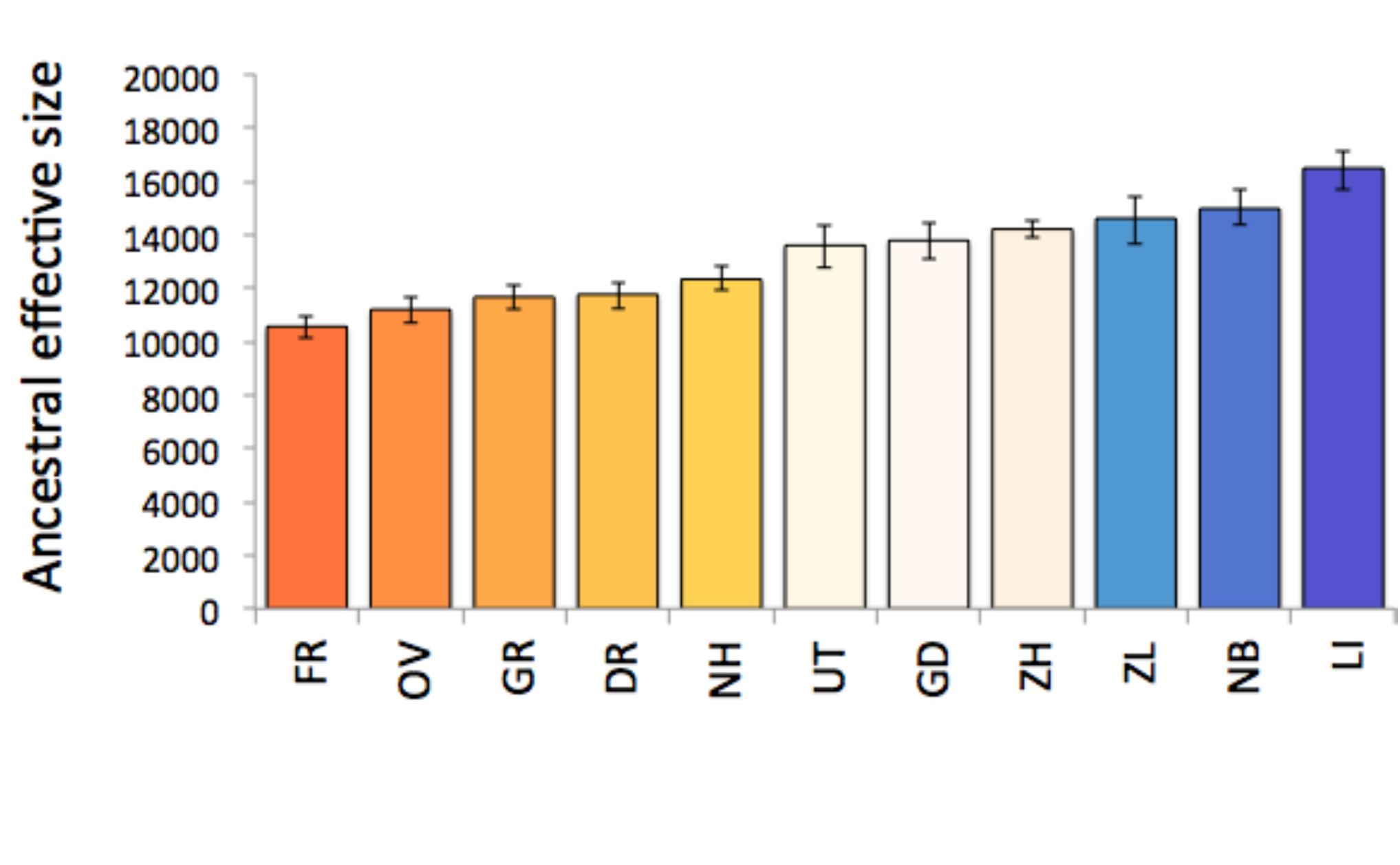}
\caption{Reconstructed ancestral population sizes.}
\label{fig:mig:GoNLAncestral}
\end{figure}

We then considered the fraction of genome shared within each province through short IBD segments ($1$ to $2$ cM), and thus inferred the ancestral effective population size per province (Figure \ref{fig:mig:GoNLAncestral}). Although these effective population sizes are rather homogeneous across the $11$ provinces, consistent with common genetic origins, we observed a South to North gradient of decreasing ancestral population size accompanied by increased homozygosity in the northern provinces (correlation between province latitude and IBD sharing $r=0.923$, $p=5.12\times10^{-5}$). Such gradient has been previously described for average inbreeding coefficients and similar metrics of genome-wide similarity across Dutch individuals \cite{lao2013clinal,abdellaoui2013population}, and interpreted as the signature of remote northwards migration during early waves of European colonization, although more complex scenarios involving recent demographic events could not be ruled out. Indeed, a model of serial founder migrations from the South to the North of the country may produce the observed pattern of increasing homozygosity towards the North, as the results of shrinking effective population sizes of smaller group that migrate away from larger populations.

\begin{figure}[!ht]
\vspace{-2mm}
\centering
\includegraphics[width=\textwidth]{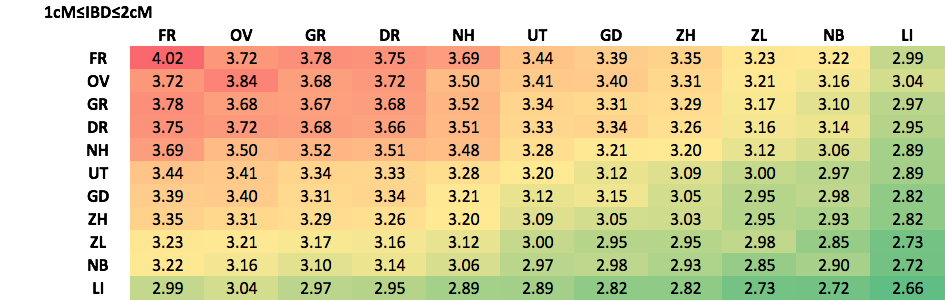}
\caption{Sharing of segments between $1$ and $2$ cM within and across provinces.}
\label{fig:mig:GoNL1-2}
\end{figure}

\begin{figure*}[!ht]
	\vspace{-2mm}
    \centering
	\vspace{-2mm}
    \centering
    \includegraphics[width=\textwidth]{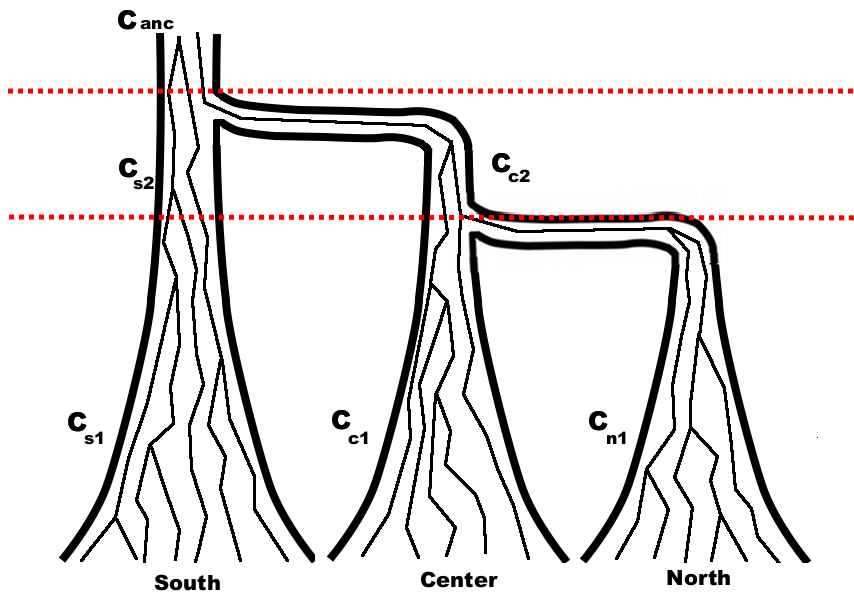}
    \caption{A model of serial migrations from the south to the north. Red lines separate periods in which lineages are allowed to coalesce. Cumulative coalescence probabilities for each region and each period are shown (e.g. $C_{anc}$ indicates coalescence probability in the ancestral period, and $C_{c2}$ indicates the probability of coalescence in the central population during the second period.)}
    \label{fig:serialSN}
\end{figure*}

In addition to the observed gradient of increasing haplotypic homozygosity within provinces, however, we observed that all GoNL samples, regardless of modern-day geographic location, share on average more IBD segments with other individuals from the North of the country than with individuals from the same province (Figure \ref{fig:mig:GoNL1-2}, off-diagonal elements, correlation between average IBD sharing and average latitude of provinces $r=0.934$, $p<10^{-5}$). This counterintuitive observation was confirmed when we grouped the $11$ provinces into three clusters, North, Center and South, based on hierarchical clustering \cite{ward1963hierarchical} of the cross-province IBD matrix for segments of length $1$ cM or more. The observation of higher sharing across provinces compared to within provinces cannot be reproduced in a simple model of serial migrations from the South to the North of the country. Consider the model shown in Figure \ref{fig:serialSN}. It easy to note that the expected number of shared common ancestors for two individuals sampled in the South is strictly more than the number of ancestors expected between an individual from the North and one from the South in this model (see Table \ref{tbl:IBDdata:GonlSerShar}).

\begin{table*}	
	\centering
	\begin{tabular}{c|c|c|c}
		& South & Center & North \\ \hline
	South & $C_{s1}+C_{s2}+C_{anc}$ & $C_{anc}$ & $C_{anc}$ \\ \hline
	Center & $C_{anc}$ & $C_{c1}+C_{c2}+C_{anc}$ & $C_{c2}+C_{anc}$ \\ \hline
	North & $C_{anc}$ & $C_{c2}+C_{anc}$ & $C_{n1}+C_{c2}+C_{anc}$ \\
	\end{tabular}
	\caption{Total coalescence rates within/across regions in the serial migration model of Figure \ref{fig:serialSN}.}
	\label{tbl:IBDdata:GonlSerShar}
\end{table*}

\begin{figure*}
	\vspace{-2mm}
    \centering
		\vspace{-2mm}
            \centering
            \includegraphics[width=\textwidth]{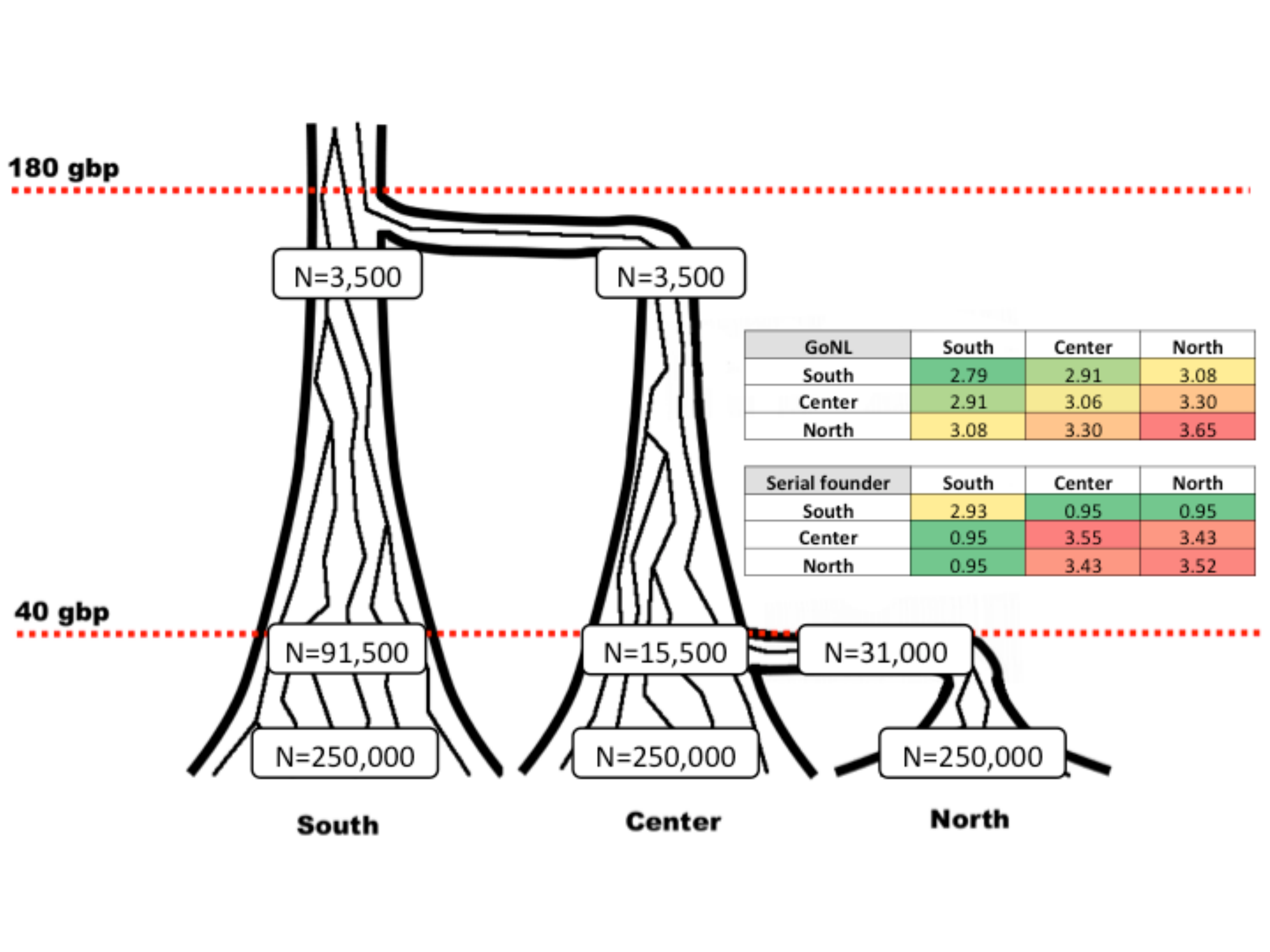}
            \caption{Simple model of serial migrations and comparison of IBD patterns in GoNL and expected in this model.}
            \label{fig:GoNLSerWithMigA}
\end{figure*}

\begin{figure*}
		\vspace{-2mm}
            \centering
            \includegraphics[width=\textwidth]{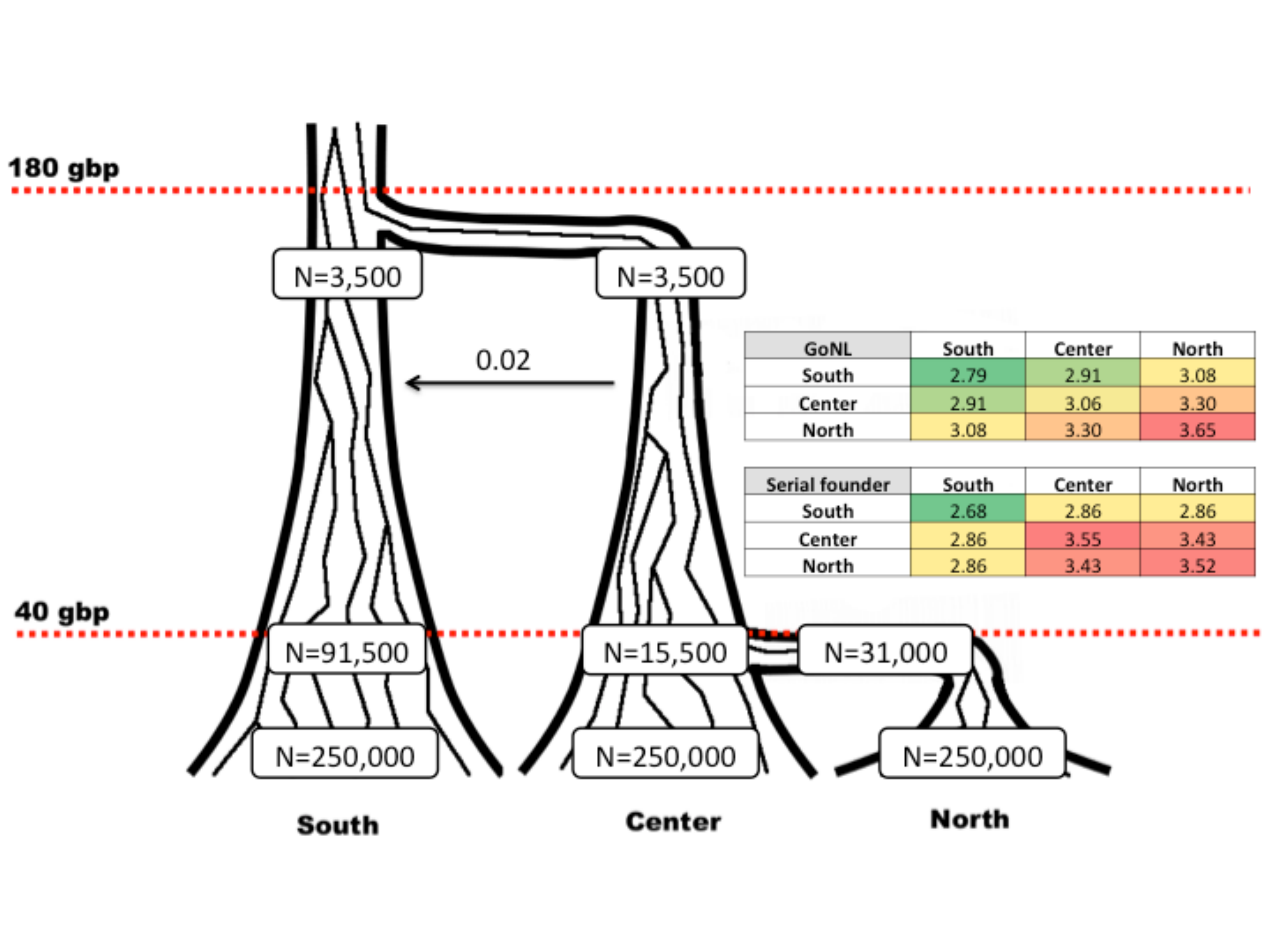}
            \caption{Model of serial migrations that includes remote migrations to the South, and comparison of IBD patterns in GoNL and expected in this model.}
            \label{fig:GoNLSerWithMigB}
\end{figure*}

\begin{figure*}
            \centering
            \includegraphics[width=\textwidth]{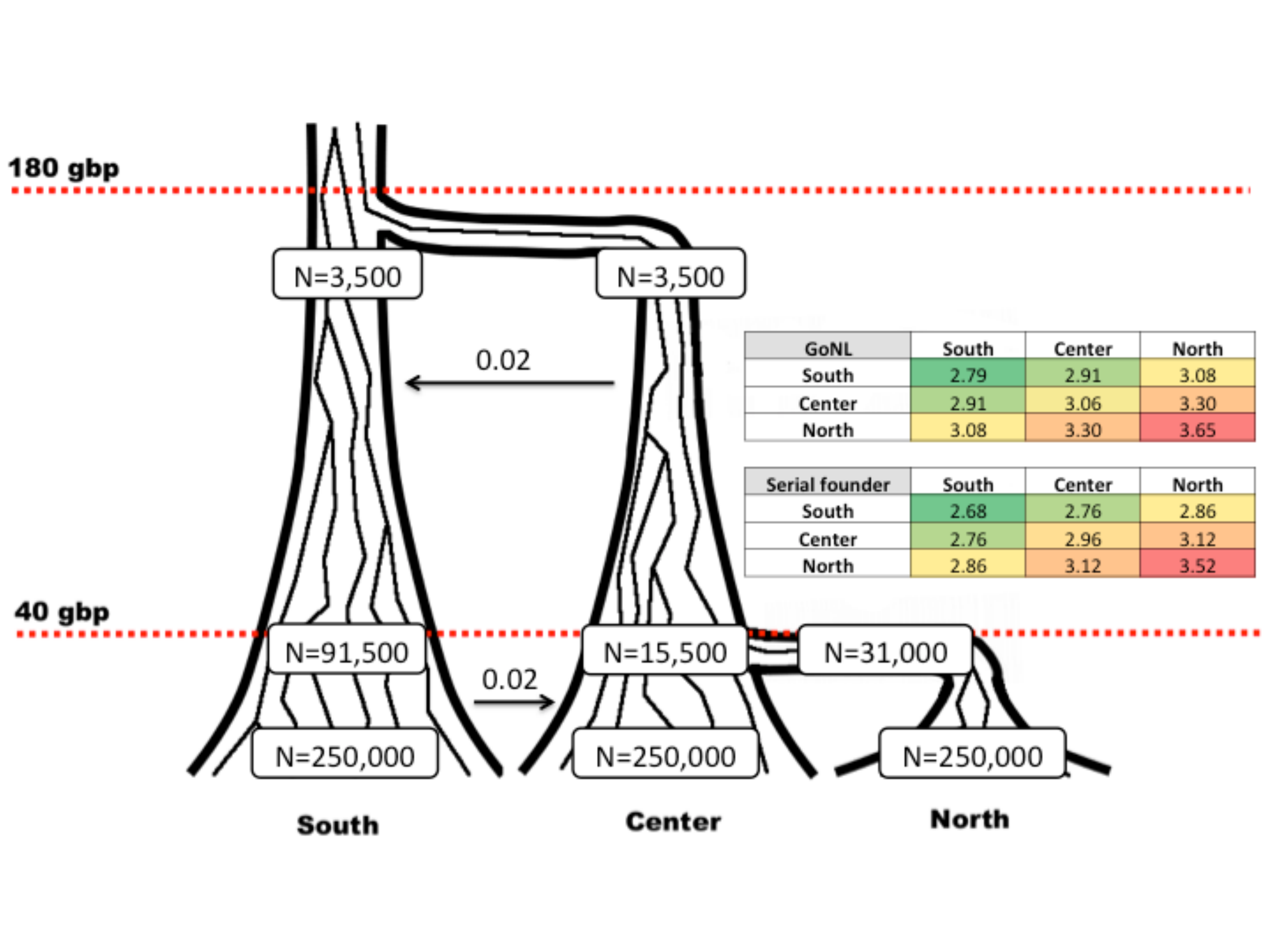}
            \caption{Model of serial migrations that includes remote migrations to the South and subsequent migrations to the Center. Comparison of IBD patterns in GoNL and expected in this model.}
            \label{fig:GoNLSerWithMigC}
\end{figure*}

However, we note that the model of simple northwards serial migrations shown in Figure \ref{fig:serialSN} may be enriched to explain the pattern observed in the GoNL data. Figure \ref{fig:GoNLSerWithMigA} shows the same simple model of subsequent population subdivisions. As predicted in Table \ref{tbl:IBDdata:GonlSerShar}, the IBD pattern computed for such model is not compatible with what is observed in the GoNL data.
However, note that increased sharing from the South to the Center and the North may be achieved by including migration from the Center to the South (forward in time) in the period preceding the formation of the Northern group (Figure \ref{fig:GoNLSerWithMigB}).
Finally, the increased sharing between Southern and Northern individuals may be obtained if individuals are allowed to migrate from the South back to the Center in the period following the creation of the Northern group, as shown in Figure \ref{fig:GoNLSerWithMigC}.

We note that while the inclusion of these two migration rates in the model recapitulates the observed pattern of IBD sharing, comprehensive demographic analysis should include further testing of several models, and formal comparison across them. Further note that in this model a migration rate of $0.02$ per individual, per generation, implies the ancestry of an individual has a relatively high $1/50$ chance of moving between regions at each generation. During the course of $140$ generations, this results in $1-(1-0.02)^{140}=0.94$ probability of migration for an ancestral lineage. For the more recent period, the chance of moving to the Center in this model is $1-(1-0.02)^{40} = 0.55$. This model therefore implies substantial ancestral contribution from the genetic pool currently represented in the north of the country, rather than a unidirectional colonization from the South.

While a conclusive analysis would imply substantial additional hypothesis testing, possibly involving the inclusion of non-Dutch samples from neighboring regions, the compatibility of the presented model is in line with the complex fine-grained migratory history expected in the Netherlands. The Dutch territory was in fact shaped by frequent sea level changes and flooding events, which caused many parts of the coastal regions to vanish and reappear, likely resulting in several pulses of colonization and admixture across demes. A notable event is the occurrence of St. Lucia's flood in $1,287$ C.E., which separated the provinces of Noord Holland and Friesland. Interestingly, these two provinces exhibit high sharing for segments longer than $5$cM, suggesting remote genetic links, which may go back to migration across these geographic regions before the occurrence of flooding.

Finally, we outline that the development of algorithms that detect short IBD segments in potentially heterogeneous groups is a subject of current research, and improvements in this direction will facilitate studying cross-population IBD sharing and strengthen the conclusions of these analyses. While the accuracy of currently available IBD detection methods in cross-population analysis is not fully understood (e.g. see Figure \ref{fig:06}), the observed enrichment for IBD sharing with Northern provinces of the GoNL dataset was also observed in independent analysis performed using Beagle's FastIBD method for detecting haplotype sharing (correlation of within-province sharing and  latitude: $r=0.784$, $p=4.27\times10^{-3}$, correlation between cross-province sharing and average latitude $r=0.882$, $p<10^{-5}$).

\section{Appendix}
\label{Appendix}
\subsection{Estimators for different $N_{e1}$ and $N_{e1}$}

When the population sizes of $N_{e1}$ and $N_{e2}$ are allowed to vary, the derivation of Section \ref{IBDMig} leads to the following closed form estimators

\begin{equation}
\begin{split}
\hat{N}_{e1}= \{9 \hat{f}_{2}^3 + 31 \hat{f}_{1}^3 + 128 \hat{f}_{d}^3 + 4 \hat{f}_{1} \hat{f}_{d} (k-18 \hat{f}_{d}) + \\
- 3 \hat{f}_{1}^2 (18 \hat{f}_{d} + k) + \hat{f}_{2}^2 (49 \hat{f}_{1} - 10 \hat{f}_{d} + 3 k) + \\
+ \hat{f}_{2} [71 \hat{f}_{1}^2 - 64 \hat{f}_{1} \hat{f}_{d} - 4 \hat{f}_{d} (22 \hat{f}_{d} + k)]\}\times\\
\times \frac{1}{2u}[\hat{f}_{1} (\hat{f}_{2} + \hat{f}_{1})^2 (9 \hat{f}_{2} + 11 \hat{f}_{1}) + 8 \hat{f}_{2} \hat{f}_{1} (\hat{f}_{2} + \hat{f}_{1}) \hat{f}_{d} + \\
- 4 (4 \hat{f}_{2}^2 + 19 \hat{f}_{2} \hat{f}_{1} + 13 \hat{f}_{1}^2) \hat{f}_{d}^2 - 16 \hat{f}_{2} \hat{f}_{d}^3 + 64 \hat{f}_{d}^4]^{-1}
\end{split}
\label{eq:23}
\end{equation}

\begin{equation}
\begin{split}
\hat{N}_{e2}=\{31 \hat{f}_{2}^3 + 9 \hat{f}_{1}^3 + 128 \hat{f}_{d}^3 - 4 \hat{f}_{1} \hat{f}_{d} (22 \hat{f}_{d} + k) + \\
+ \hat{f}_{1}^2 (3k-10 \hat{f}_{d}) + \hat{f}_{2} [49 \hat{f}_{1}^2 - 64 \hat{f}_{1} \hat{f}_{d} + \\
+ 4 \hat{f}_{d} (k-18 \hat{f}_{d})] + \hat{f}_{2}^2 [71 \hat{f}_{1} - 3 (18 \hat{f}_{d} + k)]\} \times \\
\times \frac{1}{2u}[\hat{f}_{2} (\hat{f}_{2} + \hat{f}_{1})^2 (11 \hat{f}_{2} + 9 \hat{f}_{1}) + 8 \hat{f}_{2} \hat{f}_{1} (\hat{f}_{2} + \hat{f}_{1}) \hat{f}_{d} + \\
- 4 (13 \hat{f}_{2}^2 + 19 \hat{f}_{2} \hat{f}_{1} + 4 \hat{f}_{1}^2) \hat{f}_{d}^2 - 16 \hat{f}_{1} \hat{f}_{d}^3 + 64 \hat{f}_{d}^4]^{-1}
\end{split}
\label{eq:24}
\end{equation}

\begin{equation}
\begin{split}
\hat{m}= \frac{u(k-3 \hat{f}_{2} - 3 \hat{f}_{1} + 10 \hat{f}_{d})}{8 (\hat{f}_{2} + \hat{f}_{1} - 2 \hat{f}_{d})}
\end{split}
\label{eq:25}
\end{equation}

where $\hat{f}_{1}, \hat{f}_{2}, \hat{f}_{d}$ are observed within and across populations, and

\begin{equation}
\begin{split}
k=\sqrt{[9 (\hat{f}_{1} + \hat{f}_{2}) - 14 \hat{f}_{d}] (\hat{f}_{1} + \hat{f}_{2} + 2 \hat{f}_{d})}
\end{split}
\label{eq:26}
\end{equation}

\subsection{Probability of coalescence in a two-population model with migration}

The probability that the two ancestral lineages are found in the same population can be obtained from the terms specified in equations \ref{eq:mig:05} and \ref{eq:06}. Specifically, the chance of the two ancestral lineages being in the same population is given by

\begin{equation}
\begin{split}
p(P_i(t)=P_j(t)) = \mathbf{v}_i(t)\mathbf{v}_j(t)^T \approx \frac{1+e^{-4mt}}{2}
\end{split}
\end{equation}

If both individuals are sampled from the same population, or

\begin{equation}
\begin{split}
p(P_i(t)=P_j(t)) \approx \frac{1-e^{-4mt}}{2}
\end{split}
\end{equation}

Otherwise. We will focus on the first case, as the derivation in the latter case is analogous. In order to obtain the full coalescent distribution for the ancestral lineages of the considered two individuals sampled from one of the extant populations, we need to consider the chance that their ancestral lineages coalesce while being in the same populations. To do this, we consider the expected time that these lineages spend in the same population after $t$ generation (expressed in continuous time). Such quantity can be computed as

\begin{equation}
\begin{split}
E[f|T,m,N_e] &= \frac{\int_{0}^{T} p(P_i(t)=P_j(t)) dt}{T} \\
&= \frac{\int_{0}^{T} 1-e^{-4mt} dt}{2T} \\
&= \frac{1-e^{-4mT}+4mT}{8mT}
\end{split}
\end{equation}

A constant population of effective size $N_e$ has coalescent distribution $p(t|N_e) = \frac{1}{N}e^{-t/N_e}$. Since the two populations have the same effective population size, $N_e$, we can obtain the coalescent distribution in the case of two population model by scaling the time in the distribution of single population model by the expected fraction of time spent in the same deme, which was computed above:

\begin{equation}
\begin{split}
p(T|m,N_e) &= \frac{1}{N}e^{-E[f|T,m,N_e]T/N_e} \\
&= \frac{1}{N}e^{-\frac{1-e^{-4mT}+4mT}{8mN_e}}
\end{split}
\end{equation}

With cumulative distribution

\begin{equation}
\begin{split}
P(T|m,N_e) = 1-e^{-\frac{1-e^{-4mT}+4mT}{8mN_e}}
\end{split}
\end{equation}

Note, however, that we are not interested in accurately describing this distribution for large values of $T$. In fact, we later introduce the factor $p(l|t)$, which quickly goes to $0$ for values of $u$ that are large enough to be practically considered, e.g. above 0.5 cM. For the purpose of this section, given reasonably large $u$ and $N_e$ (e.g. above $1,000$ effective individuals), the coalescent distribution can be approximated using a Taylor expansion of the kind $e^x \approx 1+x$. Applying this approximation to the cumulative distribution we just derived, we get

\begin{equation}
\begin{split}
P(T|m,N_e) &= 1-e^{-\frac{1-e^{-4mT}+4mT}{8mN_e}} \\
&\approx 1-1+\frac{1-e^{-4mT}+4mT}{8mN_e} \\
&= \frac{1-e^{-4mT}+4mT}{8mN_e}
\end{split}
\end{equation}

Whose derivative gives the approximate pdf

\begin{equation}
\begin{split}
p(t|m,N_e) \approx \frac{1+e^{-4mt}}{2N_e}
\end{split}
\end{equation}

\chapter{Leveraging whole sequence information: IBD sharing and mutation}
\label{chap:mutation}
In chapters \ref{chap:IBDmodel} and \ref{chap:migration}, we have introduced a coalescent-based framework that allows computing several quantities related to haplotype sharing in purportedly unrelated samples as a function of past demographic events. The presented framework was derived using coalescent theory to model the occurrence of recombination events at the boundary of IBD segments. We have thus far entirely neglected the occurrence of mutation in the described ancestral processes, but as we shall describe in this chapter, the joint consideration of IBD sharing and mutations may support additional analyses. As mentioned in the Introduction, ``identical-by-descent'' segments may not be strictly identical. They may in fact harbor mutations that occurred along ancestral lineages connecting a pair of individuals to their common ancestor. We here describe the distribution of these mutations on IBD segments, and discuss applications that make use of this information. We limit derivations to the case of Wright-Fisher populations. The extension to arbitrary coalescent distributions is similar to previous chapters.

\section{The number of mutations on an IBD segment}

Suppose a common ancestor that lived $g$ generations in the past transmits an IBD segment of genetic length $l$ to a pair of modern day individuals. We assume fixed recombination and mutation rates per nucleotide are provided, so that multiplying genetic length into a constant $r$ returns the number of nucleotides $n$ the segment spans (i.e. $r=\rho^{-1}$). The probability that the IBD segment has length $l$ Morgans and harbors $k$ mutated sites is

\begin{equation}
\begin{split}
p(k,l|g,r,\mu) = p(k|l,g,r,\mu)\times p(l|g),
\end{split}
\label{eq:mut:01}
\end{equation}
where $\mu$ is the mutation rate per generation, per nucleotide, per individual. As previously described, the distribution of $p(l|g)$ is exponential, with mean $1/(2g)$. Under the infinite sites assumption \cite{kimura1969number}, the probability that a single site mutates during $2g$ transmission events may be modeled as $1-(1-\mu)^{2g}$ or, for continuous $g$, $1-e^{-2g\mu}$. Since the value of $\mu$ is in the order of $10^{-8}$ \cite{roach2010analysis}, the product $2g\mu$ is also expected to be small, and we can use the approximation $1-e^{-x}\approx x$ for $x\longrightarrow 0$, to rewrite this probability as $2g\mu$. Because each site is modeled as independently mutating, the total number of mutated sites is binomial, with probability $2g\mu$ and $n=lr$ independent attempts. In addition, because $n$ is large and $2g\mu$ is small, we can model the distribution for the number of mutations on the segment as Poisson with mean $2lr g\mu$. It is interesting to note that the number of mutations that is expected to be found on an IBD segment does not depend on how long ago the common ancestor that transmitted the segment has lived. Because the mutation process is independent from the recombination process, the expected number of mutations $\operatorname{E}[k|g]$ for an IBD segment transmitted from an ancestor $g$ generations ago is obtained by taking the number of mutations that is expected for a segment of length $\operatorname{E}[l|g]=\frac{1}{2g}$. Namely

\begin{equation}
\begin{split}
	\operatorname{E}[k|g]=2g\mu r\times\frac{1}{2g}=\mu r,
\end{split}
\label{eq:mut:02}
\end{equation}
where the generation cancels out. This occurs because segments transmitted from a common ancestor decrease in length at the same rate as mutations accumulate. Substituting the Poisson and the exponential distributions into Equation \ref{eq:mut:01}, we obtain

\begin{equation}
\begin{split}
p(k,l|g,r,\mu) = \frac{(2lr g\mu)^k e^{-2lr g\mu}}{k!} \times ({2g}e^{-2gl})
\end{split}
\label{eq:mut:03}
\end{equation}

If the IBD segment is allowed to be of any length, we can integrate $l$ out, to obtain the distribution for the number of segments found on a segment from generation $g$

\begin{equation}
\begin{split}
p(k|g,r,\mu) &= \int_{0}^{\infty} p(k,l|g,r,\mu) dl \\
&= \int_{0}^{\infty} \frac{(2lr g\mu)^k e^{-2lr g\mu}}{k!} \times ({2g}e^{-2gl}) dl \\
&= (\mu r)^k (1+\mu r)^{-(k+1)} ~,
\end{split}
\label{eq:mut:04}
\end{equation}
which has expectation and variance given by

\begin{equation}
\begin{split}
	\operatorname{E}[k|g,r,\mu] &= \sum_{k=0}^{\infty} k (\mu r)^k (1+\mu r)^{-(k+1)} = \mu r \\
	\operatorname{Var}[k|g,r,\mu] &= \sum_{k=0}^{\infty} (k-\mu r)^2 (\mu r)^k (1+\mu r)^{-(k+1)} = \mu r (1 + \mu r)
\end{split}
\label{eq:mut:05}
\end{equation}

Note that this is a Negative Binomial distribution

\begin{equation}
\begin{split}
\operatorname{NB}(k,r,\theta) &= \frac{\Gamma(r+k)}{k!\;\Gamma(r)} \; \theta^k (1-\theta)^r ~,
\end{split}
\label{eq:mut:06}
\end{equation}
where $r=1$ and $\theta = (\mu r)/(1+\mu r)$ (or, equivalently, a Geometric distribution with $p=1-(\mu r)/(1+\mu r)$, since $r=1$). This distribution occurs as a result of the Gamma-Poisson mixture described in Equation \ref{eq:mut:05}, where the rate of the Poisson distribution is a random variable itself. In particular, it is an exponential random variable, i.e. it is Gamma distributed with shape parameter $k=1$ and scale parameter $1/2g$. Again, note that the distribution for the number of mutations is independent from when the transmitting common ancestor has lived, therefore this result extends to arbitrary demographic histories, as the coalescent distribution becomes irrelevant.

One can now ask what is the distribution for the number of mutations when the IBD segment can only be longer than a detectable length threshold $u$. Recall that an exponentially distributed random variable $L$ enjoys the \emph{memorylessness} property, i.e. $p(L > s + t\; |\; L > s) = p(L > t) \text{ for all } s, t \ge 0. $ This implies that if a segment has length distributed as a truncated exponential random variable, we can express the distribution of its length as a constant segment of length $u$, plus a remaining part that is itself exponentially distributed with parameter $1/(2g)$, i.e. $L = u + L_\tau$, where the tip of the segment, $L_\tau$, has exponential distribution $p(l_\tau|g)=2ge^{-2gl}$.

We have already computed the distribution for the number of mutations on $L_\tau$ in Equation \ref{eq:mut:04}, and noted it does not depend on $g$. The distribution of the number of mutations on the fixed part of length $u$, however, does depend on $g$, and can be modeled as a Poisson with mean $\lambda = 2g\mu r u$, as previously motivated. The distribution for the number of mutations on the entire segment $L$, therefore, can be expressed as the discrete convolution of this Poisson distribution and the Negative Binomial (or Geometric) distribution of Equation \ref{eq:mut:04}, with expectation and variance given by

\begin{equation}
\begin{split}
	\operatorname{E}[k|g,r,\mu,u] &= \mu r (2gu + 1) \\
	\operatorname{Var}[k|g,r,\mu,u] &= \mu r (1 + \mu r + 2 g u)
\end{split}
\label{eq:mut:07}
\end{equation}

\begin{figure*}
	\vspace{-2mm}
    \centering
    \begin{subfigure}[b]{0.7\textwidth}
		\vspace{-2mm}
            \centering
            \includegraphics[width=\textwidth]{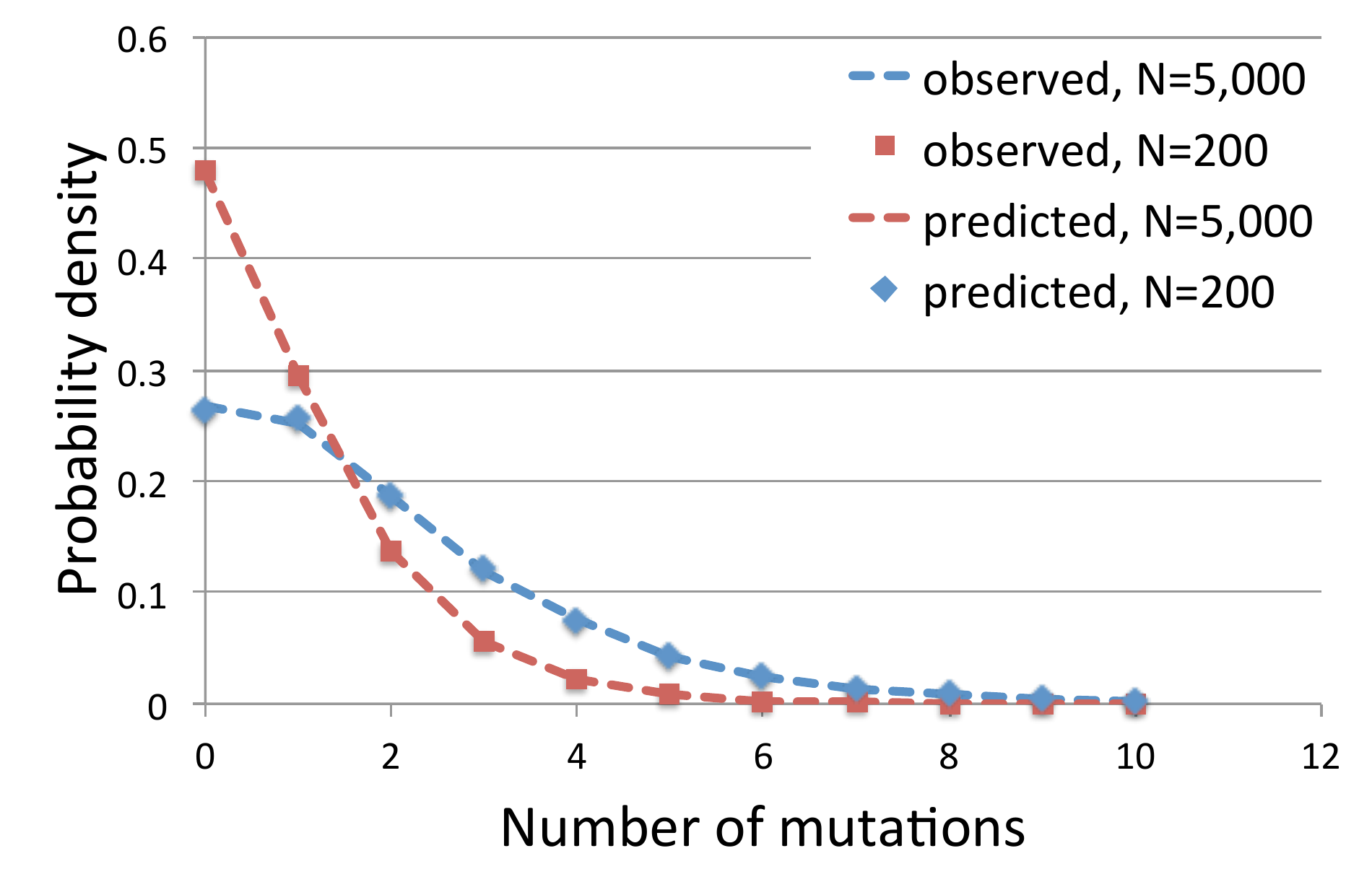}
            \caption{Predicted vs. observed distribution for the ``fixed-length'' part of the segments.}
    \end{subfigure}
    
    \begin{subfigure}[b]{0.7\textwidth}
		\vspace{-0mm}
            \centering
            \includegraphics[width=\textwidth]{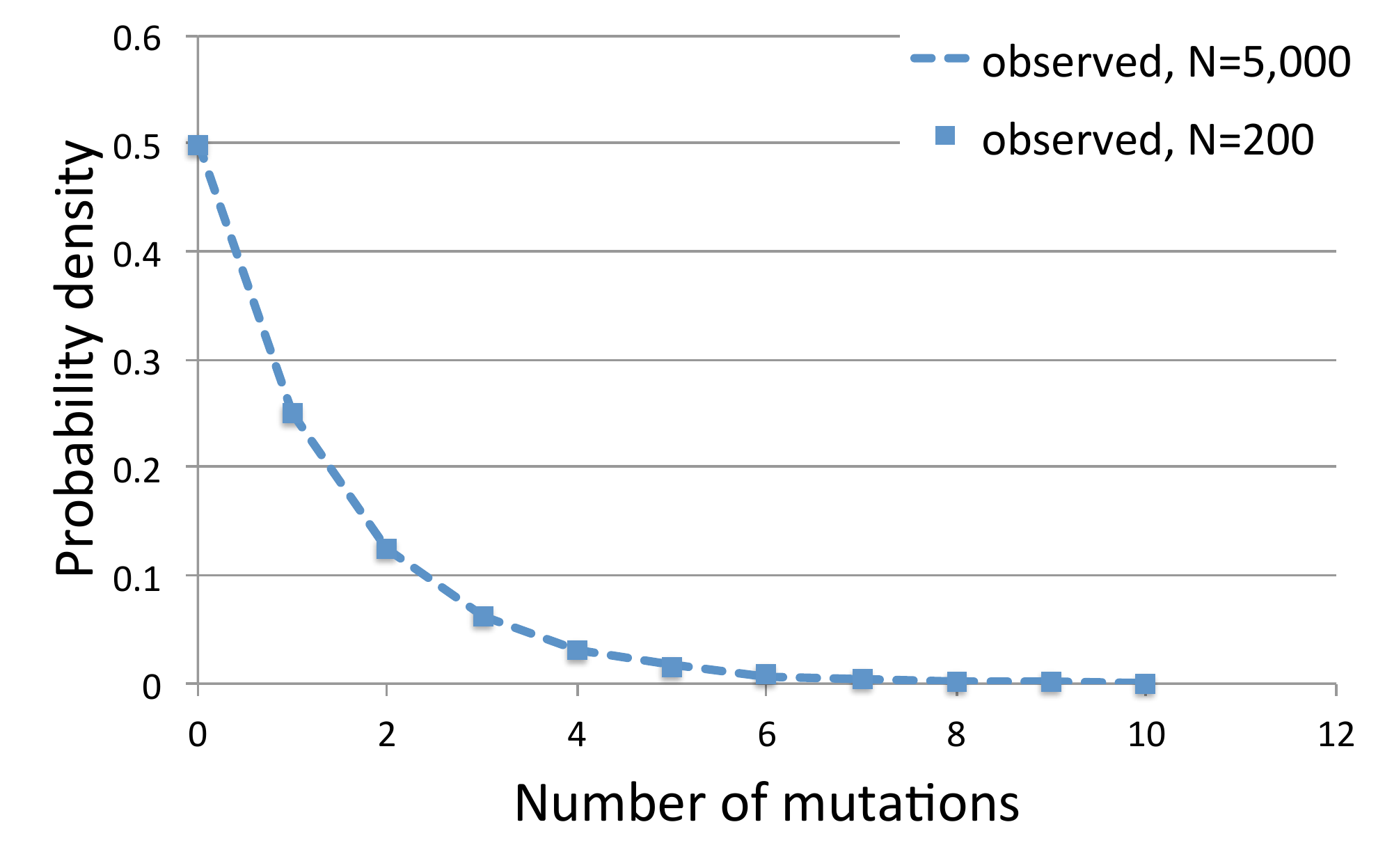}
            \caption{Observed distribution for the ``variable-length'' part of the segments, which does not depend on $N$.}
    \end{subfigure}
	        \caption{Comparison of empirical and analytical distributions for the number of mutations in the ``fixed-" and ``variable-length'' parts of IBD segments, using SMC simulations.}
			\label{fig:mut:goodFit}
	\end{figure*}

We have tested these models on empirical distributions obtained via simulations, and observed a good fit (Figure \ref{fig:mut:goodFit}).

\section{Mutations on IBD segments and demographic history}

\subsection{The age of a randomly drawn IBD segment}

We now want to move on to computing the distribution for the number of mutations found on an IBD segment of a given minimum length coming from a population of specific demographic history, rather than the exact generation of the common ancestor. We will need to express the distribution for the age of a randomly sampled IBD segment of length $l$, given the population has constant size $N$. This quantity can be computed using the results of Chapter \ref{chap:IBDmodel}. Recall that for a pair of individuals, given a common ancestor that lived $g$ generations in the past, the probability a genomic site is spanned by a segment of length $l$ is the Erlang-$2$ distribution with parameter $(2g)^{-1}$, multiplied by the chance of coalescence at generation $g$, which is $p(g|N) = N^{-1}e^{-gN^{-1}}$. We can divide this expression by $l$, switching our unit from genomic site to whole segment, and then divide by a normalizing factor to obtain the segment's age distribution as
\begin{equation}
\begin{split}
	p(g | l, N) &= \frac{4g^2 l e^{-2gl} \times l^{-1} \times N^{-1}e^{-gN^{-1}}}{\int_{0}^{\infty} 4g^2 l e^{-2gl} \times l^{-1} \times N^{-1}e^{-gN^{-1}}dg} \\
	&= \frac{g^2 (2 l N+1)^3 e^{-g \left(N^{-1}+2l\right)}}{2 N^3} \\
	&= g \left(N^{-1}+2l\right)^2 e^{-g(N^{-1}+2l)} \times \left(N^{-1}+2l\right) \frac{g}{2}
\end{split}
\label{eq:mut:age1}
\end{equation}
Note that this is closely related to computing the expected number of segments of length $l$ for a population of size $N$, as the expected number of segments of length $l$ is obtained by summing the contributions of each generation in the considered demographic history, which for a constant population size is given by $\gamma \times p(g|N)p(l|g)/l$ (see Equation \ref{eq:model:14}, which is expressed in centimorgans and averaged over all segments of length at least $u$). This approach was used in \cite{ralph2013geography} in the more general context of an arbitrary demographic history, although with a slightly different computation for the expected number of IBD segments.

To get the age of a segment of length greater than $u$, we write

\begin{equation}
\begin{split}
	p(g | l \geq u, N) &= \frac{\int_{u}^{\infty} p(g | l, N) p(l|N) dl}{\int_{0}^{\infty} \int_{u}^{\infty} p(g | l, N) p(l|N) dl ~ dg} \\
	&= \frac{\int_{u}^{\infty} g^2 (2 l N+1)^3 e^{-g \left(N^{-1}+2l\right)} \times (2 N^3)^{-1} dl}{\int_{0}^{\infty} \int_{u}^{\infty} g^2 (2 l N+1)^3 e^{-g \left(N^{-1}+2l\right)} \times (2 N^3)^{-1} dl ~ dg} \\
	&= g \times e^{-g(N^{-1}+2u)}\left(N^{-1}+2u\right)^2 ,
\end{split}
\label{eq:mut:age2}
\end{equation}
Note that this is again the Erlang-$2$ distribution, with parameter $N^{-1}+2u$, and that $\lim_{u \to 0} p(g | l \geq u, N) = N^{-2}ge^{-gN^{-1}}$.

\subsection{The number of mutations on an IBD segment with minimum length}

To compute the distribution of mutations on a segment of minimum detectable length $u$ in the population, we again separate the contributions of the segment into its two deterministic and stochastic parts, now marginalizing the time $g$ of the transmitting common ancestor for the portion of fixed length $u$, using Equation \ref{eq:mut:age2}

\begin{equation}
\begin{split}
	p(k | u, r, \mu, N) &= \int_{0}^{\infty} p(k,g | u, r, \mu, N) dg \\
	&= \int_{0}^{\infty} p(k| g, u, r, \mu, N) p(g|l\geq u,N) dg \\
	&= \int_{0}^{\infty} \frac{(2g\mu r u)^k e^{-2g\mu r u}}{k!} g e^{-g(N^{-1}+2u)}\left(N^{-1}+2u\right)^2 dg \\
	&= \frac{2^k (1 + k) (\mu N r u)^k (1 + 2 N u)^2}{[1 + 2 N (u + \mu r u)]^{k+2}}
\end{split}
\label{eq:mut:k_u}
\end{equation}

The mean for the number of mutations in this portion of the segment is
\begin{equation}
\begin{split}
	\operatorname{E}[k|g,r,\mu,l=u,N] &= \sum_{k=0}^{\infty} k \times \frac{2^k (1 + k) (\mu N r u)^k (1 + 2 N u)^2}{[1 + 2 N (u + \mu r u)]^{k+2}} \\
	&= \frac{4 \mu N r u}{1 + 2 N u},
\end{split}
\label{eq:mut:08}
\end{equation}
and the variance is
\begin{equation}
\begin{split}
	\operatorname{Var}[k|g,r,\mu,l=u,N] &= \sum_{k=0}^{\infty} \left(k-\frac{4 \mu N r u}{1 + 2 N u}\right)^2 \times \frac{2^k (1 + k) (\mu N r u)^k (1 + 2 N u)^2}{[1 + 2 N (u + \mu r u)]^{k+2}} \\
	&= \frac{4 \mu N r u [1 + 2 N (u + \mu r u)]}{(1 + 2 N u)^2}.
\end{split}
\end{equation}

The remaining part of the segment, the ``tip'', has variable length, but as we have seen the number of mutations it harbors (Equation \ref{eq:mut:04}) does not depend on the time $g$ of the transmitting common ancestor, and marginalizing it does not affect the resulting distribution. Putting together the ``tip'' distribution of Equation \ref{eq:mut:04} with the distribution for the fixed part of the segment (Equation \ref{eq:mut:k_u}), we get

\begin{equation}
\begin{split}
	p(k | u, r, \mu, N) &= p(k | l=u, r, \mu, N) \ast p(k|l>0,r,\mu)\\
	&= \sum_{n=-\infty}^{\infty} ~ \chi_a(k) \frac{2^k (1 + k) (\mu N r u)^k (1 + 2 N u)^2}{[1 + 2 N (u + \mu r u)]^{k+2}} \times \chi_0(n-k) \frac{(\mu r)^{n-k}}{(1+\mu r)^{n-k+1}} \\
	&= (2 N u+1)^2 \left\{\frac{(\mu r)^n}{(\mu r+1)^{n+1}}-\frac{2^{n+1} N u (\mu N r u)^n[2 N (\mu r u+u)+n+2]}{[2 N (\mu r u+u)+1]^{n+2}}\right\} ~ ,
\end{split}
\end{equation}
where $\chi_a(x)$ is the step function
\begin{equation}
\begin{split}
	\chi_a(x) =
	\begin{cases}
	1 & \mbox{if } x \geq a, \\
	0 & \mbox{if } x < a. \\
	\end{cases}
\end{split}
\end{equation}

The mean and the variance of this distribution can be obtained summing mean and variance of the two components of the final distribution, as

\begin{equation}
\begin{split}
	\operatorname{E}[k | u, r, \mu, N] &= \mu r + \frac{4 \mu N r u}{1 + 2 N u}\\
	\operatorname{Var}[k | u, r, \mu, N] &= \mu r (1 + \mu r) + \frac{4 \mu N r u [1 + 2 N (u + \mu r u)]}{(1 + 2 N u)^2}.
\end{split}
\label{eq:mut:09}
\end{equation}

\begin{figure}[!ht]
	\vspace{-2mm}
\centering
\includegraphics[width=0.8\textwidth]{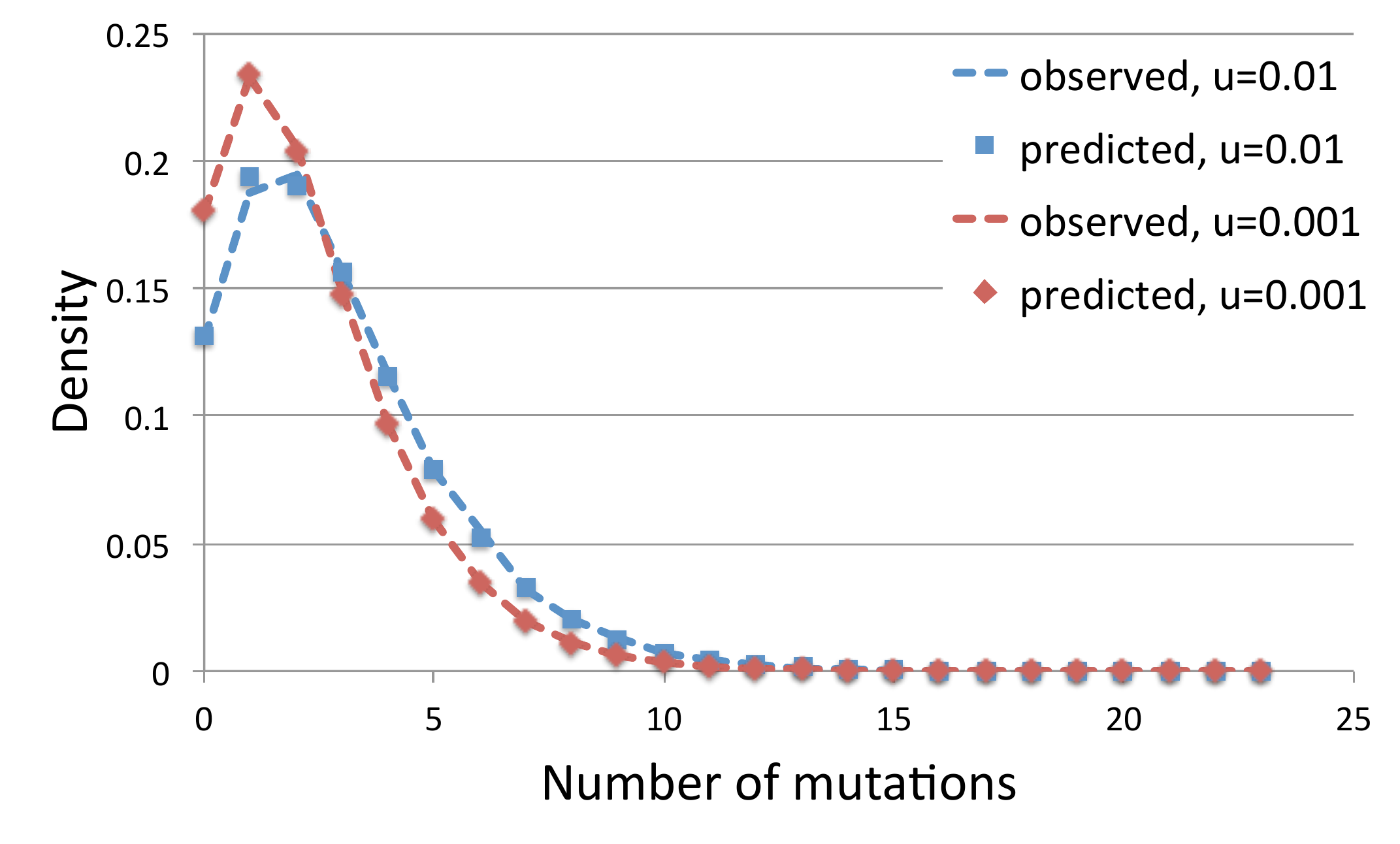}
\caption{Empirical and analytical distributions of mutation counts on IBD segments for $N=500$, using SMC simulations.}
\label{fig:mut:distributionPredictedObserved}
\end{figure}

The obtained distribution for the number of mutations on an IBD segment was found to provide a good fit for empirical distributions obtained from simulations (Figure \ref{fig:mut:distributionPredictedObserved}). Note that the distribution is overdispersed, as the variance is always larger than the mean.

\section{Using mutations on IBD segments for demographic inference}

We now turn to the problem of using knowledge on the number of mutations on IBD segments to infer demographic history. We consider the number of segments $s_u$ that are longer than a Morgan threshold $u$, and the number $k$ of mutations found on these segments in a population of size $N$. Again, we can write
\begin{equation}
\begin{split}
p(s_u,k|N)dl &= p(k,|s_u,N)p(s_u|N)
\end{split}
\label{eq:mut:10}
\end{equation}

\begin{figure}[!ht]
\centering
\includegraphics[width=0.9\textwidth]{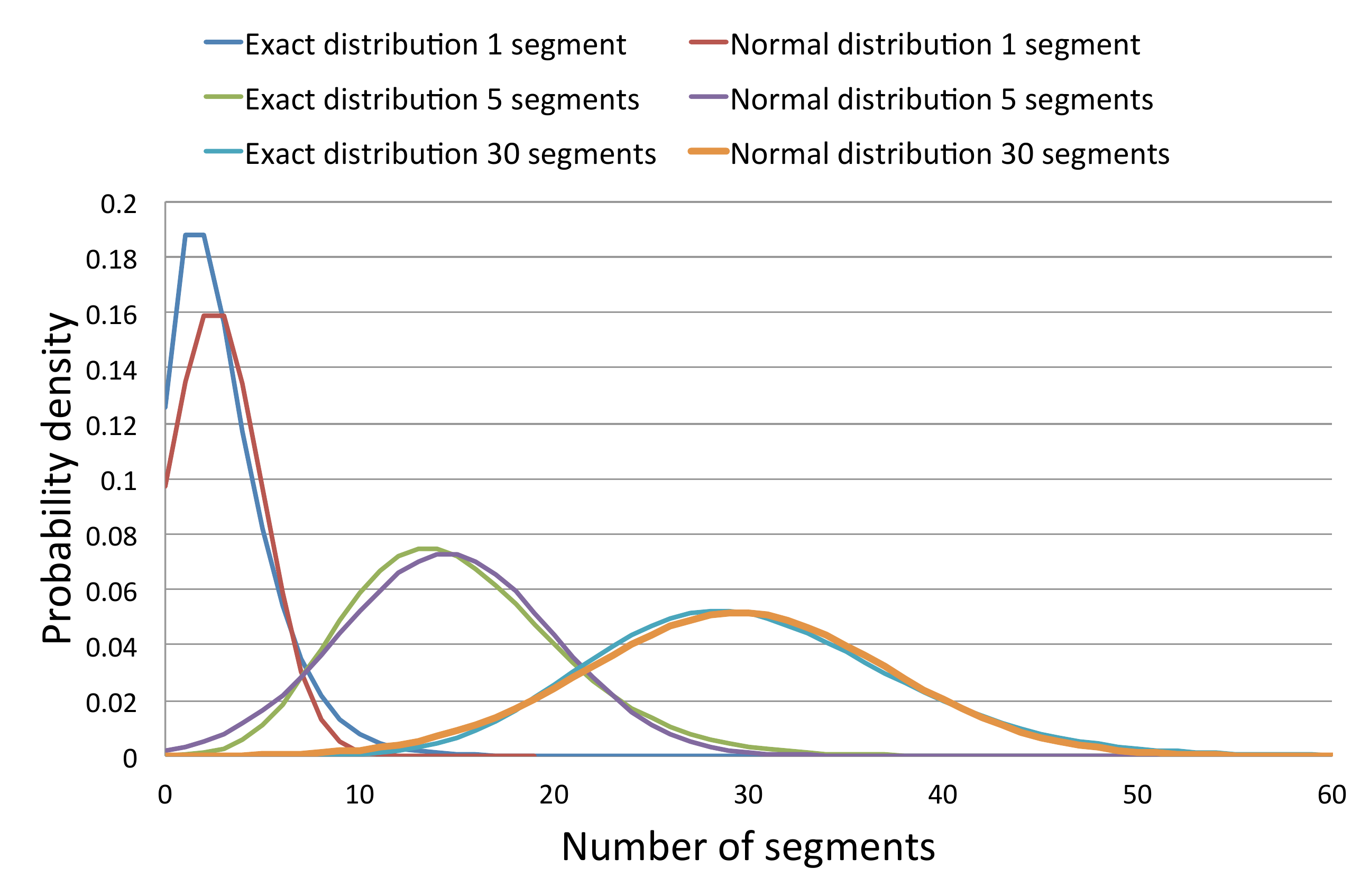}
\caption{Convergence of the distribution for the number of mutations to a Normal distribution.}
\label{fig:mut:convergence}
\end{figure}

We have previously computed an expression for $p(s_u|N)$, which is Poisson distributed with mean described in Equation \ref{eq:model:numSeg_u}, and Equation \ref{eq:mut:age1} provides the distribution for the number of mutations coming from a single segment sampled from the population. To compute the distribution for an independent number of segments, we can repeatedly take the convolution of Equation \ref{eq:mut:08}. Note, however, that this is computationally unfeasible for large $s_u$, and it is also unnecessary, as we can rely on the Central Limit Theorem and use the mean and variance of Equation \ref{eq:mut:09} in a Normal distribution. Note, furthermore, that the shape of repeated convolutions of the distribution in Equation \ref{eq:mut:08} quickly converges to a Gaussian distribution (Figure \ref{fig:mut:convergence}).

\begin{figure}[!ht]
	\vspace{-2mm}
\centering
\includegraphics[width=0.87\textwidth]{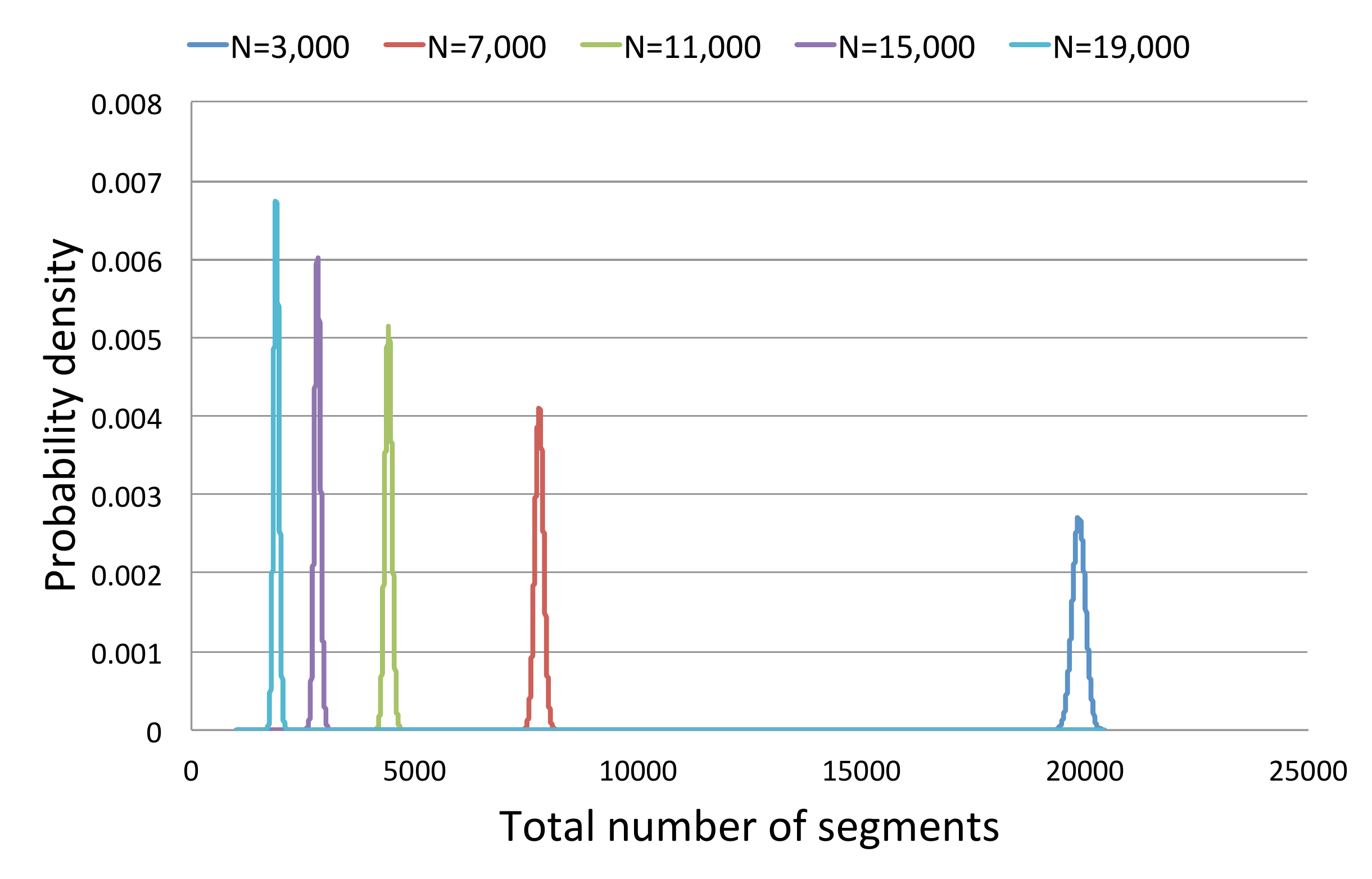}
\caption{The distribution of the number of IBD segments does not overlap across different populations ($20$ samples, segments of at least $0.5$cM in SMC simulations).}
\label{fig:mut:nooverlap}
\end{figure}
\begin{figure}[!ht]
	\vspace{-2mm}
\centering
\includegraphics[width=0.85\textwidth]{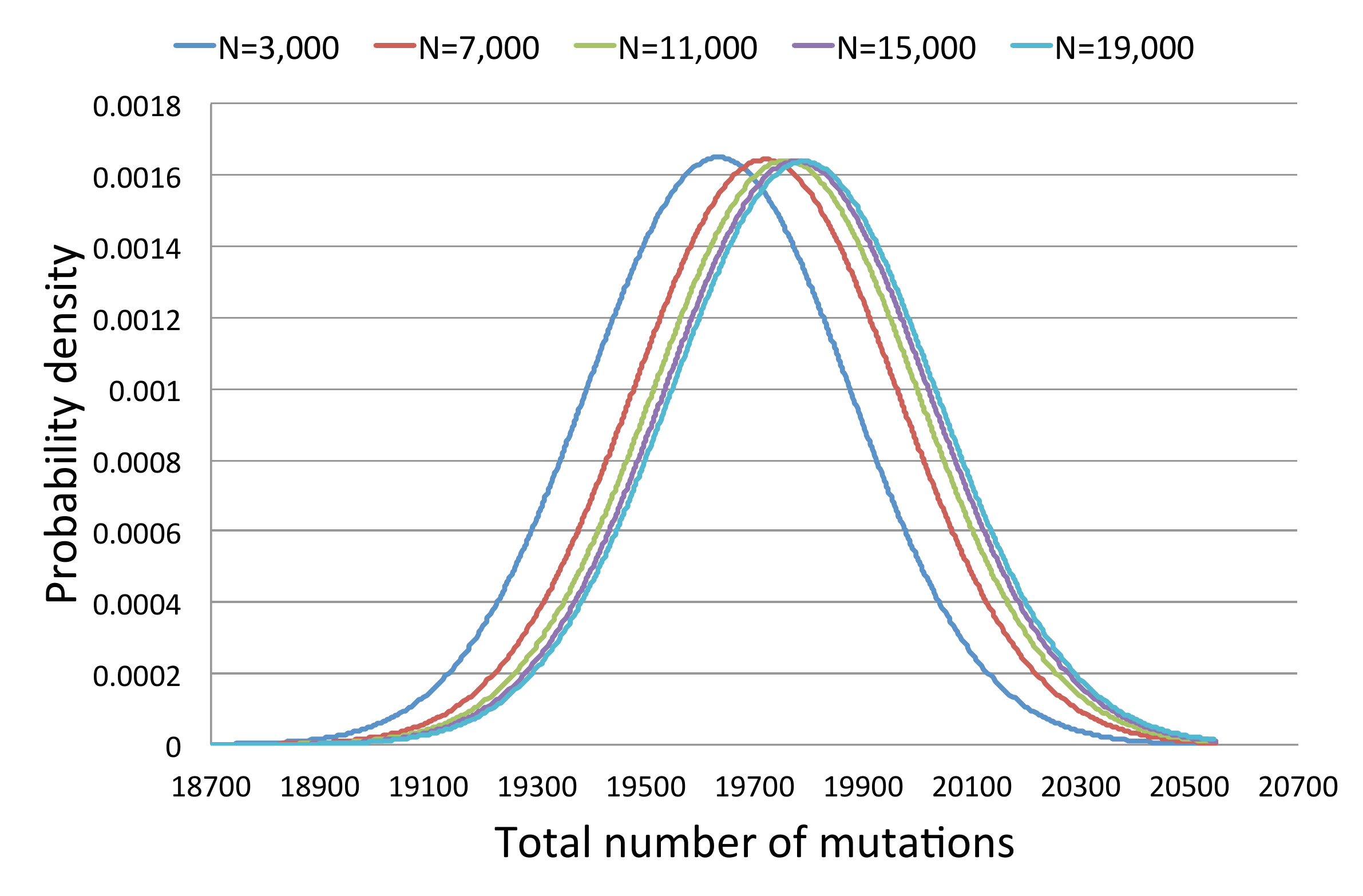}
\caption{Substantial overlap for the distribution of the total number of mutations in different populations ($20$ samples, segments of at least $0.5$cM in SMC simulations).}
\label{fig:mut:overlap}
\end{figure}

\begin{figure*}
	\vspace{-2mm}
    \centering
    \begin{subfigure}[b]{0.7\textwidth}
		\vspace{-2mm}
            \centering
            \includegraphics[width=\textwidth]{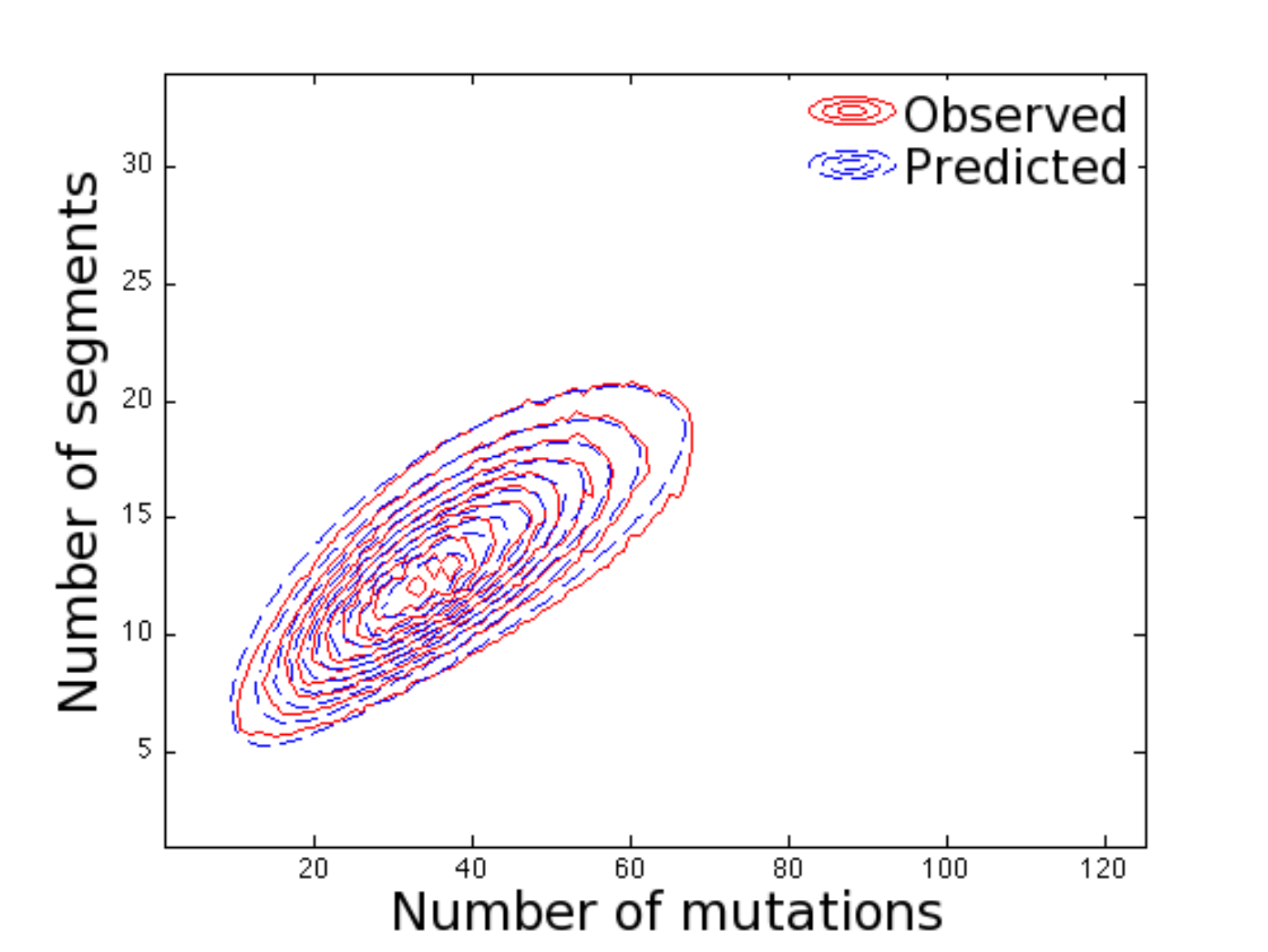}
            \caption{$N=2,000$, $\mu=10^{-8}$, $r=10^8$, $u=0.01$, genomic region of $10$M.}
            \label{fig:mut:jointContoura}
    \end{subfigure}
    
    \begin{subfigure}[b]{0.7\textwidth}
		\vspace{-0mm}
            \centering
            \includegraphics[width=\textwidth]{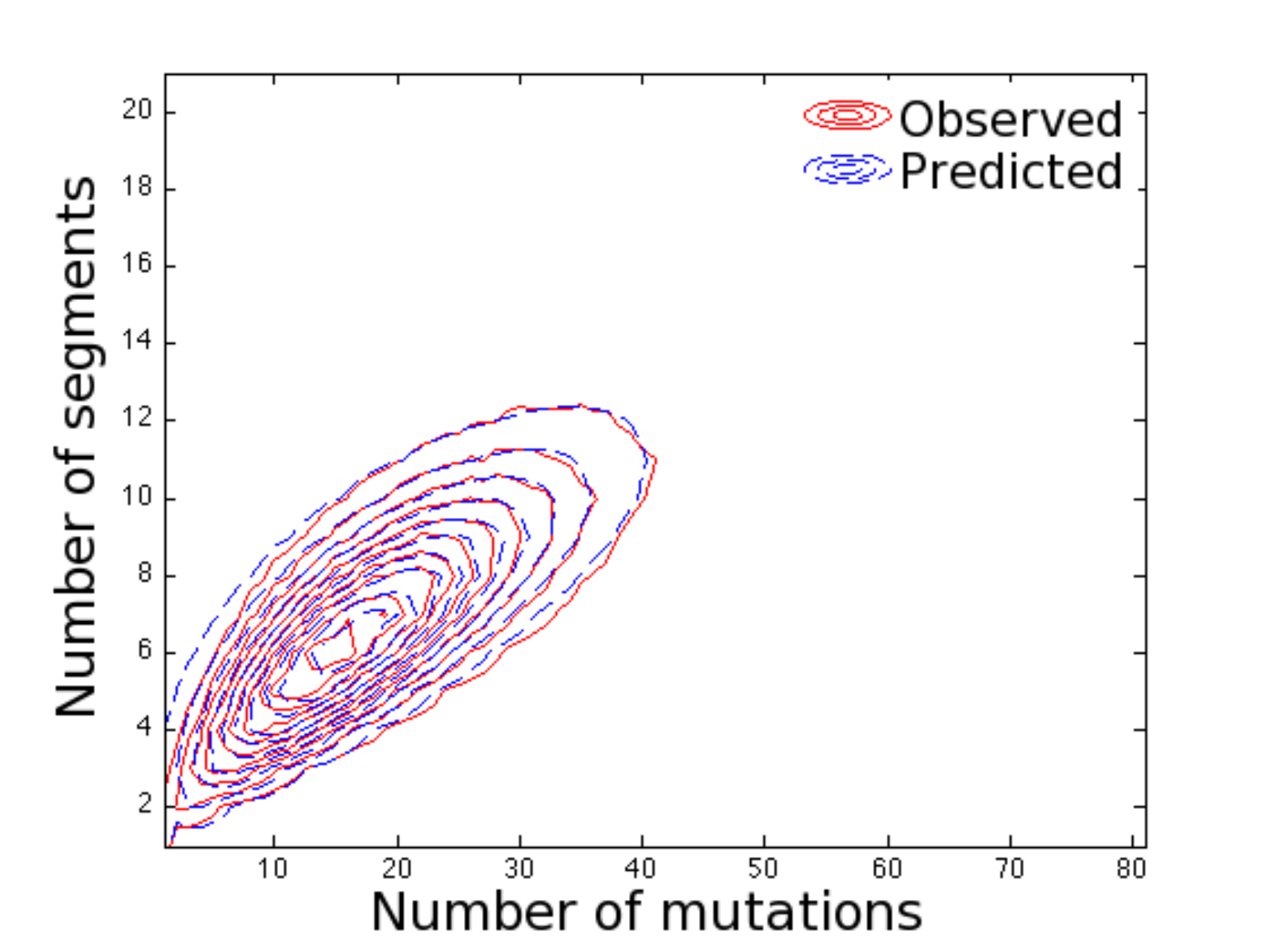}
            \caption{$N=4,000$, $\mu=10^{-8}$, $r=10^8$, $u=0.01$, genomic region of $10$M.}
            \label{fig:mut:jointContourb}
    \end{subfigure}
	        \caption{Comparison of joint distribution for the number of segments and the number of mutations and empirical distribution obtained from GENOME simulations.}
			\label{fig:mut:jointContour}
\end{figure*}

After substituting the previously discussed distributions into Equation \ref{eq:mut:10}, we tested the prediction of this expression against several synthetic datasets (Figure \ref{fig:mut:jointContour}), and we observed good correspondence between analytical and empirical distributions (note that some of the approximations done in the GENOME simulator software, e.g. the minimum size of recombinant blocks, may distort the empirical distribution).

\begin{figure}[!ht]
	\vspace{-2mm}
\centering
\includegraphics[width=0.9\textwidth]{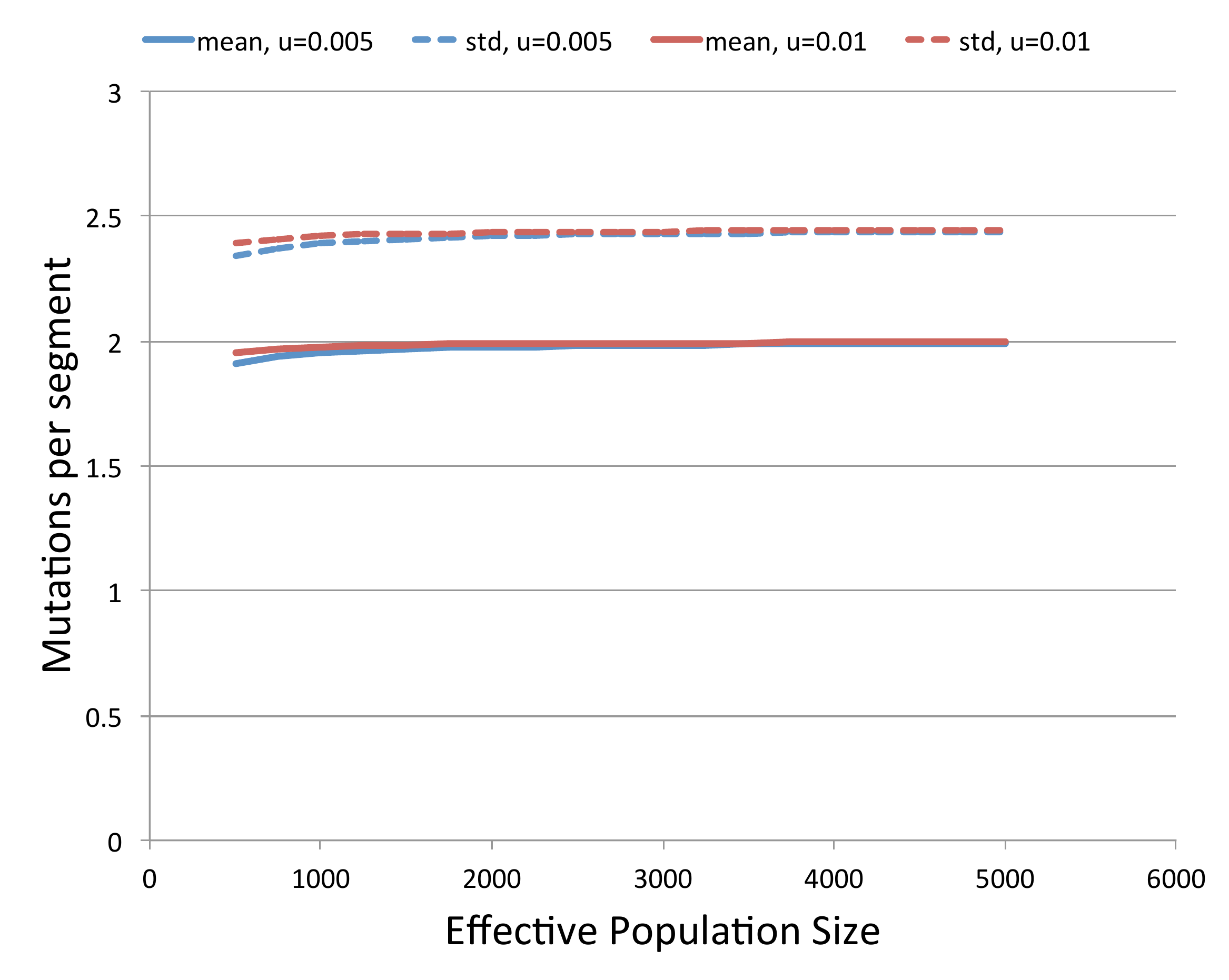}
\caption{Mean and standard deviation (Equation \ref{eq:mut:09}) for several values of $N$.}
\label{fig:mut:meanStd}
\end{figure}

Compared to the distribution for the number of IBD segments (Figure \ref{fig:mut:nooverlap}), the distributions for the number of mutations on IBD segments overlap substantially across different values of the effective population size $N$ (Figure \ref{fig:mut:overlap}), suggesting that knowledge about the distribution of mutations over IBD segments does not result in substantial improvements for demographic inference. Inspecting the expression for the mean number of mutations in an IBD segment in Equation \ref{eq:mut:09}, it is clear that variations of $N$, the effective population size, result in minimal variation, as can be observed in Figure \ref{fig:mut:meanStd}.

\section{Inferring mutation rates using IBD}

\begin{figure}[!ht]
	\vspace{-2mm}
\centering
\includegraphics[width=0.9\textwidth]{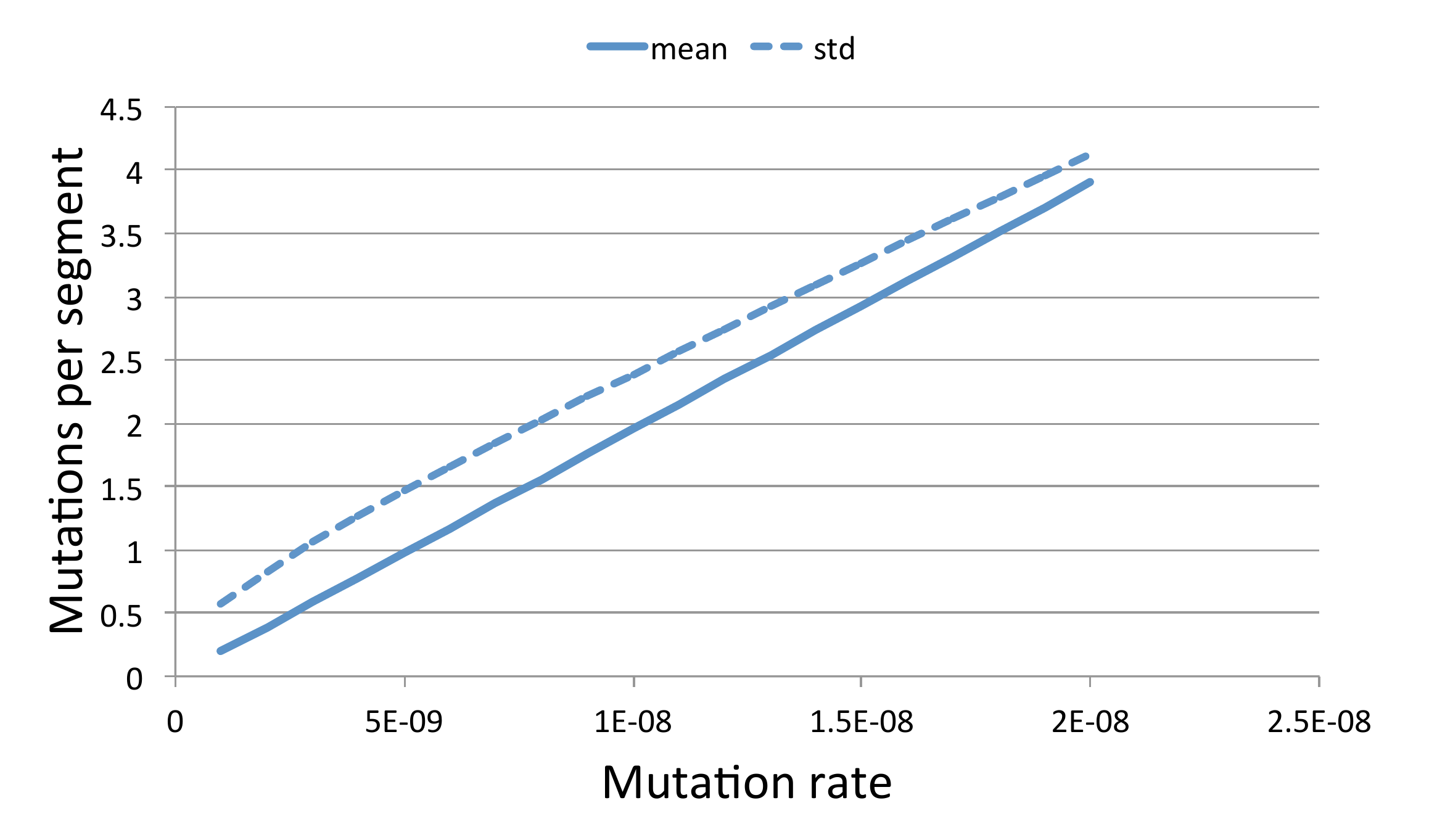}
\caption{Number of mutations per IBD segment as a function of mutation rate.}
\label{fig:mut:mutInfluence}
\end{figure}

The mean and variance of Equation \ref{eq:mut:09} show a weak connection between demography and mutations in IBD segments. While this does not facilitate using this feature for demographic inference, it does provide support for other analyses. The mean and variance are in fact strongly influenced by the mutation rate (Figure \ref{fig:mut:mutInfluence}), suggesting this parameter may be inferred based on the derived distributions of mutation on IBD segments. Furthermore, we note that it is possible to entirely remove the dependence on demographic history through very simple manipulations of the observed IBD segments. We have previously noted that due to the memorylessness property of the exponential distribution, the length of a detected IBD segment longer than a detectable threshold $u$ may be seen as the sum of two parts: $L = u + L_\tau$, the fixed length, determined by the threshold of detectable IBD segments, and a stochastic part, $L_\tau$. As previously described, the latter is not affected by coalescent times, while the ``fixed'' portion of the segment is. This property can be exploited to isolate summary statistics that are informative about mutation rates, while not being confounded by the presence of latent demographic history. Recall that the sum of two Poisson random variables is also Poisson distributed, with rate given by the sum of the addends' rates. The number of mutations on the segment $L = u + L_\tau$ that are found on the $L_\tau$ portion are therefore expected to be a fraction $L_\tau/L$ of those observed on the entire segment. To estimate the total number of such mutations, it is therefore sufficient to multiply the number of mutations observed on each segment of the detected set $S_u$ by the factor $(l-u)/l$.

\begin{equation}
\begin{split}
	\hat{k} = \sum_{s \in S_u} k_s \times [(l_s-u)/l_s]
\end{split}
\label{eq:mut:11}
\end{equation}

As shown in Equation \ref{eq:mut:05}, the expected number of mutations occurring in the stochastic portion of each segment is simply $\mu r$. Because the distribution for the number of mutations found on a large number of segments $n_s=\left\vert{S_u}\right\vert$ is well described by a Normal distribution with mean $n_s \mu r$, we can derive the maximum likelihood estimator

\begin{equation}
\begin{split}
	\hat{\mu} = \frac{\hat{k}}{r n_s},
\end{split}
\label{eq:mut:12}
\end{equation}
which is an unbiased estimator. If we denote the set of segments coming from generation $g$ as $S_{ug}$, then

\begin{equation}
\begin{split}
	\operatorname{E}[\hat{\mu}] &= \operatorname{E}\left[\frac{\hat{k}}{r n_s}\right] = \frac{1}{r n_s}\operatorname{E}\left[\hat{k}\right] \\
	&= \frac{1}{r n_s} \operatorname{E}\left[\sum_{s \in S_u} \frac{l_s-u}{l_s} \times k_s \right] \\
	&= \frac{1}{r n_s} \operatorname{E}\left[\sum_{g=1}^{\infty} \sum_{s \in S_{ug}} \frac{l_s-u}{l_s} \times k_{s}\right] \\
	&= \frac{1}{r n_s} \sum_{g=1}^{\infty} \sum_{s \in S_{ug}} \operatorname{E}\left[\frac{l_s-u}{l_s} \times k_{s}\right] \\
	&= \frac{1}{r n_s} \sum_{g=1}^{\infty} \sum_{s \in S_{ug}} \int_{u}^{\infty} P(l|g) \operatorname{E}\left[\frac{l-u}{l} \times k \middle| l \right] dl \\
	&= \frac{1}{r n_s} \sum_{g=1}^{\infty} \sum_{s \in S_{ug}} \int_{u}^{\infty} 2ge^{-2g(u-l)} \left( \frac{l-u}{l} \times 2g\mu l r \right) dl \\
	&= \frac{1}{r n_s} \sum_{g=1}^{\infty} \sum_{s \in S_{ug}} \mu r \\
	&= \frac{1}{r n_s}\times n_s \mu r = \mu\\
\end{split}
\label{eq:mut:13}
\end{equation}

\begin{figure*}
	\vspace{-2mm}
    \centering
    \begin{subfigure}[b]{0.7\textwidth}
		\vspace{-2mm}
            \centering
            \includegraphics[width=\textwidth]{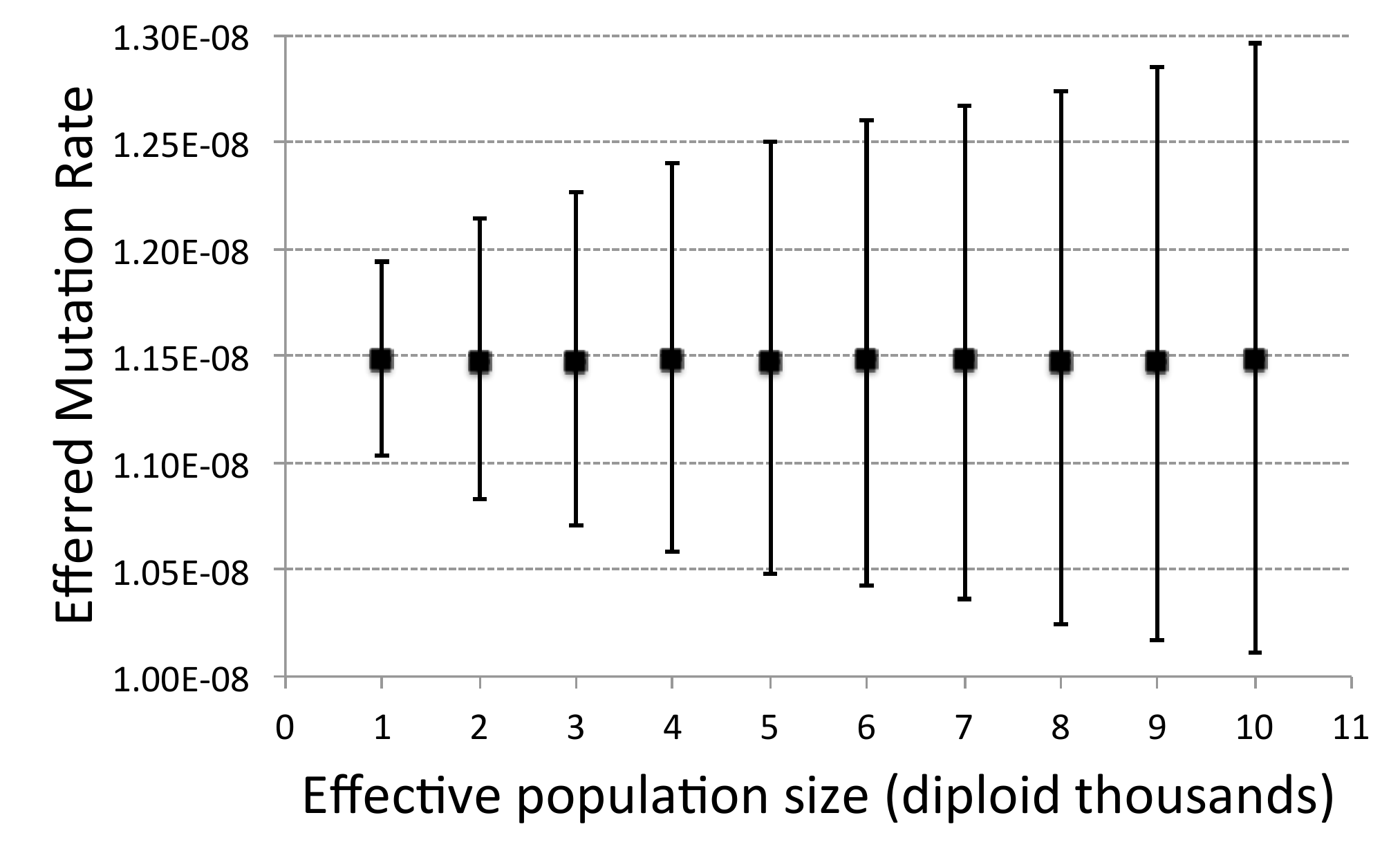}
            \caption{Inferred values of $\mu$ for different population sizes.}
            \label{fig:mut:muInfa}
    \end{subfigure}

    \begin{subfigure}[b]{0.7\textwidth}
		\vspace{-0mm}
            \centering
            \includegraphics[width=\textwidth]{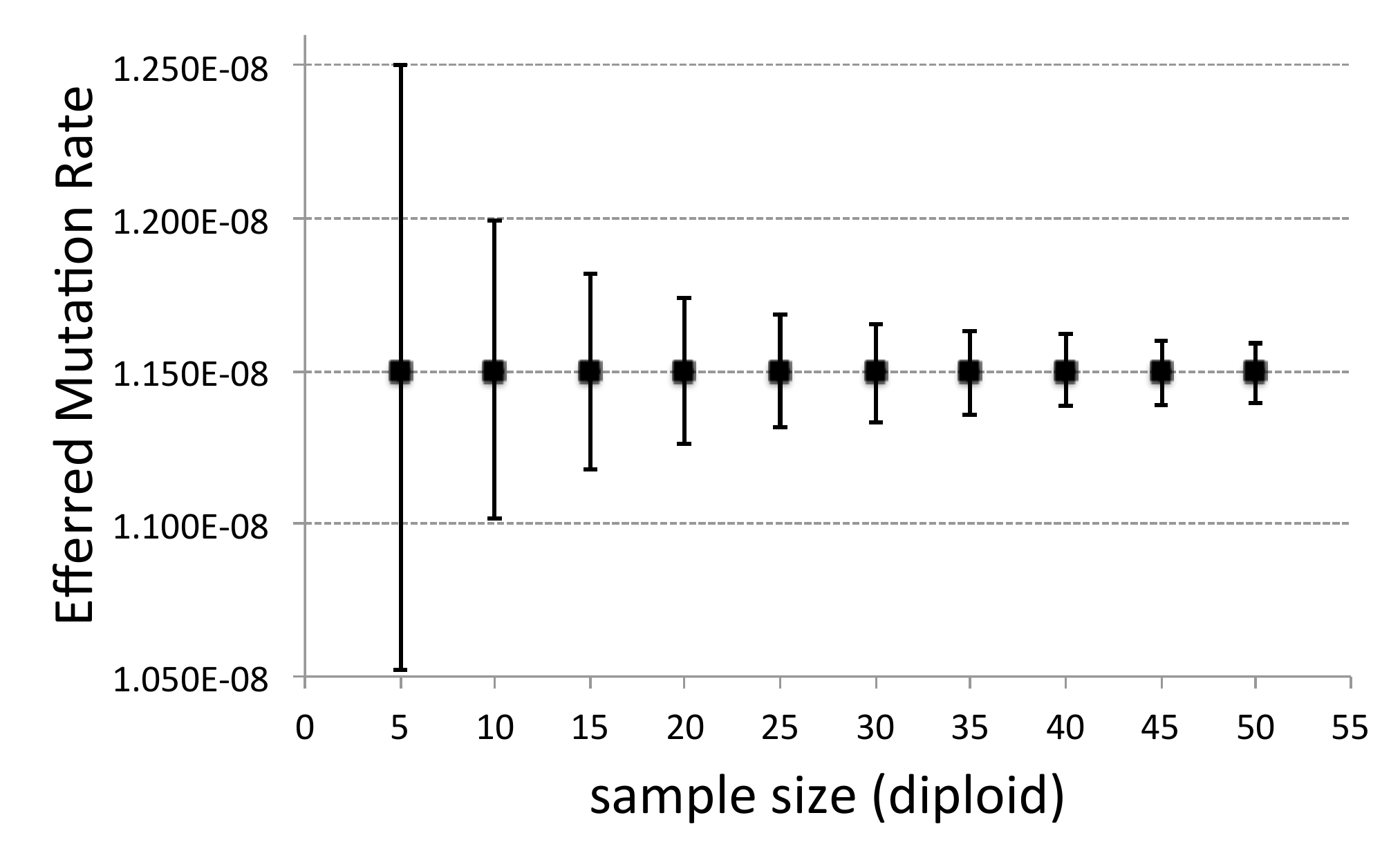}
            \caption{Inferred mutation rate as a function of sample size.}
            \label{fig:mut:muInfb}
    \end{subfigure}
\caption{Inference of mutation rates via IBD sharing in SMC simulations.}
		\label{fig:mut:muInf}
\end{figure*}

We tested this estimator on simulated data using the SMC algorithm \cite{mcvean2005approximating}, therefore using definition (c) of Section \ref{intro:subsec:IBD_def} to generate shared segments, and we observed good performance. By analyzing IBD segments of length at least $1$ cM for only $5$ diploid samples, assuming a genome of $35$ Morgans and $\mu=1.15$, $r=0.83$, we attained tight $95\%$ confidence intervals around the true mutation rate (Figure \ref{fig:mut:muInfa}). Variations in the effective population size had a moderate influence on the width of confidence intervals, as larger population sizes result in fewer IBD segments. Because a sample of $n$ individuals contains ${2n\choose 2}$ haploid chromosomes that may contain IBD segments, there is a quadratic gain of statistical power when more samples are analyzed (Figure \ref{fig:mut:muInfb}). A sample of $50$ independent individuals provide extremely tight confidence intervals in a population of $N=10,000$ effective individuals, using the above listed parameters.

Back-of-the-envelope calculations suggest that the statistical power of IBD-based inference of mutation rates is extremely high compared to a trio-based analysis. When mutation rates are estimated based on observed de-novo mutations in transmitted haplotypes of sequenced trio individuals, the number of ``effective haplotypes'' that are compared is $n/3$, for $n$ sequenced individuals, i.e. the two transmitted haplotypes. In a population of effective size $10,000$, individuals will share on average $0.5\%$ of their genome through IBD haplotypes longer than $1$cM. The number of individuals tested for IBD is however ${2n\choose 2}$, as previously mentioned, resulting in ${2n\choose 2}\times 0.05$ effective comparisons across haplotypes, assuming that ancestral lineages do not overlap significantly. Using the method previously described, we use on average half of the expected length of each IBD haplotype, which is ${\sim}2u$ Morgans. In this scenario, for a sample of $60$ individuals, about ${120\choose 2}\times 0.005 \times 0.5 \approx 18$ haploid genomes are effectively compared to estimate mutation rates using IBD, against the $40$ used in the trio-based approach. However, the number of mutations found on a nucleotide spanned by these IBD segments is higher than the average number of mutations for a nucleotide transmitted by a parent in a trio-based analysis. The average age number of meioses for a nucleotide on these IBD segments can be computed based on Equation \ref{eq:model:4} (here expressed in Morgans)

\begin{equation}
\begin{split}
	\operatorname{E}[t|u,N] &= \int_{0}^{\infty} t \times \frac{N^{-1} e^{-tN^{-1}} \left[ \int_{u}^{\infty} (2t)^2le^{-2tl} dl\right]}{\int_{0}^{\infty} N^{-1} e^{-tN^{-1}} \left[ \int_{u}^{\infty} (2t)^2le^{-2tl} dl\right]} dt \\
	&= \frac{N (6 N u+1)}{8 N^2 u^2+6 N u+1} \approx \frac{3}{4u}.
\end{split}
\end{equation}

If segments longer than $1$cM are considered, this results in an average $75$ meioses to a common ancestor, or $150$ counting the length of the ancestral path between the sharing individuals. This results in roughly $\frac{18\times 150}{40} = 67.5$ times the statistical power of a trio-based analysis. If more isolated populations are considered, and shorter segments can be detected, this gain is substantially further increased. For a population of effective size $N=1,000$, the fraction of genome shared for segments of at least $1$ cM is about $4.8\%$, with a gain of $\frac{{120\choose 2}\times 0.048 \times 0.5\times 150}{40} = 642.6$, and if segments of $0.5$ cM and longer can be detected, the gain would reach roughly $1,245$.

Such increase in statistical power enables further analyses, such as the inference of locus-specific mutation rates. Note, however, that some of the assumptions that were made in this derivation, such as the uniformity of recombination rates and selective neutrality, will need to be addressed. Furthermore, this analysis relies on accurate IBD detection, and IBD segments are defined as non-recombinant chromosomal regions transmitted from a common ancestor (see definition (c) of Section \ref{intro:subsec:IBD_def}). Refinements of these models to account for mismatches between the described quantities and what is realistically possible to infer using available IBD detection algorithms will improve this analysis. This approach also assumes lack of genotyping errors, however note that because IBD detection is feasible using only high-frequency markers, typically less prone to genotype errors, there is no substantial interference in using shared haplotypes in this scenario. A related application is the fine-tuning of genotype-calling parameters, as prior knowledge on a plausible range for mutation rates enables detecting a component of error in the inferred mutation parameter. Furthermore, note that assuming the mutation rate is known, this approach may be used to study recombination rates.

It is interesting to note that this approach measures mutation rates observing mutation events that potentially occurred several generations in the past. This method can therefore be used to asses variation in historical mutation rates. A simple test involves obtaining estimates based on different cutoffs for the minimum length of the observed IBD segments. Because shorter haplotypes tend to be transmitted from more remote common ancestors, estimates based on smaller values of $u$ reflect mutation rates at more remote time scales in the population. If IBD detection is accurate and a demographic model has been reconstructed (using the methods of Chapter \ref{chap:IBDmodel} and \ref{chap:migration}, or others that more suitable for remote time scales e.g. \cite{li2011inference}), it is possible to estimate the time of these variations using the described methods in conjunction with the segment age distributions of Equation \ref{eq:mut:age2}.

\section{Appendix}
\subsection{Efficient computation of mean and variance for the number of mutations in an arbitrary demography}
If a population of arbitrary demographic history $\theta$ has population size $N(g,\theta)$ at time $g$, the coalescent distribution is described by

\begin{equation}
\begin{split}
	\left(\prod_{j=1}^{g-1}1-\frac{1}{N(j,\theta)}\right) \frac{1}{N(g,\theta)} = C(g,\theta).
\end{split}
\end{equation}

The probability of seeing $k$ mutations on an IBD segment in this population is
\begin{equation}
\begin{split}
	p(k|u,r,\mu,\theta) &= \sum_{g=1}^{\infty} \left[ C(g,\theta)\operatorname{Poiss}(k,2 \mu r u g) \right].
\end{split}
\end{equation}

For the mean of the constant part:

\begin{equation}
\begin{split}
	\operatorname{E}[k|u,r,\mu,\theta] &= \sum_{k=0}^{\infty} \left\{ k ~ p(k|u,r,\mu,\theta) \right\} \\
	&= \sum_{g=1}^{\infty} \left\{ \sum_{k=0}^{\infty} \left[ k ~ C(g,\theta)\operatorname{Poiss}(k,2 \mu r u g) \right]\right\} \\
	&= \sum_{g=1}^{\infty} \left\{ C(g,\theta) \sum_{k=0}^{\infty} \left[ k \operatorname{Poiss}(k,2 \mu r u g) \right]\right\} \\
	&= \sum_{g=1}^{\infty} \left\{ 2 \mu r u g ~ C(g,\theta) \right\}.
\end{split}
\end{equation}

For the variance of the constant part (using $\operatorname{E}_K = \operatorname{E}[k|u,r,\mu,\theta]$):

\begin{equation}
\begin{split}
	\operatorname{Var}[k|u,r,\mu,\theta] &= \sum_{k=0}^{\infty} \left\{ (k - \operatorname{E}_K )^2 ~ p(k|u,r,\mu,\theta) \right\} \\
	&= \sum_{k=0}^{\infty} \left\{ (k - \operatorname{E}_K )^2 \sum_{g=1}^{\infty} \left[ C(g,\theta)\operatorname{Poiss}(k,2 \mu r u g) \right]\right\} \\
	&= \sum_{g=1}^{\infty} \left\{ C(g,\theta) \sum_{k=0}^{\infty} \left[ (k - \operatorname{E}_K )^2 \operatorname{Poiss}(k,2 \mu r u g) \right]\right\} \\
	&= \sum_{g=1}^{\infty} \left\{ \left[ (2 \mu r u g - \operatorname{E}_K)^2 + 2 \mu r u g \right] C(g,\theta) \right\}.
\end{split}
\end{equation}

Where the last step is obtained considering that, for any random variable $K$ with discrete probability distribution $f(k)$, having defined $b=\operatorname{E}_K-c$, where $\operatorname{E}_K$ is the expectation of the distribution, then

\begin{equation}
\begin{split}
	\sum_{k=-\infty}^{\infty} \left[ (k-c)^2 f(k) \right] &= \sum_{k=-\infty}^{\infty} \left[(k-\operatorname{E}_K+b)^2 f(k) \right] \\
	&= \sum_{k=-\infty}^{\infty}  \left\{ \left[ (b^2 - 2 b \operatorname{E}_K + 2 b k) + (k - \operatorname{E}_K)^2 \right] f(k) \right\} \\
	&= \sum_{k=-\infty}^{\infty} \left\{ \left[ b^2 - 2 b \operatorname{E}_K + 2 b k \right] ~ f(k) \right\} + \sum_{k=-\infty}^{\infty}  \left\{ (k - \operatorname{E}_K)^2 f(k) \right\} \\
	&= b^2 - 2 b \operatorname{E}_K + 2 b \sum_{k=-\infty}^{\infty} \left[ k f(k) \right] + \sum_{k=-\infty}^{\infty} \left\{ (k - \operatorname{E}_K)^2 f(k) \right\} \\
	&= b^2 + \operatorname{Var}_K = (\operatorname{E}_K-c)^2 + \operatorname{Var}_K.
\end{split}
\end{equation}

\chapter{Conclusions}
\label{chap:conclusions}
In this thesis we presented several new methodologies for population genetics analysis based on the sharing of IBD haplotypes across purportedly unrelated individuals from one or multiple populations. Specifically,

\begin{itemize}
\item In Chapter \ref{chap:IBD_in_data}, by analyzing several world-wide populations from the HapMap 3 dataset and the Jewish Hapmap dataset, we demonstrated that IBD sharing is informative about demographic events, revealing past migrations and population size fluctuations, carries the signature of evolutionary events, demonstrating enrichment of loci under positive selection for commonly shared regions, and reflects recent stratification, in some cases more accurately than standard methodologies.
\item In Chapter \ref{chap:IBDmodel}, motivated by the results of Chapter \ref{chap:IBD_in_data}, we used coalescent theory (see Chapter \ref{chap:intro}) to derive a theoretical framework that allows quantitatively describing the relationship between IBD sharing and demography. We demonstrated these methods by inferring the occurrence of recent demographic events in two populations of distinctive recent demographic profiles: Ashkenazi Jews, exhibiting evidence for a recent founder event followed by substantial expansion and isolation, and the Maasai from Kenya, where haplotype sharing is compatible with a societal structure of several small demes interacting through high migration rates.
\item In Chapter \ref{chap:migration}, we extended this framework to enable simultaneous analysis of several populations, inferring both migration and population size fluctuation. We showed this approach can be used for the analysis of recently diverged populations, where state-of-the-art methods based on ancestry deconvolution using a panel of reference ancestral populations are complicated due to the limiting assumption of strongly diverged ancestral groups. We used these models to study recent demographic history in the Netherlands, showing that IBD-based analysis reveals demographic structure even at fine-grained geographic scales.
\item Chapter \ref{chap:mutation} discussed utilizing IBD segments in studies where whole sequence information is available. We derived distributions for the number of mutated sites on shared haplotypes, and shown that while this information does not provide substantial improvements for demographic inference, it can be used for inferring additional parameters, such as mutation rate, increasing the statistical power by orders of magnitude compared to classical family-based methods. This boost in statistical power enables further applications, such as the inference of a map of locus-specific mutation rates, studying recombination rates, and studying historical variations of these quantities.
\end{itemize}

During the development of this work, several other methodologies related to demographic inference were published. We here provide a brief overview of their advantages and limitations. Methods for demographic inference available when this thesis work begun (reviewed in \cite{pool2010population}) often relied on the simplifying but limiting assumption of unlinked genetic markers, or modeled the linkage induced by the lack of historical recombination using measures of local correlation such as linkage disequilibrium. In the following years, sustained by the increasingly dense genomic datasets, several haplotype-based methods were proposed. In \cite{pool2009inference}, subsequently extended by \cite{gravel2012population}, an approach based on the frequency and length of migrant tracts was proposed for the inference of migration rates. While effectively recovering migration rates in several demographic scenarios, however, these methods do not model population size fluctuations, and are dependent on the possibility of reliably performing ancestry deconvolution to assign chromosomal tracts to a set of reference populations. These populations may not be available and, more importantly, need to be substantially divergent to attain high-quality deconvolution, as shown in our analysis. Although whole sequence datasets and methodological developments may improve the performance of deconvolution methods, this limitation may prevent methods based on migrant tracts from being effectively employed in the reconstruction of fine-scale migration patterns of the recent millennia. Methods based on ancestry deconvolution, however, may in some scenarios be used in concert with methods based on IBD sharing. Knowing whether an IBD tract was co-inherited from a specific population, in fact, may provide information on the directionality of migration, and also offer further insight into deeper time scales, as shown in \cite{velez2012impact} and \cite{campbell2012north} and discussed in Chapter \ref{chap:IBD_in_data}. These methods may be further explored in light of the recently developed analytical model for migrant tracts and the presented model for IBD. While the methods based on IBD detection, presented in this work, provide some advantages over ancestry deconvolution, it should be noted that these involve dealing with increasingly complex demographic models, and because the ancestors of IBD segments tend to be more remote than those of migrant tracts, larger sample sizes may be required for this analysis.

As mentioned in the Introduction (Section \ref{subsec:intro:approx}), a recently developed Markovian approximation of the coalescent process (Sequentially Markovian Coalescent, SMC, \cite{mcvean2005approximating}) resulted in the development of several new population genetics methods, including methods for inferring demographic history. We note that because we relied on definition (c) of Section \ref{intro:subsec:IBD_def}, the methods described in this thesis are also intrinsically linked to the SMC framework, as IBD segments are defined as being delimited by any recombination event. Future developments include dealing with the potential discrepancies of this definition and the output of IBD detection algorithms, potentially incorporating these and other calculations in new methods for IBD discovery. In \cite{li2011inference}, a method based on the SMC model studied population size fluctuations affecting human populations following the out-of-Africa migrations. This approach was limited to pairs of phased whole-sequence haploid individuals from different populations, or single individuals from a population, resulting in limited power for the inference of recent demographic events. Recently, these methods have been extended to allow analyzing multiple phased individuals simultaneously, using a composite likelihood approach \cite{steinrucken2012sequentially,sheehan2013estimating}, thus gaining insight into more recent demographic events. Using similar techniques, a recently published pre-print relies on approximated extensions of the SMC to the analysis of multiple individuals, providing insight into the ancestral recombination graph of groups of individuals \cite{rasmussen2013genome}. These methods are extremely appealing, as they make use of almost all available genomic information, but may be suffering from computational limitations. Exploiting the Markovian properties of the SMC model typically leads to algorithms that scan all $s$ sites for all pairs of $n$ samples resulting in at least $O(n^2s)$ complexity. When the analysis of thousands of samples across millions of markers is required, these methods scale poorly compared to methods that rely on summary statistics that can be obtained through methods that are sub-quadratic in the number of samples and may not require analyzing all available genomic markers. A method recently proposed in \cite{harris2013inferring} relies on summary statistics of IBS tract length, decreasing the computational burden but still requiring full pair-wise analysis of sequences to extract summary statistics, also requiring $O(n^2s)$ computation. In addition to these recently developed methods for inferring recent demographic history, a work developed in parallel to the methods described in Chapter \ref{chap:migration}, \cite{ralph2013geography}, infers historical demographic changes from length distributions of IBD segments, using the principles described in Chapter \ref{chap:IBDmodel} within a less parametric approach to the inference of coalescent distributions across different populations, thereby allowing increased flexibility compared to the model-based inference procedure of Chapter \ref{chap:IBDmodel}, but without providing explicit inference of migration and population size changes, described in Chapter \ref{chap:migration}.

The models that we proposed in this thesis assume selective neutrality. Although the distribution of haplotype sharing is likely to be affected by localized natural selection \cite{bamshad2003signatures}, the extent to which the human genome has been shaped by selective forces has yet to be quantified \cite{hernandez2011classic}. The proposed model of IBD sharing can be locally used to test deviations from neutrality and can be improved to explicitly handle the presence of selective forces. Further enhancements of the proposed methodology include improved approaches to demographic model optimization and selection, which may lead to automatic clustering of analyzed individuals into subpopulations. Furthermore, as described in Chapter \ref{chap:mutation} the proposed framework allows studying basic biological parameters such as mutation and recombination rates, potentially providing insights into questions regarding locus- or population-specific differences, and historical variation of these quantities (\cite{scally2012revising,coop2007evolutionary})

The proposed methodology facilitates tackling questions beyond demographic inference from genotype data; such questions include those that arise when phenotype data are also considered. A problem that has recently received much attention is that of estimating heritability with the use of large samples of unrelated individuals. Haplotype sharing across purportedly unrelated individuals has been used in this context \cite{zuk2012mystery,zaitlen2013using,price2011single}, and the proposed model for IBD sharing across unrelated samples can be used for improving such analysis.

On the applied side, genome-wide association studies have taught us the lesson of needing to know the demographic makeup of a study population. Although linear-trend analysis has been shown to capture population stratification when common genomic variants are considered \cite{price2006principal}, methods for association of rare variants are an active field of investigation \cite{li2008methods,madsen2009groupwise,price2010pooled} in which recent stratification poses new challenges \cite{mathieson2012differential}. The reconstruction of a fine-grained picture of population stratification thus gains importance in the context of full sequence data. Stratification might in fact occur at different historical timescales, and statistical indicators designed to account for ancient diversification trends might not reveal signatures of recent demographic events.

The reported analysis of HapMap's MKK samples provides an example of this phenomenon. This sample exhibits high levels of endogamy through ubiquitous shared long-range haplotypes, suggesting a small population size, but it appears to have an outbred profile when the decay of LD is analyzed \cite{mcevoy2011human}. As discussed in Chapter \ref{chap:IBDmodel}, a plausible reason for the observed data might in this case be found in the societal structure of the MKK people. We hypothesize that this ``village effect'' will be established in other modern populations that are commonly considered outbred on the basis of their ancient-timescale characteristics. Several genetic surveys have in fact outlined surprisingly high levels of runs of homozygosity in a number of outbred populations worldwide \cite{henn2011hunter,henn2011hunter,broman1999long,gibson2006extended}. When migration events are included in the model, long runs of homozygous haplotypes in otherwise outbred populations are plausibly interpreted as reflecting a genetic pool of several small demes that slowly but constantly intermix. The ability to reconstruct recent demographic events will enable the analysis of these phenomena. Combined with prior knowledge of a population's history, this analysis will provide a useful tool for describing the fine-grained evolutionary context in which recent genetic variation arose.

\appendix

\addcontentsline{toc}{chapter}{Bibliography}
\bibliography{LATEX_SOURCE/refs}
\bibliographystyle{named} 

\end{document}